\documentclass[a4paper,10pt]{elsarticle}
\usepackage{jcomp}
\usepackage{framed,multirow}
\usepackage[utf8]{inputenc}
\usepackage{bm}
\usepackage{amsmath}
\usepackage{mathrsfs}
\usepackage{graphicx}
\usepackage{epsfig, setspace}
\usepackage{url}
\usepackage{graphicx, transparent, color}
\usepackage{algorithm} 
\usepackage{algorithmicx}
\usepackage{etex}
\usepackage[bookmarks=true]{hyperref}
\usepackage{amsmath}
\usepackage{xcolor, soul}
\usepackage{rotating} 
\usepackage{pdflscape} 
\usepackage{silence}
\usepackage{ulem,cancel}
\newtheorem{remark}{\bf Remark}[section]
\newtheorem{example}{Example}[section]
\usepackage{graphicx}
\usepackage{float}
\usepackage{subfigure}
\usepackage{subfigmat}

\usepackage{tikz}
\usepackage{capt-of}
\usepackage{setspace,lipsum}

\usepackage{setspace,lipsum}
\usepackage{listings}
\usepackage{xcolor}
\definecolor{aquamarine}{rgb}{0.5, 1.0, 0.83}
\definecolor{OliveGreen}{rgb}{0,0.6,0}

\sethlcolor{aquamarine}

\definecolor{codegreen}{rgb}{0,0.6,0}
\definecolor{codegray}{rgb}{0.5,0.5,0.5}
\definecolor{codepurple}{rgb}{0.58,0,0.82}
\definecolor{backcolour}{rgb}{0.95,0.95,0.92}

\lstdefinestyle{mystyle}{
    backgroundcolor=\color{backcolour},   
    commentstyle=\color{codegreen},
    keywordstyle=\color{magenta},
    numberstyle=\tiny\color{codegray},
    stringstyle=\color{codepurple},
    basicstyle=\ttfamily\footnotesize,
    breakatwhitespace=false,         
    breaklines=true,                 
    captionpos=b,                    
    keepspaces=true,                 
    numbers=left,                    
    numbersep=5pt,                  
    showspaces=false,                
    showstringspaces=false,
    showtabs=false,                  
    tabsize=2
}
\begin{document}

\begin{frontmatter}
\title{\textcolor{black}{Efficient high-order Gradient-based Reconstruction for compressible flows}}

\author{Amareshwara Sainadh Chamarthi \cortext[cor1]{Corresponding author. \\ 
E-mail address: skywayman9@gmail.com (Amareshwara Sainadh  Ch.)}}
\address{Faculty of Mechanical Engineering, Technion - Israel Institute of Technology, Haifa, Israel}

\begin{abstract}
This paper extends the gradient-based reconstruction approach of Chamarthi \cite{chamarthi2023gradient} to genuine high-order accuracy \textcolor{black}{for inviscid test cases involving smooth flows}. A seventh-order \textcolor{black}{accurate scheme} is derived using the \textcolor{black}{the same stencil as of the explicit fourth-order scheme proposed in Ref. \cite{chamarthi2023gradient}, which also has low dissipation properties.} \textcolor{black}{The proposed method is seventh-order accurate under} the assumption that the variables at the \textit{cell centres are point values}. A problem-independent discontinuity detector is used to obtain high-order accuracy. Accordingly, primitive or conservative variable reconstruction is performed around regions of discontinuities, whereas smooth solution regions apply flux reconstruction. The proposed approach can still share the derivatives between the inviscid and viscous fluxes, which is the main idea behind the gradient-based reconstruction.		 Several standard benchmark test cases are presented. \textcolor{black}{The proposed method is more efficient than the seventh-order weighted compact nonlinear scheme (WCNS) for the test cases considered in this paper}.
\end{abstract}

\begin{keyword}
Gradient based reconstruction, Shock-capturing, Compressible flows, \textcolor{black}{High-order}
\end{keyword}

\end{frontmatter}
\section{Introduction}
The physical behaviour of the viscous, compressible flow considered in this work is mathematically modelled by the Navier Stokes equations \cite{pletcher2012computational}. The two-dimensional formulation in the Cartesian coordinate system can be expressed as:

\begin{equation}\label{CNS-base}
\frac{\partial \mathbf{Q}}{\partial t}+\frac{\partial \mathbf{F^c}}{\partial x}+\frac{\partial \mathbf{G^c}}{\partial y}+\frac{\partial \mathbf{F^v}}{\partial x}+\frac{\partial \mathbf{G^v}}{\partial y}=0,
\end{equation}
where  $\textbf{Q}$ is the conservative variable vector $(\rho, \rho u, \rho v, E)$, where $\rho$ is the density and $u$ and $v$ are velocity components in the $x-$ and $y-$ directions, respectively. \textcolor{black}{The fluid's total energy per unit volume is given as $E = \rho (e + (u^2+v^2)/2)$}, where $e$ is the specific internal energy.  The system is closed using the ideal gas equation of state relating the thermodynamic pressure $p$ and the total energy given by

\begin{equation}\label{eqn:pressure}
p = (\gamma -1) (E - \rho \frac{(u^2+v^2)}{2}),
\end{equation}
where $\gamma$ is the ratio of specific heats, this paper takes the value of $\gamma$ as 1.4. The fluxes, $\mathbf{F}^c$, $\mathbf{G}^c$, and $\mathbf{F}^v$, $\mathbf{G}^v$, are the convective (superscript $c$) and viscous (superscript $v$) flux vectors in each coordinate direction and are defined as,
\begin{equation}
\mathbf{F^c}=\left(\begin{array}{c}
\rho u \\
\rho u^{2}+p \\
\rho u v \\
\rho u w \\
u(E+p)
\end{array}\right), 
\mathbf{G^c}=\left(\begin{array}{c}
\rho v \\
\rho u v \\
\rho v^{2}+p \\
\rho v w \\
v(E+p)
\end{array}\right),
\mathbf{F^v}=\left(\begin{array}{c}
0 \\
-\tau_{x x} \\
-\tau_{y x} \\
- \phi_x+q_{x}
\end{array}\right),\; 
\mathbf{G^v}=\left(\begin{array}{c}
0 \\
-\tau_{x y} \\
-\tau_{y y} \\
-\phi_y+q_{y}
\end{array}\right),
\end{equation}
where $\phi_x = \tau_{x x} u+\tau_{x y} v$ and $\phi_y = \tau_{y x} u+\tau_{y y} v$. The components of the viscous stress tensor $\tau$ and the heat flux $q$ are as follows:

\begin{equation}\label{eqn:visc}
\begin{aligned}
\tau_{x x}=\frac{2}{3} \mu\left(2 \frac{\partial u}{\partial x}-\frac{\partial v}{\partial y}\right),& \quad \tau_{x y}=\tau_{y x}=\mu\left(\frac{\partial u}{\partial y}+\frac{\partial v}{\partial x}\right), \quad \tau_{y y}=\frac{2}{3} \mu\left(2 \frac{\partial v}{\partial y}-\frac{\partial u}{\partial x}\right), \\
\\
q_{x} &=-k \frac{\partial T}{\partial x}, \quad q_{y} =-k \frac{\partial T}{\partial y}, \quad T=\frac{\gamma p}{(\gamma-1) C_{p} \rho},
\end{aligned}
\end{equation}
where $\mu$ is the dynamic viscosity, $T$ is the temperature, and the heat conductivity is computed by $\kappa$ = $C_p\mu/Pr$, $Pr$ is the Prandtl number, and $C_p$ is the specific heat at constant pressure. 

\begin{figure}[H]
    \centering
    \includegraphics[width=0.35\textwidth]{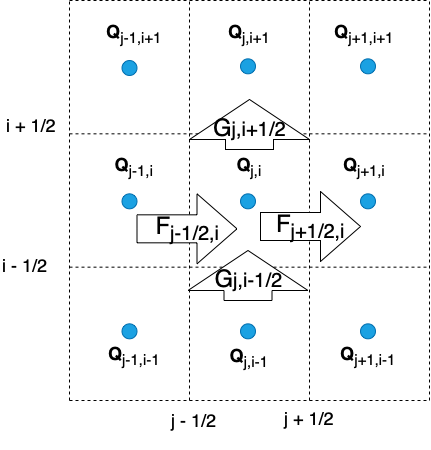}
    \caption{Schematic representation of the computational grid in two dimensions with cell centre points (blue) and relevant fluxes for the cell of interest. }
    \label{fig:grid}
\end{figure}

The Navier Stokes equations (\ref{CNS-base}) are discretized using a conservative numerical method on a uniform Cartesian grid of cell sizes $\Delta x$ and $\Delta y$. The domain of interest is discretized by non-overlapping cells, and the schematic representation of the computational grid is shown in Fig. \ref{fig:grid}. The conservative variables $\mathbf{ Q}$ are stored at the center of cell $I_{j,i} = [x_{j-\frac{1}{2}}, x_{j+\frac{1}{2}}] \times [y_{i-\frac{1}{2}}, y_{i+\frac{1}{2}}]$ and the indices $i$ and $j$ denote the $j-$th cell in $x-$ direction and $i-$th cell in $y-$ direction. The cell interfaces in $x-$ direction are denoted by ${j+\frac{1}{2}}$ and by ${i+\frac{1}{2}}$ in $y-$ direction, respectively.
 
 The time evolution of the cell-centered conservative variables $\mathbf{Q_{j, i}}$ is given by the following semi-discrete equation:
 
 \begin{equation}\label{eqn-differencing_residual}
\frac{\mathrm{d}}{\mathrm{dt}} {\mathbf{ Q}}_{j,i} = \mathbf{Res}_{j,i} = - \frac{d \mathbf F}{dx}_{j,i} - \frac{d \mathbf G}{dy}_{j,i},	
 \end{equation}
where $\frac{d \mathbf F}{dx}_{j,i}$ and $\frac{d \mathbf G}{dy}_{j,i}$ are approximations to the flux derivatives at the cell center, and we seek their high-order approximations in the conservative form:


\begin{equation}\label{deriv}
\begin{aligned}
	\frac{d \mathbf F}{dx}_{j,i}  = \left[ (\mathbf {\hat{F}^c}\textcolor{black}{+}\mathbf {\hat{F}^v})_{j+\frac{1}{2},i} -  (\mathbf {\hat{F}^c}\textcolor{black}{+}\mathbf {\hat{F}^v})_{j-\frac{1}{2},i} \right]\frac{1}{\Delta x}, \quad \frac{d \mathbf G}{dy}_{j,i}  = \left[ (\mathbf {\hat{G}^c}\textcolor{black}{+}\mathbf {\hat{G}^v})_{j,i+\frac{1}{2}} - (\mathbf {\hat{G}^c}\textcolor{black}{+}\mathbf {\hat{G}^v})_{j,i-\frac{1}{2}} \right]\frac{1}{\Delta y},	
\end{aligned}
\end{equation}
where $\Delta x=x_{j+1 / 2}-x_{j-1 / 2}$, $\Delta y=y_{i+1 / 2}-y_{i-1 / 2}$, $\mathbf {\hat{F}^c}$, $\mathbf {\hat{G}^c}$ and $\mathbf {\hat{F}^v}$, $\mathbf {\hat{G}^v}$ are the numerical approximation of the convective and viscous fluxes in the $x-$, and $y-$ directions, respectively. The viscous fluxes are discretized using the $\alpha$-damping approach \textcolor{black}{\cite{chamarthi2023gradient,nishikawa:AIAA2010,chamarthi2022}}, and the details are presented later. The spatial discretization of the convective fluxes, $\mathbf {\hat{F}^c}$, in Equation (\ref{eqn-differencing_residual}) are determined by a Riemann solver \cite{ivings1998riemann,roe1981approximate, einfeldt1988godunov, toro1994restoration, osher1982upwind} which can be written in canonical form as follows:

\begin{equation}
\mathbf {\hat{F}^c}_{j+ 1 / 2}=\mathbf{F}^{\rm Riemann}_{j+\frac{1}{2}}
= \frac{1}{2}
\left[
\textcolor{black}{\mathbf{F}^{c}_{L}}
+ 
\textcolor{black}{\mathbf{F}^{c}_{R}}
\right]
-
 \frac{1}{2} | {\mathbf{A}_{j+\frac{1}{2}}}|({\mathbf{Q}^R_{j+\frac{1}{2}}}-{\mathbf{Q}^L_{j+\frac{1}{2}}}),
\label{eqn:Riemann}
\end{equation}
where ``L'' and ``R'' denotes the reconstructed or interpolated states from the left and right side of a cell interface, respectively, and $|{\mathbf{A}_{j+\frac{1}{2}}}|$ denotes the inviscid Jacobian of the Euler equations. In this paper, the HLLC Riemann solver  \cite{toro1994restoration} is used for the flux evaluation.

\textcolor{black}{The objective of the paper is to obtain the values of the fluxes, $\mathbf{F}^{c}_{L}$ and $\mathbf{F}^{c}_{R}$ and the conservative variables, $\mathbf{Q}_{j+\frac{1}{2}}^{L}$ and $\mathbf{Q}_{j+\frac{1}{2}}^{R}$, at the left and right interfaces of the cell interfaces, $x_{j+\frac{1}{2}}$, $\forall j \in \{ 0, \: 1, \: 2, \: \dots, \: N \}$, in the Equation (\ref{eqn:Riemann}) in a novel way.}

\textcolor{black}{The novel approach has two components. The first one is regarding the computation of $\mathbf{Q}_{j+\frac{1}{2}}^{L}$ and $\mathbf{Q}_{j+\frac{1}{2}}^{R}$. This procedure uses the values of the conservative variable at the cell centers and their gradients (derivatives of the variables of interest) as in Ref. \cite{chamarthi2023gradient}}. \textcolor{black}{The variables of interest can be not only conservative variables ($\mathbf Q$) but also primitive variables ($\mathbf U$ = ($\rho, u, v, p)$) at the cell centers themselves. If primitive variables are used for reconstruction, then the values of $\mathbf{Q}_{j+\frac{1}{2}}^{L}$ and $\mathbf{Q}_{j+\frac{1}{2}}^{R}$ are obtained from $\mathbf{U}_{j+\frac{1}{2}}^{L}$ and $\mathbf{U}_{j+\frac{1}{2}}^{R}$, respectively.} \\

The second component of the procedure is regarding the computation of $\mathbf{F}^{c}_{L}$ and $\mathbf{F}^{c}_{R}$. These values can be obtained as a function of the conservative variables, denoted as $\mathbf{F(Q_L)}$ for the left interface. As a function of the primitive variables denoted as $ \mathbf{F(U_L)}$ for the left interface. Or as a function of fluxes at the cell centers themselves, denoted as $ \mathbf{F(f_L)}$ \textcolor{black}{(where $\mathbf{f_L}$ is the value of the convective flux $F^c$ reconstructed at the interface from the left using $F^c$ evaluated at cell centers, $f_{j-1}$, $f_j$, $f_{j+1}$, etc. Further details will be given in Section \ref{subsec:high}).} This entire procedure will lead to a seventh-order accurate scheme when differenced over the cell, and the details will be presented in the later sections.

\subsection{Literature review}

There are two popular ways to obtain the interface values, interpolation and reconstruction \cite{Shu2009}. The author has used both reconstruction \cite{chamarthi2018high}\textcolor{black}{,} and interpolation polynomials \cite{chamarthi2019first} in the past and know the differences, advantages\textcolor{black}{,} and disadvantages of both these methods. MUSCL scheme of van Leer is a third\textcolor{black}{-}order reconstruction polynomial \cite{van1977towards,van2021towards} and can be expressed in Legendre basis. Balsara et al. have presented polynomials for greater than third order, also in the Legendre basis \cite{balsara2016efficient}. Detailed derivations of these polynomials can be found in \cite{Shu1997}. These reconstruction polynomials are sometimes called essentially non-oscillatory (ENO) reconstructions, which were first developed by Harten et al. \cite{harten1987uniformly}, according to chapter 9 of Ref. \cite{laney1998computational}. For discontinuous flows, there has been a great deal of interest in ENO and WENO (Weighted ENO) approaches as they can obtain high-order accuracy (greater than second-order) in the vicinity of smooth flows. The variables that can be reconstructed at the interface can be conservative variables, as in the original ENO schemes of Harten et al. \cite{harten1987uniformly}, or the fluxes themselves, as in \cite{Shu1988}. The key idea of the WENO scheme is to express a high-order polynomial as a combination of multiple low-order polynomials. Jiang and Shu have expressed the fifth-order polynomial as a combination of three third-order polynomials, WENO-JS \cite{Jiang1995}. Smoothness indicators based on gradients are proposed to detect discontinuous and smooth flow regions. Despite the enormous success of this scheme, several improvements have been proposed in the literature. WENO-M scheme of Henrick et al. \cite{Henrick2005} improved the order of accuracy at critical points using a mapping procedure. WENO-Z Borges et al. \cite{Borges2008} have proposed a global smoothness indicator that improved the WENO scheme's accuracy and the overall resolution characteristics. Hu and Adams \cite{Hu2010} further improved the approach by proposing a central-upwind WENO scheme that can capture shocks and small-scale turbulent flow features. There are many contributions with incremental improvements of the WENO methodology in the literature, and it is increasingly hard to follow their development \cite{martin2006bandwidth, Fan2014a,fu2018new}. These reconstruction polynomials can also be implicit in space with superior spectral properties but require a tridiagonal matrix inversion \cite{Pirozzoli2002,chamarthi2021high, Ghosh2014}. Apart from WENO, The fifth-order monotonicity-preserving (MP) scheme proposed by Suresh and Hyunh \cite{suresh1997accurate} also can effectively suppress numerical oscillations across discontinuities and preserve accuracy. WENO schemes of extreme high-order often lack robustness and are typically combined with the monotonicity preserving approach of Suresh and Hyunh \cite{balsara2000monotonicity,li2021low,zhao2019general}.

On the other hand, interpolation schemes are based on Lagrange polynomials. Lele presented explicit and implicit interpolation polynomials \cite{lele1992compact}. The corresponding shock capturing schemes that use interpolation polynomials are called Weighted compact nonlinear schemes (WCNS), which are presented first by \cite{Deng1997, Deng2000}. WCNS is also similar to that of the WENO scheme as it uses the multi-stencil weighting technique. The key difference between the interpolation and reconstruction polynomials is that the interpolation polynomials use high-order accurate flux derivatives instead of Equation (\ref{deriv}), readers can refer \cite{Deng2000,Nonomura2013,nagarajan2003robust,deng2015family}. Several authors have proposed high-order variations of the WCNS schemes \textcolor{black}{\cite{Nonomura2013,Zhang2008,Wong2017,Nonomura2009}}. The main advantage of the WCNS approach over WENO is that high-order accuracy can be achieved even with primitive and conservative variable interpolation. \textcolor{black}{On the other hand, the} reconstruction polynomials require expensive integral evaluation over the cell interface, which includes multiple reconstructed values, to maintain the formal order of the scheme \cite{titarev2004finite}. For three-dimensional simulations, the WENO approach with primitive variable reconstruction can be 3-20 times more expensive than the WCNS approach \cite{Nonomura2012}. Liao and He \cite{liao2020high} have proposed adapter schemes that can convert reconstruction and interpolation schemes into each other. Nevertheless, Reconstruction schemes are more widely used than interpolation schemes.

\subsection{Novelty of the present paper}
Unlike the interpolation and reconstruction polynomials, Chamarthi \cite{chamarthi2023gradient} has proposed a slightly different approach for obtaining the interface values called \textcolor{black}{Gradient-based Reconstruction (GBR)}. The key advantage of using gradients for reconstruction is that they can be used in viscous flux computations, shock capturing, computing turbulence statistics like enstrophy, etc., which makes the scheme efficient. Gradient based-reconstruction is similar to that of the $FV^2S$ scheme proposed by Sengupta et al. \cite{sengupta2005new} in 2005. Readers can also refer chapter 12 of \cite{sengupta2013high} regarding the $FV^2S$ scheme. Sengupta et al. have used upwind compact finite difference schemes to compute the first derivatives and, therefore, may not be reused for the viscous fluxes. \textcolor{black}{For one-dimensional scenario, GBR method employ the first two moments of the Legendre polynomial \cite{van1977towards} evaluated on $x_{i-\frac{1}{2}} \leq x \leq x_{i+\frac{1}{2}}$ for interpolation. This may be written as (assuming primitive variable reconstruction):}

\begin{equation}\label{eqn:legendre}
\mathbf{U}(x)=\mathbf{{U}}_{j}+\frac{1}{\textcolor{black}{\Delta x}}\left(x-x_{j}\right)\left(\frac{\partial \mathbf{U}}{\partial x}\right)_{j}+\frac{3\kappa}{2 \textcolor{black}{\Delta x^{2}}}\left(\frac{\partial^2 \mathbf{U}}{\partial x^2}\right)_{j} \left[\left(x-x_{j}\right)^{2}-\frac{\Delta \textcolor{black}{x^{2}}}{12}\right],
\end{equation}
where $\mathbf{{U}}_{j}$ is the primitive variable vector and $\left(\frac{\partial \mathbf{U}}{\partial x}\right)_{j}$, $\left(\frac{\partial^2 \mathbf{U}}{\partial x^2}\right)_{j}$ are the corresponding first- and second-derivatives within cell $j$.  As we seek only the cell-interface values, By setting $x = x_j \pm \Delta x/2$ and within a cell $j$ and $\kappa=\frac{1}{3}$ the following expression results:
\begin{equation}\label{eqn:3linear}
\begin{aligned}
\mathbf{U}_{j+ \frac{1}{2}}^{L} &=\mathbf{{U}}_{j}+\frac{\Delta x}{2} \left(\frac{\partial \mathbf{U}}{\partial x}\right)_{j}+\frac{\Delta x^2}{12} \left(\frac{\partial^2 \mathbf{U}}{\partial x^2}\right)_{j} \\
\mathbf{U}_{j+ \frac{1}{2}}^{R} &=\mathbf{{U}}_{j+1}-\frac{\Delta x}{2}  \left(\frac{\partial \mathbf{U}}{\partial x}\right)_{j+1}+\frac{\Delta x^2}{12}  \left(\frac{\partial^2 \mathbf{U}}{\partial x^2}\right)_{j+1} \quad 
\end{aligned}
\end{equation}

 \textcolor{black}{By substituting higher-order difference formulas (sixth-order explicit and fourth-order optimized implicit \cite{lele1992compact} schemes), Chamarthi has obtained fourth-order schemes in Ref. \cite{chamarthi2023gradient}. Ref. \cite{chamarthi2023gradient} is inspired by the work of Van Leer \cite{van1977towards} and the recent paper by Nishikawa \cite{nishikawa2018green}. In this paper, the approach is generalized and improved as follows:}

\begin{enumerate}
\item Two parameter gradient-based reconstruction has been proposed that leads to seventh-order accuracy. The fourth-order scheme presented in \cite{chamarthi2023gradient} can be considered a subset of the current approach.
\item For nonlinear test cases, the proposed scheme in \cite{chamarthi2023gradient} is only second-order accurate. A problem-independent discontinuity detector is used to obtain high-order accuracy. Accordingly, primitive or conservative variable reconstruction is performed around regions of discontinuities, whereas smooth solution regions apply flux reconstruction.
\end{enumerate}

The rest of the paper is organized as follows. The discretization of viscous fluxes and the seventh-order gradient-based reconstruction for the convective fluxes are presented in Section \ref{sec:num}. Several one- and two-dimensional test cases for Euler and Navier-Stokes equations are presented in Section \ref{sec:results}, and the  Section \ref{sec:finish} presents the conclusions.

\section{Numerical discretization}\label{sec:num}

In this section, the details of the numerical discretization are presented. In subsection \ref{subsec:visc} the viscous flux discretization is presented. In subsections \ref{subsec:conv} and \ref{subsec:high} the seventh-order gradient-based reconstruction and the algorithm to obtain high-order accuracy are presented. Finally, temporal discretization is discussed in subsection \ref{subsec:time}.

\subsection{Spatial discretization of viscous fluxes}\label{subsec:visc}
In this subsection, the spatial discretization of viscous fluxes is presented. For simplicity, a one-dimensional scenario is considered as extending it to multi-dimensional problems is straightforward via a dimension-by-dimension approach. The viscous flux at the interface in the one-dimensional scenario can be expressed as follows:
\begin{equation}
\mathbf{\hat F^v}_{j+\frac{1}{2}}=\left[\begin{array}{c}
0 \\
-\tau_{j+\frac{1}{2}} \\
-\tau_{j+\frac{1}{2}} u_{j+\frac{1}{2}}+q_{j+\frac{1}{2}}
\end{array}\right],
\end{equation}
where
\begin{equation}\label{eqn:visc-interface}
\tau_{j+1 / 2}=\frac{4}{3} \mu_{j+1 / 2}\left(\frac{\partial u}{\partial x}\right)_{j+1 / 2}, \quad q_{j+1 / 2}=-\frac{\mu_{j+1 / 2}}{\operatorname{Pr}(\gamma-1)}\left(\frac{\partial T}{\partial x}\right)_{j+1 / 2}.
\end{equation}

To evaluate the viscous fluxes at the cell interfaces ${j+\frac{1}{2}}$, the velocity and temperature gradients at the interface in the Equation (\ref{eqn:visc-interface}), $\left(\frac{\partial u}{\partial x}\right)_{j+1 / 2}$ and $\left(\frac{\partial T}{\partial x}\right)_{j+1 / 2}$, should be computed. These gradients are computed using the $\alpha$-damping approach of Nishikawa \cite{nishikawa:AIAA2010}. It has been shown recently by Chamarthi et al. \cite{chamarthi2022} that the $\alpha$-damping approach will prevent odd-even decoupling and improve the solution quality of the simulations. The interface gradients, for velocity, are evaluated as follows:
\begin{equation}\label{eqn:alphad}
\begin{aligned}
\left(\frac{\partial u}{\partial x}\right)_{j+1 / 2}&=\frac{1}{2}\left( \left(\frac{\partial u}{\partial x}\right)_{j}+ \left(\frac{\partial u}{\partial x}\right)_{j+1}\right)+\frac{\alpha_v}{2 \Delta x}\left( u_{j+1}-\frac{\Delta x}{2}  \left(\frac{\partial u}{\partial x}\right)_{j+1}- u_{j}-\frac{\Delta x}{2}  \left(\frac{\partial u}{\partial x}\right)_{j}\right),
\end{aligned}
\end{equation}
where $\alpha_v$ is taken as 4 and the velocity gradients, $\left(\frac{\partial {u}}{\partial x}\right)_{j}$, at the cell-centers are computed by the following sixth-order finite-difference formula:

\begin{eqnarray}\label{sixth-ordergrad}
\left(\frac{\partial u}{\partial x}\right)_{j} 
 =
 \frac{  1 }{60} \left[ 
         45  \frac{  u_{j+1} - u_{j-1}   }{\Delta x}
        -  9  \frac{  u_{j+2} - u_{j-2}   }{\Delta x}
     +   {\color{black}   \frac{  u_{j+3} - u_{j-3}   }{\Delta x} }
 \right].
 \end{eqnarray}
 
 It has been shown that using the above sixth-order formula, Equation (\ref{sixth-ordergrad}) and for $\alpha_v$ = 4, will lead to a fourth-order viscous scheme which has superior spectral properties than the sixth-order $\alpha$-damping scheme presented in \cite{chamarthi2022}. The temperature gradients at the cell interface, $\left(\frac{\partial T}{\partial x}\right)_{j+1 / 2}$, are also computed similar to the velocity gradients using the $\alpha$-damping approach. The parameter $\alpha_v$ is not only a key parameter associated with the damping property of a viscous scheme but also for viscous time step computations; See \cite{nishikawa:AIAA2010,chamarthi2022} and temporal discretization in subsection \ref{subsec:time}. As explained in the introduction, the main advantage of the gradient-based reconstruction is the reuse of gradients, and these velocity gradients, computed by Equation (\ref{sixth-ordergrad}), will be reused in the convective flux discretization, which is presented next.
 \begin{remark}
 \normalfont \textcolor{black}{It is important to note that the viscous flux discretization considered in this paper is not ''truly'' high-order accurate non-linear test cases. It is also true for the viscous flux discretization presented in \cite{chamarthi2023gradient,chamarthi2022,chamarthi2022gradient}. For the test cases considered in this paper, the value of $\mu_{j+\frac{1}{2}}$ is constant and is not spatially varying. If the value $\mu_{j+\frac{1}{2}}$ is computed by Sutherland's law (as a function of temperature), then the proposed approach is only second-order accurate. However, the readers need to note that the proposed $\alpha$-damping scheme still gave superior results over a truly high-order accurate scheme with poor spectral properties. Readers can refer to such results presented in Ref. \cite{chamarthi2023gradient,chamarthi2022,chamarthi2022gradient}.}
\end{remark}

\subsection{Spatial discretization of convective fluxes}\label{subsec:conv}
Here, we present the spatial discretization of convective fluxes using the gradient-based reconstruction approach. For clarity, \textcolor{black}{we once again write the the first two moments of the Legendre polynomial \cite{van1977towards}, Equation (\ref{eqn:legendre}), valid for $x_{j-1 / 2} \leq x \leq x_{j+1 / 2}$} 

\begin{equation}\label{eqn:legendre_gen}
\mathbf{\Phi}(x)=\mathbf{{\Phi}}_{j}+\frac{1}{\textcolor{black}{\Delta x}}\left(x-x_{j}\right)\left(\frac{\partial \mathbf{\Phi}}{\partial x}\right)_{j}+\frac{3\kappa}{2 \textcolor{black}{\Delta x^{2}}}\left(\frac{\partial^2 \mathbf{\Phi}}{\partial x^2}\right)_{j} \left[\left(x-x_{j}\right)^{2}-\frac{\Delta \textcolor{black}{x^{2}}}{12}\right], 
\end{equation}
where $\mathbf{{\Phi}}_{j}$ is the \textit{point value} at the cell-centers and $\left(\frac{\partial \mathbf{\Phi}}{\partial x}\right)_{j}$, $\left(\frac{\partial^2 \mathbf{\Phi}}{\partial x^2}\right)_{j}$ are the first- and second-derivatives within cell $j$ and  by setting $x = x_j \pm \Delta x/2$ and within a cell $j$ and $\kappa=\frac{1}{3}$ one obtains the following expressions
\begin{equation}\label{eqn:lingrb}
\begin{aligned}
\mathbf{\Phi}_{j+ \frac{1}{2}}^{L} &=\mathbf{{\Phi}}_{j}+\frac{\Delta x}{2} \left(\frac{\partial \mathbf{\Phi}}{\partial x}\right)_{j}+\frac{\Delta x^2}{12} \left(\frac{\partial^2 \mathbf{\Phi}}{\partial x^2}\right)_{j}, \\
\mathbf{\Phi}_{j+ \frac{1}{2}}^{R} &=\mathbf{{\Phi}}_{j+1}-\frac{\Delta x}{2}  \left(\frac{\partial \mathbf{\Phi}}{\partial x}\right)_{j+1}+\frac{\Delta x^2}{12}  \left(\frac{\partial^2 \mathbf{\Phi}}{\partial x^2}\right)_{j+1} \quad.
\end{aligned}
\end{equation}
The first-order derivatives, $\left(\frac{\partial \mathbf{\Phi}}{\partial x}\right)_{j}$, in Equation (\ref{eqn:3linear}) are computed by using following the sixth-order finite difference formula

\begin{eqnarray}\label{eqn:six}
\left(\frac{\partial \mathbf{\Phi}}{\partial x}\right)_{j} 
 =
 \frac{  1 }{60} \left[ 
         45  \frac{  \mathbf{\Phi}_{j+1} - \mathbf{\Phi}_{j-1}   }{\Delta x}
        -  9  \frac{  \mathbf{\Phi}_{j+2} - \mathbf{\Phi}_{j-2}   }{\Delta x}
     +   {\color{black}   \frac{  \mathbf{\Phi}_{j+3} - \mathbf{\Phi}_{j-3}   }{\Delta x} }
 \right],
 \end{eqnarray}
 which is essentially the same as that of Equation (\ref{sixth-ordergrad}). The variable $\mathbf{\Phi}$ can be either the conservative variable ($\mathbf Q$) or the primitive variable ($\mathbf U$) at the cell centres themselves. In the literature both conservative \cite{titarev2004finite} and primitive variables \cite{deng2019fifth} are used for reconstruction. In this paper, we consider both primitive and conservative reconstruction. Depending on the choice of the variables, corresponding gradients of the variable should be computed. For brevity, only the gradients in the $x-$ direction are shown below:
 
\begin{gather}
 \textbf{Primitive variable reconstruction}  \rightarrow \frac{\partial \rho}{\partial x}, \frac{\partial u}{\partial x}, \frac{\partial v}{\partial x}, \frac{\partial p}{\partial x}  \nonumber\\
  \textbf{Conservative variable reconstruction}  \rightarrow \frac{\partial \rho}{\partial x}, \frac{\partial \rho u}{\partial x}, \frac{\partial \rho v}{\partial x}, \frac{\partial E}{\partial x}  \nonumber .
\end{gather} 
If the primitive variables are used for reconstructing the inviscid fluxes, the necessary velocity gradients for viscous fluxes are readily available. On the other hand, if the conservative variables are used for reconstruction, then the velocity gradients can be calculated from the following relations:
\begin{equation}
\begin{aligned}
\frac{\partial u}{\partial x}&=\frac{1}{{\rho}^{2}}\left[{\rho} \frac{\partial \rho u}{\partial x}-{\rho} {u} \frac{\partial \rho }{\partial x} \right], \frac{\partial v}{\partial x}=\frac{1}{{\rho}^{2}}\left[{\rho} \frac{\partial \rho v}{\partial x}-{\rho} {v} \frac{\partial \rho }{\partial x} \right].
\end{aligned}
\end{equation}

The key modification in this paper for obtaining seventh-order accuracy using the Equation (\ref{eqn:lingrb}), along with sixth-order first derivative, is in the computation of $\left(\frac{\partial^2 \mathbf{\Phi}}{\partial x^2}\right)_{j}$. The second derivative can be explicitly written as follows:

{\begin{equation}\label{eqn:how-second}
\begin{aligned}
\left(\frac{\partial^2 \mathbf{\Phi}}{\partial x^2}\right)_{j} =  \frac{\left(\frac{\partial \mathbf{\Phi}}{\partial x}\right)_{j+1 / 2} - \left(\frac{\partial \mathbf{\Phi}}{\partial x}\right)_{j-1 / 2}} {\Delta x}
\end{aligned}
\end{equation}
Inspired by the Ref. \cite{chamarthi2022} the gradients at the interface, $\left(\frac{\partial \mathbf{\Phi}}{\partial x}\right)_{j+1 / 2}$, are written as follows
\begin{equation} \label{good_6tth-order_scheme_flux}
\begin{aligned}
  \left(\frac{\partial \mathbf{\Phi}}{\partial x}\right)_{j+1 / 2}=\frac{1}{2}\left[\left(\frac{\partial \mathbf{\Phi}}{\partial x}\right)_{j}+\left(\frac{\partial \mathbf{\Phi}}{\partial x}\right)_{j+1}\right]+\frac{\alpha}{ 2 \Delta x}\left(\mathbf{\Phi}_R- \mathbf{\Phi}_L\right), \\
\end{aligned}
\end{equation}
where $ \mathbf{\Phi}_L$ and $ \mathbf{\Phi}_R$ are given by the following equations

\begin{eqnarray}\label{alpha-beta}
 \mathbf{\Phi}_L = \mathbf{\Phi}_j + \left(   \frac{ \partial \mathbf{\Phi} }{  \partial x }   \right)_{j} \frac{\Delta x}{2} + \beta  \left({ {\mathbf{\Phi}}}_{j+1} - 2{ {\mathbf{\Phi}}}_{j} + {{\mathbf{\Phi}}}_{j-1}\right) , \quad \\
 \mathbf{\Phi}_R =  \mathbf{\Phi}_{j+1}  - \left(   \frac{ \partial \mathbf{\Phi} }{  \partial x }   \right)_{j+1} \frac{\Delta x}{2} + \beta  \left({ {\mathbf{\Phi}}}_{j+2} - 2 \textcolor{black}{{ {\mathbf{\Phi}}}}_{j+1} + { {\mathbf{\Phi}}}_{j}\right),
 \label{uLuR_damp_sixth}
\end{eqnarray}
and $\alpha$ and $\beta$ are free parameters. By using Fourier analysis \cite{lele1992compact} for the linear advection equation, which is also helpful in analyzing a scheme's dispersion and dissipative properties, one would measure or compute the order of accuracy of a scheme. By carrying out Fourier analysis, it has been found that for $\alpha=38/7$ and $\beta=3/190$, one would obtain a seventh-order upwind scheme. The seventh-order leading error is shown in the following equation
\begin{equation}
\textcolor{black}{\Delta x \left(i +\frac{\Delta x^7}{280}-\frac{i \Delta x^8}{360}-\frac{\Delta x^{9}}{560}+O\left(\Delta x^{10}\right)\right)}.
\end{equation}

Details of the Fourier analysis can be found in Ref. \textcolor{black}{\cite{chamarthi2023gradient,chamarthi2022,lele1992compact,nishikawa2018green,moin2010fundamentals}} and the MATHEMATICA workbook that is used to derive the seventh-order scheme is given in Appendix A. By substituting $\alpha=38/7$ and $\beta=3/190$ in Equation (\ref{alpha-beta}) one would obtain the following expression for the second-order derivatives which will be substituted in Equation (\ref{eqn:lingrb})

\begin{equation}\label{eqn:grb-sec}
\left(\frac{\partial^2 \mathbf{\Phi}}{\partial x^2}\right)_{j}=\frac{1}{70} (60 \mathbf{\Phi}'_{j-1}-60 \mathbf{\Phi}'_{j+1}-362 \mathbf{\Phi}_{j}+178 \mathbf{\Phi}_{j-1}+3 \mathbf{\Phi}_{j-2}+178 \mathbf{\Phi}_{j+1}+3 \mathbf{\Phi}_{j+2}),
\end{equation}
where $ \mathbf{\Phi}'_j$ = ${\Delta x}\left(\frac{\partial \mathbf{\Phi}}{\partial x}\right)_{j}$. One may choose to compute and store the second-derivatives using the Equation (\ref{eqn:grb-sec}) and substitute those values in the Equations (\ref{eqn:lingrb}) in order to obtain the upwind scheme. It is also possible to use the following expressions, which do not require computation of second-derivatives and only depend on the first-derivatives and cell-center values of variables, for reconstruction:

\begin{equation}\label{eqn:grb}
\begin{aligned}
\mathbf{\Phi}^{L,GRB} &=\frac{1}{840} (420 \mathbf{\Phi}'_j+60 \mathbf{\Phi}'_{j-1}-60 \mathbf{\Phi}'_{j+1}+478 \mathbf{\Phi}_{j}+178 \mathbf{\Phi}_{j-1}+3 \mathbf{\Phi}_{j-2}+178 \mathbf{\Phi}_{j+1}+3 \mathbf{\Phi}_{j+2})\textcolor{black}{,}\\
\mathbf{\Phi}^{R,GRB} &=\frac{1}{840} (60 \mathbf{\Phi}'_j-420 \mathbf{\Phi}'_{j+1}-60 \mathbf{\Phi}'_{j+2}+178 \mathbf{\Phi}_{j}+3 \mathbf{\Phi}_{j-1}+478 \mathbf{\Phi}_{j+1}+178 \mathbf{\Phi}_{j+2}+3 \mathbf{\Phi}_{j+3})\textcolor{black}{.}
\end{aligned}	
\end{equation}

\textcolor{black}{The above Equations (\ref{eqn:grb}) will lead to a seventh-order gradient-based reconstruction scheme (denoted as GRB) when \textit{differenced} over a cell, and this completes the scheme's derivation.}

\begin{remark}\label{eqn:accuracy}
\normalfont  The fourth-order linear explicit scheme presented in Ref. \cite{chamarthi2023gradient} is a subset of the present scheme. By setting $\alpha = 4$ and $\beta=0$ in the Equation (\ref{alpha-beta}) one would obtain the fourth-order scheme.
 \end{remark}
 
 \begin{remark}\label{eqn:no-compact}
\normalfont  If the first-order derivatives, $\left(\frac{\partial \mathbf{\Phi}}{\partial x}\right)_{j}$, in Equation (\ref{eqn:lingrb}) are computed by using the sixth-order implicit gradient (popularly known as compact schemes) schemes of Lele \cite{lele1992compact} one would also obtain a seventh-order scheme for the same the values of the parameters, $\alpha=38/7$ and $\beta=3/190$. Such a scheme will have superior spectral properties but requires a tridiagonal matrix inversion.
\end{remark}
 
 \begin{remark}
\normalfont \textcolor{black}{Unlike the interpolation and reconstruction polynomials, the present gradient-based reconstruction is more versatile because one would obtain different schemes by changing the order of gradients. These gradients are also required for the viscous fluxes. It may also be possible to obtain new schemes by changing the gradients in the Equation (\ref{eqn:lingrb}) and is left to the readers.}
\end{remark}

 \begin{remark}
 \normalfont \textcolor{black}{The proposed is different from the third-order reconstruction approach of van Leer \cite{van1977towards}. Van Leer's approach is a finite-volume scheme, whereas the current approach is a finite-difference scheme. Even though both approaches use the same Legendre polynomial (Equation (\ref{eqn:legendre})), they are fundamentally different from each other.} 
 \end{remark}

\subsection{Shock-capturing}

The proposed upwind scheme in the earlier subsection is linear and, therefore, will lead to oscillations in the presence of shocks and material interfaces. The MP scheme of Suresh and Hyunh \cite{suresh1997accurate} is used to prevent such oscillations in this paper. After forming the upwind scheme using the gradient-based reconstruction given by Equation (\ref{eqn:grb}), the following condition is checked to determine the necessity of applying the MP limiter:

\begin{equation} \label{eqn:hocus/mp5Condition}
    \left( {\mathbf{\Phi}}^{L,GRB}_{j+1/2} - {{\mathbf{\Phi}}}_j \right) \left( {\mathbf{\Phi}}^{L,GRB}_{j+1/2} - {\mathbf{\Phi}}^{MP} \right) \leq 10^{-40},
\end{equation}
where ${\mathbf{\Phi}^{MP}}$ is given by the following equation:
\begin{equation} \label{eqn:alpha}
\begin{aligned}
 &{\mathbf{\Phi}}^{M P} ={\mathbf{\Phi}}_{j}+\operatorname{minmod}\left[{\mathbf{\Phi}}_{j+1}-{\mathbf{\Phi}}_{j}, {\xi}\left({\mathbf{\Phi}}_{j}-{\mathbf{\Phi}}_{j-1}\right)\right]\,\,,\\
\text{and,} &\operatorname{minmod}(a,b) = \frac{1}{2} \left(\operatorname{sign}(a)+\operatorname{sign}(b)\right)\min(|a|,|b|)\,\,,
\end{aligned}
\end{equation}
where ${\xi}=7$ as in \cite{chamarthi2021high}. If the linear scheme fails to satisfy the Equation (\ref{eqn:hocus/mp5Condition}) the procedure of the MP limiter described through the following set of equations is applied
 
\begin{equation}\label{eqn:mp-procedure}
\begin{aligned}
 d_{j} &=2\left({ {\mathbf{\Phi}}}_{j+1}-2 { {\mathbf{\Phi}}}_j+{ {\mathbf{\Phi}}}_{j-1}\right)-0.5\left({ {\mathbf{\Phi}}}_{j+1}^{\prime}-{ {\mathbf{\Phi}}}_{j-1}^{\prime}\right).\\
{\mathbf{\Phi}}_{j+\frac{1}{2}}^{M D} &=\frac{1}{2}\left({{\mathbf{\Phi}}}_{j}+{{\mathbf{\Phi}}}_{j+1}\right)-\frac{1}{2} \operatorname{minmod}\left(d_{j},d_{j+1}\right) \\
{\mathbf{\Phi}}_{j+\frac{1}{2}}^{UL} &={{\mathbf{\Phi}}}_{j}+\xi \left({{\mathbf{\Phi}}}_{j}-{{\mathbf{\Phi}}}_{j-1}\right) \\
{\mathbf{\Phi}}_{j+\frac{1}{2}}^{L C} &=\frac{1}{2}\left(3 {{\mathbf{\Phi}}}_{j}-{\hat{\mathbf{\Phi}}}_{j-1}\right)+\frac{4}{3} \operatorname{minmod}\left(d_{j},d_{j-1}\right)\\
{\mathbf{\Phi}}_{j+\frac{1}{2}}^{\min } &=\max \left[\min \left({{\mathbf{\Phi}}}_{j}, {{\mathbf{\Phi}}}_{j+1}, {\Phi}_{i+\frac{1}{2}}^{M D}\right), \min \left({{\mathbf{\Phi}}}_{j}, {\mathbf{\Phi}}_{j+1 / 2}^{U L}, {\mathbf{\Phi}}_{j+1 / 2}^{L C}\right)\right] \\
{\mathbf{\Phi}}_{i+\frac{1}{2}}^{\max } &=\min \left[\max \left({{\mathbf{\Phi}}}_{j}, {{\mathbf{\Phi}}}_{j+1}, {\Phi}_{j+\frac{1}{2}}^{M D}\right), \max \left({{\mathbf{\Phi}}}_{j}, {\mathbf{\Phi}}_{j+\frac{1}{2}}^{U L}, {\mathbf{\Phi}}_{j+\frac{1}{2}}^{L C}\right)\right]\\
{\mathbf{\Phi}}_{j+\frac{1}{2}}^{\text {MEG7 }} &={\mathbf{\Phi}}^{L,GRB}_{j+\frac{1}{2}}+\operatorname{minmod}\left({\mathbf{\Phi}}_{j+\frac{1}{2}}^{\min }-{\mathbf{\Phi}}^{L,GRB}_{j+\frac{1}{2}}, {\mathbf{\Phi}}_{j+\frac{1}{2}}^{\max }-{\mathbf{\Phi}}^{L,GRB}_{j+\frac{1}{2}}\right).
\end{aligned}
\end{equation}

The scheme thus obtained from the above procedure is defined as Monotonicity-preserving explicit gradient reconstruction and is denoted as \textbf{MEG7} in this paper. If conservative variables are used for reconstruction, the scheme is denoted as MEG7-cons, and if primitive variables are used for reconstruction, then the corresponding scheme is denoted as MEG7-prim. It is essential to carry out shock-capturing using characteristic variables, which is pretty straightforward. The variables of choice and their corresponding gradients are transformed into the characteristic space, and details are presented in Ref. \cite{chamarthi2023gradient,chamarthi2021high}.

\subsection{Algorithm for genuine high-order accuracy}\label{subsec:high}

 Fourier analysis is carried out for a linear advection equation as is typically done in the literature \textcolor{black}{\cite{chamarthi2023gradient,van2021towards,deng2019fifth}}. Through Fourier analysis, it has been shown in the earlier subsection that the newly derived gradient-based reconstruction is seventh-order accurate. Unfortunately, \textcolor{black}{when} the proposed scheme is implemented for Euler equations, specifically for non-linear test cases, it will only be second-order accurate \cite{van2021towards}. As explained in the introduction, the convective fluxes are computed using a Riemann solver and the generic equation is written below

\begin{equation}
\mathbf{F}^{\rm Riemann}_{j+\frac{1}{2}}
= \frac{1}{2}
\left[
\textcolor{black}{\mathbf{F}^{c}_{L}}
+ 
\textcolor{black}{\mathbf{F}^{c}_{R}}
\right]
-
 \frac{1}{2} | {\mathbf{A}_{j+\frac{1}{2}}}|({\mathbf{Q}^R_{j+\frac{1}{2}}}-{\mathbf{Q}^L_{j+\frac{1}{2}}}),
\label{eqn:Riemann_revisit}
\end{equation}

The above \textcolor{black}{equation} is deliberately written in the above form. Typically, the interface fluxes \textcolor{black}{$\mathbf{F}^{c}_{L}$ and $\mathbf{F}^{c}_{R}$} are computed from the corresponding $\mathbf{Q_L}$ and $\mathbf{Q_R}$ (or from $\mathbf{U_L}$ and $\mathbf{U_R}$ if primitive variables are used for reconstruction). However, interface fluxes thus obtained will lead to loss of high-order accuracy and will only be second-order accurate for reconstruction type polynomials \cite{van2021towards}, Equation (\ref{second-cry}).

\begin{equation}\label{second-cry}
\textcolor{black}{\mathbf{F}^{c}_{L}} \quad \text{computed as} \quad \mathbf{F(Q_L)} \quad \text{or} \quad \mathbf{F(U_L)} \rightarrow \text{Second-order accurate}	
\end{equation}

If the interface fluxes are directly computed using the fluxes at the cell-centres, one will attain the desired order of accuracy, Equation (\ref{high-happy}).

\begin{equation}\label{high-happy}
\textcolor{black}{\mathbf{F}^{c}_{L}} \quad \text{computed as} \quad  \mathbf{F(f_L)} \rightarrow \text{High-order accurate}	
\end{equation}
\textcolor{black}{It is important to note that, To obtain genuine higher-order accuracy, both {$\mathbf{F}^{c}_{L}$ and $\mathbf{F}^{c}_{R}$} and $\mathbf{Q_L}$ and $\mathbf{Q_R}$  must be computed independently.}
A naive approach would be to compute the $\textcolor{black}{\mathbf{F}^{c}_{L}}$ by using the fluxes at the cell centers, but that would be incompatible when the flow has discontinuities and will lead to oscillations. When there are discontinuities in the flow, the $\mathbf Q_L$ will be computed by the MP scheme to prevent oscillations. Therefore it would be appropriate to compute the fluxes $\textcolor{black}{\mathbf{F}^{c}_{L}}$ as a function of $\mathbf Q_L$ obtained from the MP limiter. In order to overcome this issue, a hybrid approach is considered in this paper, which is as follows:
\begin{equation}
\textcolor{black}{\mathbf{F}^{c}_{L}}= \sigma_{j+1 / 2}\mathbf{F(f_L)}+(1-\sigma_{j+1 / 2}) \mathbf{F(Q_L)}
\end{equation}
\begin{equation}
\textcolor{black}{\mathbf{F}^{c}_{R}= \sigma_{j+1 / 2}\mathbf{F(f_R)}+(1-\sigma_{j+1 / 2}) \mathbf{F(Q_R)}}
\end{equation}
where
 \begin{equation}\label{flux-full}
 \mathbf{F(f_L)} = \frac{1}{840} (568 \mathbf{f}_j-146 \mathbf{f}_{j-1}+22 \mathbf{f}_{j-2}+2 \mathbf{f}_{j-3}-\mathbf{f}_{j-4}+484 \mathbf{f}_{j+1}-104 \mathbf{f}_{j+2}+16 \mathbf{f}_{j+3}-\mathbf{f}_{j+4}),
 \end{equation}
 and
  \begin{equation}\label{flux-right}
 \textcolor{black}{\mathbf{F(f_R)} =  \frac{1}{840}(-\mathbf{f}_{i+5}+2\mathbf{f}_{i+4}+22\mathbf{f}_{i+3}-146\mathbf{f}_{i+2}+568\mathbf{f}_{i+1}+484f_{i}-104\mathbf{f}_{i-1} +16\mathbf{f}_{i-2}-\mathbf{f}_{i-3})},
  \end{equation}
  and $\sigma_{j+1 / 2}$ is a discontinuity detector that is close to zero near the discontinuities and unity in the smooth regions. \textcolor{black}{One can also use the Equation (\ref{eqn:lingrb}) for computing the interface flux using the fluxes at the cell-centers. Equation (\ref{flux-full}) is the simplified form of Equation (\ref{eqn:grb}) obtained after substituting the first- and second-derivatives and is used only for flux reconstruction.} This approach makes it possible to obtain high-order accuracy in the vicinity of smooth regions and also capture discontinuities without oscillations. In this paper, the discontinuity detector proposed by Li et al. \cite{li2012high} is used and is as follows:

\begin{equation}
\psi_{j}=\frac{2}{\left(\frac{a}{b}\right)+\left(\frac{b}{a}\right)}\textcolor{black}{.}
\end{equation}

\begin{equation}\label{detector}
\begin{aligned}
&a=\left|\rho_{j}-\rho_{j-1}\right|+\left|\rho_{j}-2 \rho_{j-1}+\rho_{j-2}\right|\textcolor{black}{,} \\
&b=\left|\rho_{j}-\rho_{j+1}\right|+\left|\rho_{j}-2 \rho_{j+1}+\rho_{j+2}\right|\textcolor{black}{.}
\end{aligned}
\end{equation}

\begin{equation}\label{buffer}
\begin{aligned}
&\psi_{j+1 / 2}=\min \left(\psi_{j-2},\psi_{j-1}, \psi_{j}, \psi_{j+1}, \psi_{j+2},\psi_{j+3}\right)\textcolor{black}{,} \\
&\varepsilon=\frac{0.9 \psi_{c}}{1-0.9 \psi_{c}} \xi^{2}, \quad \xi=10^{-3}, \quad \psi_{c}=0.5\textcolor{black}{,} \\
&\sigma_{j+1 / 2}= 
\begin{cases}0 & \psi_{j+1 / 2}<\psi_{c} \\
1 & \text { otherwise. }
\end{cases}
\end{aligned}
\end{equation}

The discontinuity detector used in the present paper differs from the original paper of Li et al. \cite{li2022class} in two aspects. First, Li et al. used the fluxes at the cell centers as the variables in Equation (\ref{detector}). In contrast, in this paper, only density is used. Second, Li et al. have proposed the discontinuity detector for the hybrid scheme, but in this paper, the detector is used to obtain high-order accuracy. In addition, similar discontinuity detectors are also used in the TENO approach to improve the scheme's efficiency, and resolution characteristics \textcolor{black}{\cite{fu2018new,peng2021efficient}}. 

The motivation behind using density as the variable for the discontinuity detector is that both contact discontinuity and shockwaves can be identified by monitoring density alone. In contrast, monitoring pressure or velocity can only detect shockwaves. Sufficient buffer points are used such that the approach can be used for a broad class of test cases, three on either side of the cell interface, as shown in Equation (\ref{buffer}) $\psi_{j+1 / 2}$. The proposed approach works with both conservative and primitive variable reconstruction. Finally, \textcolor{black}{increasing} the $\psi_{c}$ value above 0.5 will reduce the accuracy, and decreasing the value will lead to oscillations. The proposed approach can also be used for the explicit fourth-order schemes presented in \cite{chamarthi2023gradient}. The current approach is more suitable for the explicit schemes than the implicit schemes presented in \cite{chamarthi2023gradient} as implicit schemes are global in nature and, therefore, will be expensive. \textcolor{black}{The proposed hybrid algorithm can still share the derivatives between the inviscid and viscous fluxes, which is the main idea behind the gradient-based reconstruction.	}

\subsection{Temporal discretization}\label{subsec:time}

Finally, The conserved variables are integrated in time using the third-order TVD Runge–Kutta scheme \cite{Jiang1995}:
\begin{eqnarray}\label{rk}
\mathbf{ Q}_{j, i}^{(1)}&=&\mathbf{ Q}_{j, i}^{n}+\Delta t \mathbf{Res}\left(\mathbf{ Q}_{j, i}^{n}\right) \\
\mathbf{ Q}_{j, i}^{(2)}&=&\frac{3}{4} \mathbf{ Q}_{j, i}^{n}+\frac{1}{4} \mathbf{ Q}_{j, i}^{(1)}+\frac{1}{4} \Delta t \mathbf{Res}\left(\mathbf{ Q}_{j, i}^{(1)}\right) \\
\mathbf{ Q}_{j, i}^{n+1}&=&\frac{1}{3} \mathbf{ Q}_{j, i}^{n}+\frac{2}{3} \mathbf{ Q}_{j, i}^{(2)}+\frac{2}{3} \Delta t \mathbf{Res}\left(\mathbf{ Q}_{j, i}^{(2)}\right),
\end{eqnarray}
where $\mathbf{Res}$ denote residuals given by the right-hand side of Equation (\ref{eqn-differencing_residual}), respectively. The superscripts ${n}$ and ${n+1}$ denote the current and the next time-steps, and superscripts ${(1)-(2)}$ correspond to intermediate steps. The time-step $\Delta t$ is computed as:

\begin{equation}
\Delta t= \text{CFL} \cdot \min \left(\frac{1}{\alpha_v} \min _{j, i}\left(\frac{\Delta \textcolor{black}{x^{2}}}{\nu_{j, i}}, \frac{\Delta \textcolor{black}{y^{2}}}{\nu_{j, i}}\right),  \min _{j, i}\left(\frac{\Delta\textcolor{black}{x}}{\left|u_{j, i}\right|+c_{j, i}}, \frac{\Delta \textcolor{black}{y}}{\left|v_{j, i}\right|+c_{j, i}}\right)\right),
\end{equation}
where $c$ is the speed of sound and given by $c=\sqrt{\gamma{p}/\rho}$, $\alpha_v$ is associated with the damping property of a viscous scheme (see Equation (\ref{eqn:alphad})), and $\nu$ is the kinematic viscosity defined as $\nu = \mu / \rho$. For all the simulations, the value of CFL is taken as 0.2.

\section{Numerical Results}\label{sec:results}
This section presents the numerical results of some benchmark tests for Euler and Navier-Stokes equations to illustrate the performance of the seventh-order gradient-based reconstruction described above.  For the MEG7 scheme, both primitive and conservative reconstruction is considered. The present scheme, MEG7, is compared with the seventh-order weighted nonlinear compact scheme (WCNS7) \cite{Nonomura2012,jin2018optimized}. \textcolor{black}{The details of the WCNS7 are presented here. The interpolations in WCNS are constructed from four-points substensils as follows: }
\textcolor{black}{
\begin{equation}
\begin{aligned}
Q_{j+1 / 2}^{L, 1} & =-\frac{5}{16} Q_{j-3}+\frac{21}{16} Q_{j-2}-\frac{35}{16} Q_{j-1}+\frac{35}{16} Q_j \\
Q_{j+1 / 2}^{L, 2} & =\frac{1}{16} Q_{j-2}-\frac{5}{16} Q_{j-1}+\frac{15}{16} Q_j+\frac{5}{16} Q_{j+1} \\
Q_{j+1 / 2}^{L, 3} & =-\frac{1}{16} Q_{j-1}+\frac{9}{16} Q_j+\frac{9}{16} Q_{j+1}+\frac{1}{16} Q_{j+2} \\
Q_{j+1 / 2}^{L, 4} & =\frac{5}{16} Q_j+\frac{15}{16} Q_{j+1}-\frac{5}{16} Q_{j+2}+\frac{1}{16} Q_{j+3}
\end{aligned}
\end{equation}
}
\textcolor{black}{Next, the $Q_{j+1 / 2}^{L}$ is computed as a weighted average as follows:}
\textcolor{black}{
\begin{equation}
\begin{aligned}
Q_{j+1 / 2}^{L}  =w_1 Q_{j+1 / 2}^{L, 1}+w_{2} Q_{j+1 / 2}^{L, 2}  +w_3 Q_{j+1 / 2}^{L, 3}+w_{4} Q_{j+1 / 2}^{L, 4}
\end{aligned}
\end{equation}}
\textcolor{black}{where $w_k$ is a nonlinear weight computed as follows:}
\textcolor{black}{
\begin{equation}
w_{k}=\frac{\alpha_{k}}{\sum_l \alpha_{l}}
\end{equation}
}

\textcolor{black}{\begin{equation}
\alpha_{k, m}=\frac{C_k}{\left(I S_{k}+\varepsilon\right)^2}
\end{equation}}

\textcolor{black}{In the above equation, (C1, C2, C3, C4) = (1/64, 21/64, 35/64, 7/64) are called the ideal weights, and $IS_k$ is known as smooth indicator which is computed as follows.}
\textcolor{black}{
\begin{equation}
I S_{k}=\sum_{n=1}^4\left(\sum_{l=1}^4 c_{n, k, l} Q_{j+k+l-5}\right)
\end{equation}}

As the nonlinear weights in WCNS7 are not always appropriate in the smooth region, the mapping technique of \cite{Henrick2005,jin2018optimized} is used, and the scheme is denoted as WCNS7M. The mapping technique efficiently recovers the formal order of accuracy of the scheme. \textcolor{black}{The $Q_{j+1 / 2}^{L}$ computed using the mapping technique is as follows:}

\textcolor{black}{\begin{equation}
Q_{j+1 / 2}^{L}=w_{1}^M Q_{j+1 / 2}^{L, 1}+w_{2}^M Q_{j+1 / 2}^{L, 2}+w_{3, m}^M Q_{j+1 / 2}^{L, 3}+w_{4}^M q_{j+1 / 2}^{L, 4}
\end{equation}}

\textcolor{black}{
\begin{equation}
w_{k}^M=\frac{w_{k}\left(C_k+\left(C_k\right)^2-3 C_k w_{k}+\left(w_{k}\right)^2\right)}{\left(C_k\right)^2-\left(1-2 C_k\right) w_{k}}
\end{equation}
}

Furthermore, the WCNS7M scheme is implemented with conservative variable interpolation, and the eighth-order scheme is used to compute the flux derivatives as in Ref. \cite{Nonomura2013}.
\textcolor{black}{
\begin{equation}
\begin{array}{r}
\frac{d \mathbf F}{dx}_{j}=\frac{1}{\Delta x} \frac{1225}{1024}\left(F_{j+1 / 2}^{C, W C N S}-F_{j-1 / 2}^{C, W C N S}\right) \\
-\frac{1}{\Delta x} \frac{245}{3072}\left(F_{j+3 / 2}^{C, W C N S}-F_{j-3 / 2}^{C, W C N S}\right) \\
+\frac{1}{\Delta x} \frac{49}{5120}\left(F_{j+5 / 2}^{C, W C N S}-F_{j-5 / 2}^{C, W C N S}\right) \\
-\frac{1}{\Delta x} \frac{5}{7168}\left(F_{j+7 / 2}^{C, W C N S}-F_{j-7 / 2}^{C, W C N S}\right)
\end{array}
\end{equation}
}

\textcolor{black}{It is essential to note the present scheme is compared with the WCNS7M approach for two reasons. a) WCNS7M is a genuinely high-order accurate approach, and b) the advantage of the WCNS7 approach is that any approximate Riemann solver can be used, unlike the WENO schemes \cite{nonomura2010freestream}. The second advantage is important as one consistent Riemann solver can be used for comparison.} \textcolor{black}{The boundary conditions for all the test cases are implemented using ghost cell approach as in Ref. \cite{fu2021very}.}

\begin{example}\label{ex:acc} {Order of accuracy}
\end{example}

In this first test case, the order accuracy of the proposed scheme is demonstrated. For this purpose the two-dimensional vortex evolution problem \textcolor{black}{\cite{balsara2000monotonicity,yee1999low}} is considered. The test case is initialized on a computation domain of [-5, 5] $\times$ [-5, 5] with periodic boundaries on all sides. To the mean flow, an isentropic vortex is added, and the initial flow field is initialized as follows:

\begin{equation}
p=\rho^{\gamma}, T=1-\frac{(\gamma-1) \varepsilon^{2}}{8 \gamma \pi^{2}} e^{\left(1-r^{2}\right)}, u=1-\frac{\varepsilon}{2 \pi} e^{\frac{1}{2}\left(1-r^{2}\right)} y, v=1+\frac{\varepsilon}{2 \pi} e^{\frac{1}{2}\left(1-r^{2}\right)} x, 
\end{equation}

where $r^2$ = $x^2$ + $y^2$ and the vortex strength $\epsilon$ is taken as 5. The computations are carried out until a final time $t$ =10. The $L_2$ density errors and the convergence rates of the MEG7 scheme using both primitive and conservative variables are shown in Table \ref{tab:accu_2_isen}. It can be seen that both MEG7-cons and MEG7-prim achieved seventh-order accuracy. WCNS7M scheme also showed a similar convergence rate for this test case.

\begin{table}[H]
  \centering
  \caption{Order of accuracy for various schemes, Example \ref{ex:acc}. $L_2$ density error norms and convergence rate are shown.}
    \begin{tabular}{||c |c|c|c|c|c|c||}
    \hline
    N & WCNS7M  &  Order     & MEG7-cons &   Order    & MEG7-prim  &  Order \\
    \hline
    $25^2$ & 2.45E-03 & - & 2.55E-03 & - & 2.38E-03 & - \\
    \hline
    $50^2$ & 9.34E-05 & 4.71  & 6.68E-05 & 5.25  & 6.77E-05 & 5.13 \\
    \hline
    $100^2$ & 1.14E-06 & 6.35  & 1.02E-06 & 6.03  & 1.06E-06 & 6.00 \\
    \hline
    $200^2$ & 1.51E-08 & 6.24  & 1.06E-08 & 6.59  & 1.03E-08 & 6.68 \\
    \hline
    \end{tabular}%
  \label{tab:accu_2_isen}%
\end{table}%

Now that it has been established that the proposed scheme can attain high-order accuracy for the nonlinear test case, thanks to the hybrid approach presented in Section \ref{subsec:high}, in the following test cases, it will be shown that it can also capture discontinuities without oscillations. 

\begin{example}\label{sod}{Shock tube problems}
\end{example}
First, we show that the proposed scheme MEG7 passes the Sod \cite{sod1978survey} and Lax \cite{lax1954weak} shock-tube problems. For the Sod problem the initial conditions are as follows:

\begin{equation}
        \left( \rho,u,p \right) =
        \begin{cases}
            (1,0,1), & \text{if } 0 \leq x < 0.5, \\
            (0.125,0,0.1), & \text{if } 0.5 \leq x \leq 1,
        \end{cases}
\end{equation}
and the simulation is carried out until a final time $t$ = 0.2. For the Lax shock-tube problem, the initial conditions are
\begin{equation}\label{lax}
        \left( \rho,u,p \right) =
        \begin{cases}
            (0.445,0.698,3.528), & \text{if } 0 \leq x < 0.5, \\
            (0.5,0,0.571), & \text{if } 0.5 \leq x \leq 1,
        \end{cases}
\end{equation}
and the final time $t$ = 01.4. The computational domain for both the cases is $x = [0,1]$ and simulation is carried out using $N = 200$ uniformly distributed grid points. The exact solution for this test case is computed \textcolor{black}{by an} exact Riemann solver \cite{toro2009riemann}. Fig. \ref{fig_sod} shows the results for the Sod problem, and all employed schemes capture the rarefaction wave, contact discontinuity, and the shock wave without obvious oscillations. Fig. \ref{fig_lax} shows the results for the Lax problem, and again all the schemes captured the discontinuities without oscillations.

\begin{figure}[H]
\centering
\subfigure[Computed density profiles.]{\includegraphics[width=0.4\textwidth]{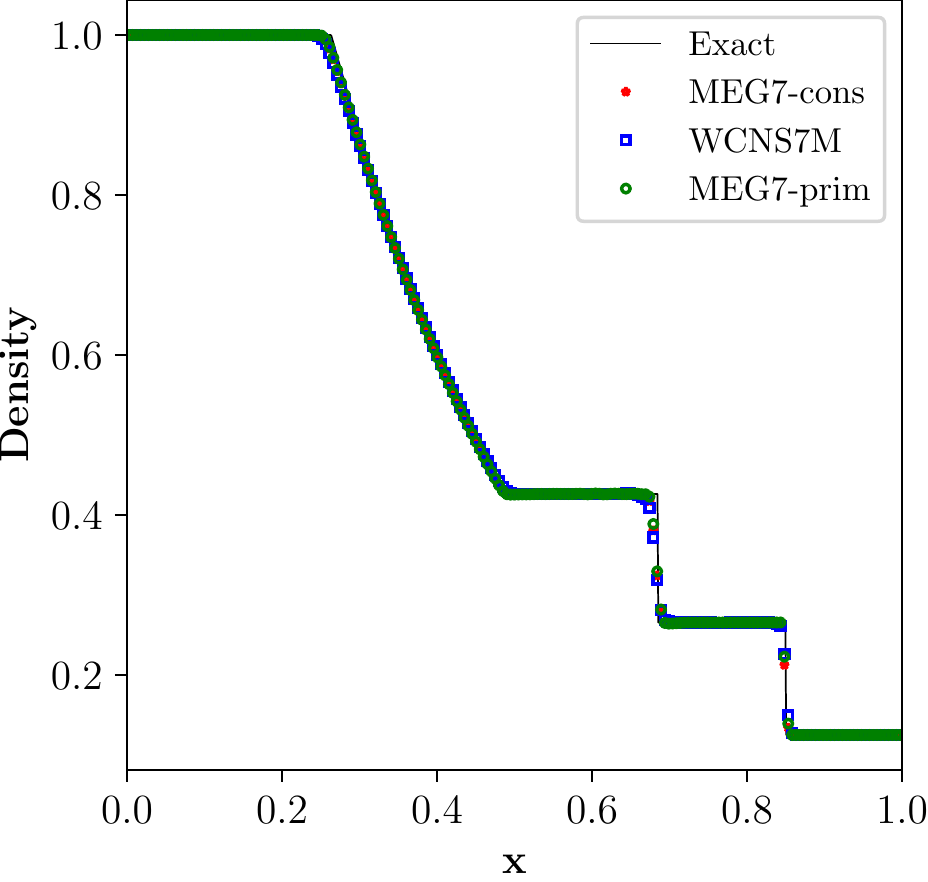}
\label{fig:sod-den}}
\subfigure[Computed velocity profiles.]{\includegraphics[width=0.4\textwidth]{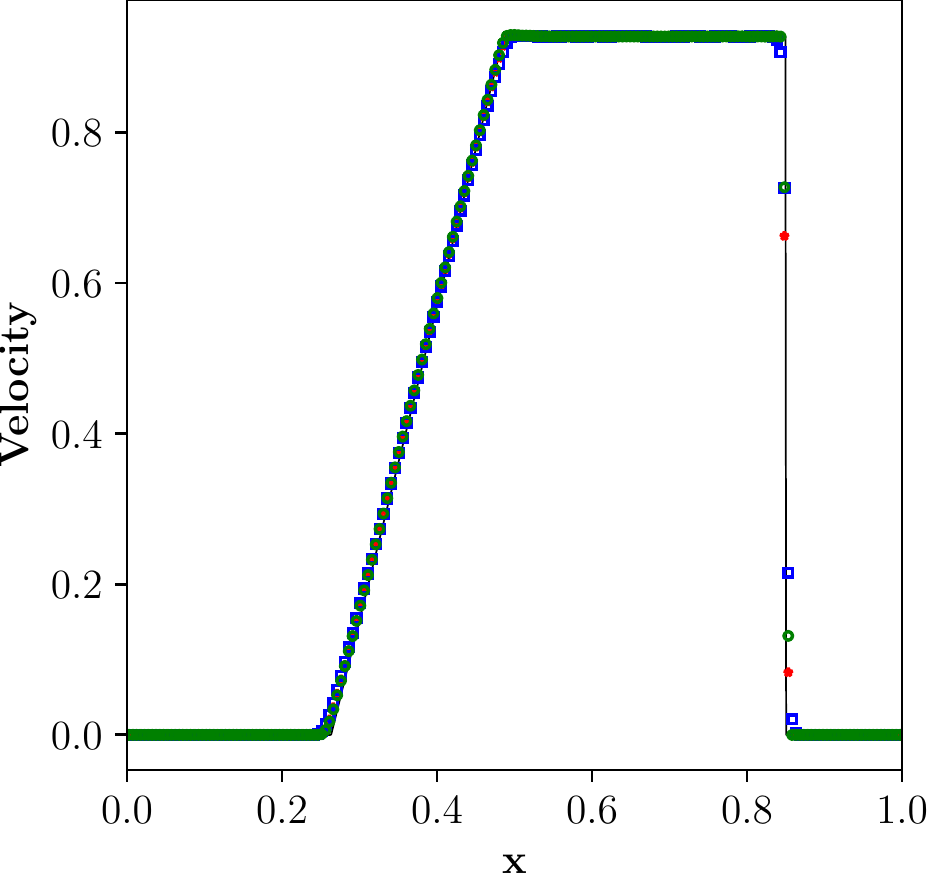}
\label{fig:sod-vel}}
\caption{Numerical solution for Sod shock tube problem using $N = 200$ grid points at $t = 0.2$, for Sod test case of Example \ref{sod}, where solid line: exact solution; blue squares: WCNS7M; red stars: MEG7-cons and green circles: MEG7-prim.}
\label{fig_sod}
\end{figure}

\begin{figure}[H]
\centering
\subfigure[Computed density profiles. ]{\includegraphics[width=0.45\textwidth]{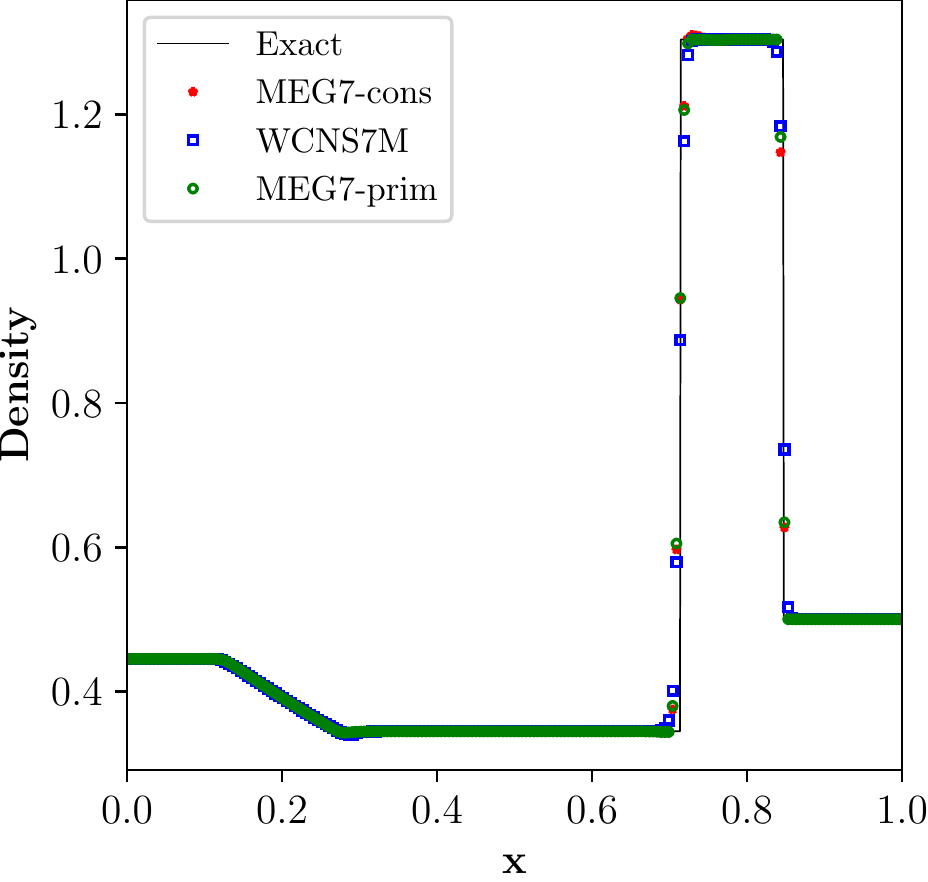}
\label{fig:lax-den}}
\subfigure[Computed velocity profiles.]{\includegraphics[width=0.45\textwidth]{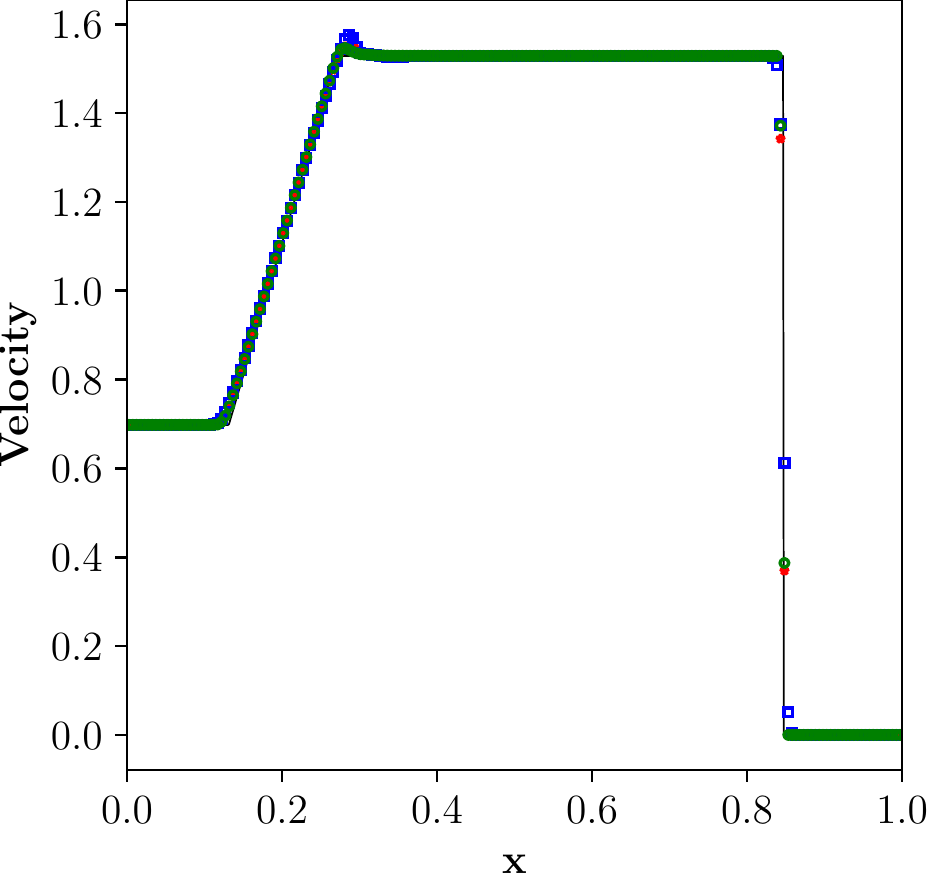}
\label{fig:lax-vel}}
\caption{Numerical solution for Lax problem using $N = 200$ grid points at $t = 0.14$, for Lax test case of Example \ref{sod}, where solid line: exact solution; blue squares: WCNS7M; red stars: MEG7-cons and green circles: MEG7-prim.}
\label{fig_lax}
\end{figure}

\begin{example}\label{leblanc}{Extreme test cases}
\end{example}

Here, we consider the Le Blanc \cite{loubere2005subcell} and Sedov \cite{zhang2012positivity} test cases which are considered as an extreme shock-tube problems. The initial conditions for the Le Blanc problem are as follows:

\begin{align}\label{blanc_prob}
(\rho,u,p)=
\begin{cases}
&(1.0\ \ ,\ \ 0,\ \ \frac{2}{3}\times10^{-1}\ \ ),\quad 0<x<3.0,\\
&(10^{-3},\ \ 0,\ \ \frac{2}{3}\times10^{-10}),\quad 3.0<x<9,
\end{cases}
\end{align}
and the final time of simulation is $t$=6. The specific heat ratio of this test case is taken as $\frac{5}{3}$. Numerical results for density and pressure obtained for WCNS7M, MEG7-cons and MEG7-prim schemes computed on a grid size of $N=900$ are shown in Fig. \ref{fig_blanc}. Density profiles obtained by the WCN7M scheme show is slightly dissipative than the profiles obtained by the MEG7-cons and -prim, respectively. The discontinuity at the location $x$ = 8 is closer to the exact solution by MEG7 scheme, unlike the WCNS7M scheme for both density and pressure and profiles, shown in Figs. \ref{fig:blanc-den} and \ref{fig:blanc-pres}.

\begin{figure}[H]
\centering
\subfigure[Density]{\includegraphics[width=0.42\textwidth]{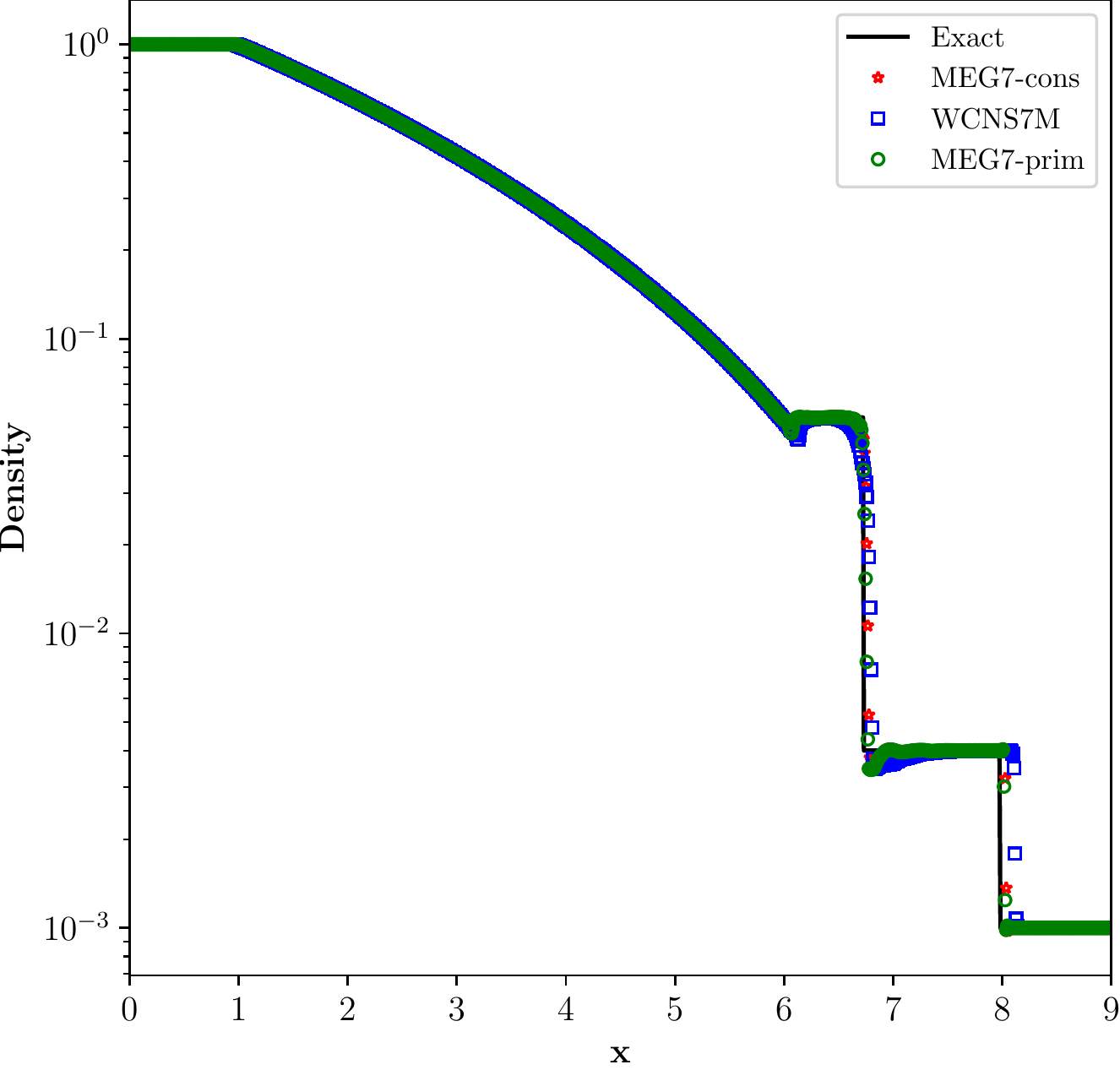}
\label{fig:blanc-den}}
\subfigure[Pressure]{\includegraphics[width=0.42\textwidth]{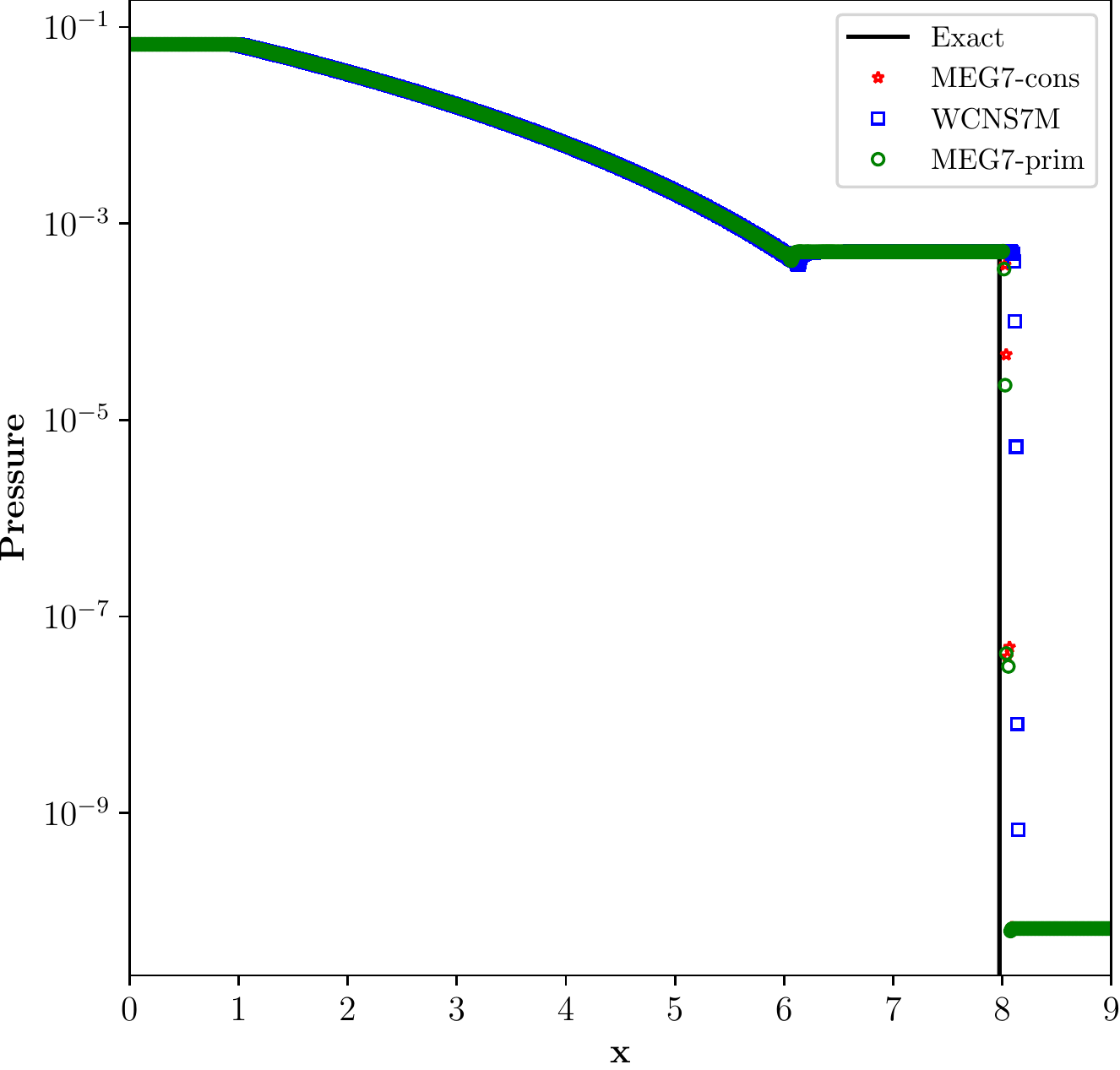}
\label{fig:blanc-pres}}
\caption{\textcolor{black}{Numerical solution for Le Blanc problem with the initial conditions given by Equation (\ref{blanc_prob}}), Example (\ref{leblanc}). Solid line: exact solution; blue squares: WCNS7M; red stars: MEG7-cons and green circles: MEG7-prim.}
\label{fig_blanc}
\end{figure}

The initial conditions for the Sedov problem are as follows:

\begin{equation}\label{sedov}
(\rho, u, p)= \begin{cases}(1,0,4.0 \mathrm{e}-13), & x<2-0.5 \Delta x, x>2+0.5 \Delta x \\ \left(1,0, \frac{1.28 \mathrm{e} 6}{\Delta x}\right), & 2-0.5 \Delta x \leq x \leq 2+0.5 \Delta x\end{cases},
\end{equation}
and the computational domain is [0,4] with the final time $t$ = 1.0e-3. This test case is also carried out using 900 grid points, and the results are shown in Fig. \ref{fig_sedov}. Figs. \ref{fig:bsedov-den} and \ref{fig:sedov-vel} show the density and pressure profiles respectively. It can be seen that the MEG7 scheme with both conservative and primitive variable reconstruction passes this test case, and there are no noticeable oscillations. WCNS7M scheme failed to pass this test. The MEG7-cons approach required a positivity preserving approach of \cite{zhang2012positivity}. These results indicate that the proposed high-order algorithm can simulate extreme test cases and is robust.
\begin{figure}[H]
\centering
\subfigure[Density]{\includegraphics[width=0.38\textwidth]{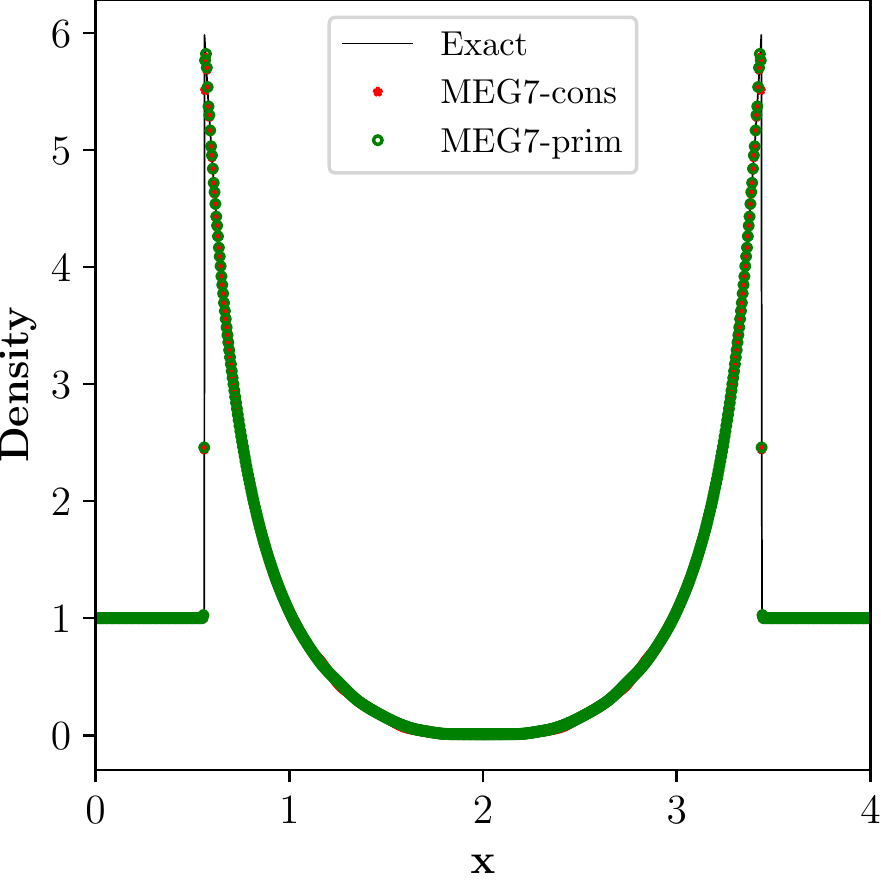}
\label{fig:bsedov-den}}
\subfigure[Pressure]{\includegraphics[width=0.42\textwidth]{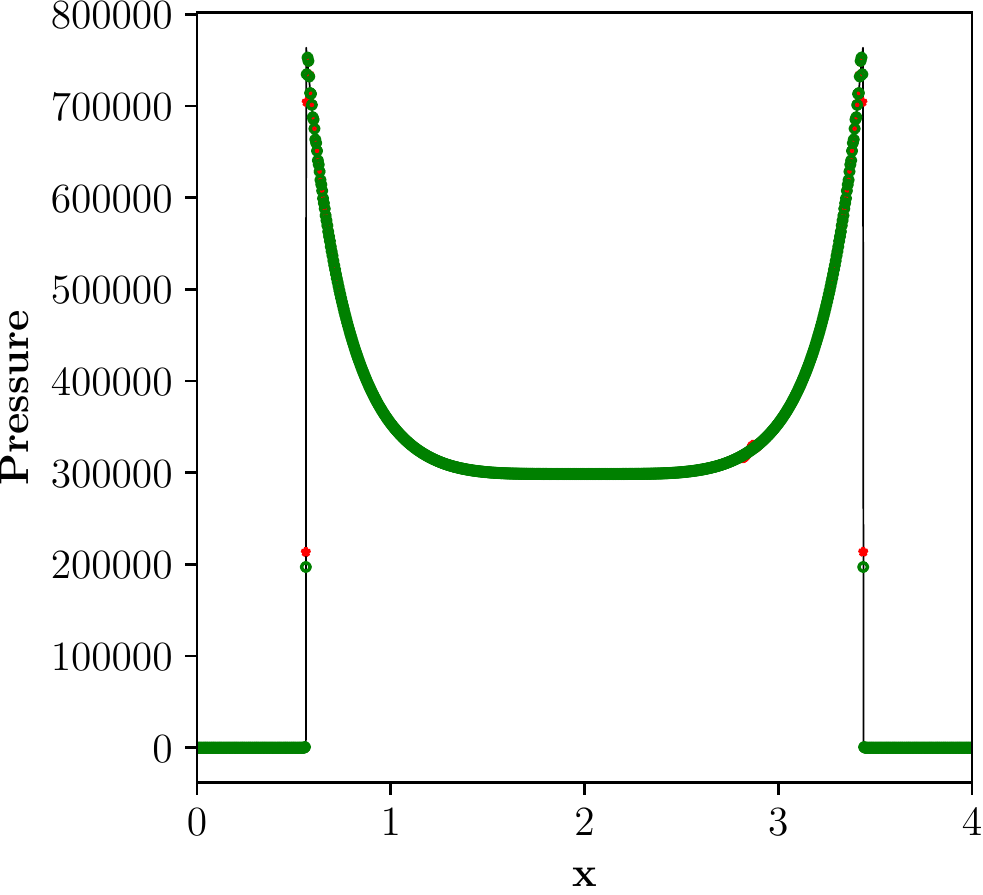}
\label{fig:sedov-vel}}
\caption{Numerical solution for Sedov problem with the initial conditions given by Equation (\ref{sedov}), Example \ref{leblanc}. Solid line: exact solution; red stars: MEG7-cons and green circles: MEG7-prim.}
\label{fig_sedov}
\end{figure}

\begin{example}\label{blast}{Blast wave problem}
\end{example}
In this test case, the blast wave problem of Ref. \cite{woodward1984numerical}, with the following initial conditions

\begin{equation}
        \left( \rho,u,p \right) =
        \begin{cases}
            (1,0,1000), & \text{if } 0.0 \leq x < 0.1, \\
            (1,0,0.01), & \text{if } 0.1 \leq x < 0.8, \\
            (1,0,100), & \text{if } 0.8 \leq x \leq 1.0,
        \end{cases}
\end{equation}
is considered. The computational domain for this test case is $x = [0,1]$ and the final time is $t = 0.038$. The simulation is carried out with $N = 600$ uniformly distributed grid points and the results are shown in Fig.\ref{fig_blast}. From Fig. \ref{fig:blast-400}, it can be seen that all considered schemes perform well and are without oscillations. When looking at the local density profile in Fig. \ref{fig:blast-400}, the density extrema is best predicted by MEG7-prim.
\begin{figure}[H]
\centering
\subfigure[Global density profiles.]{\includegraphics[width=0.4\textwidth]{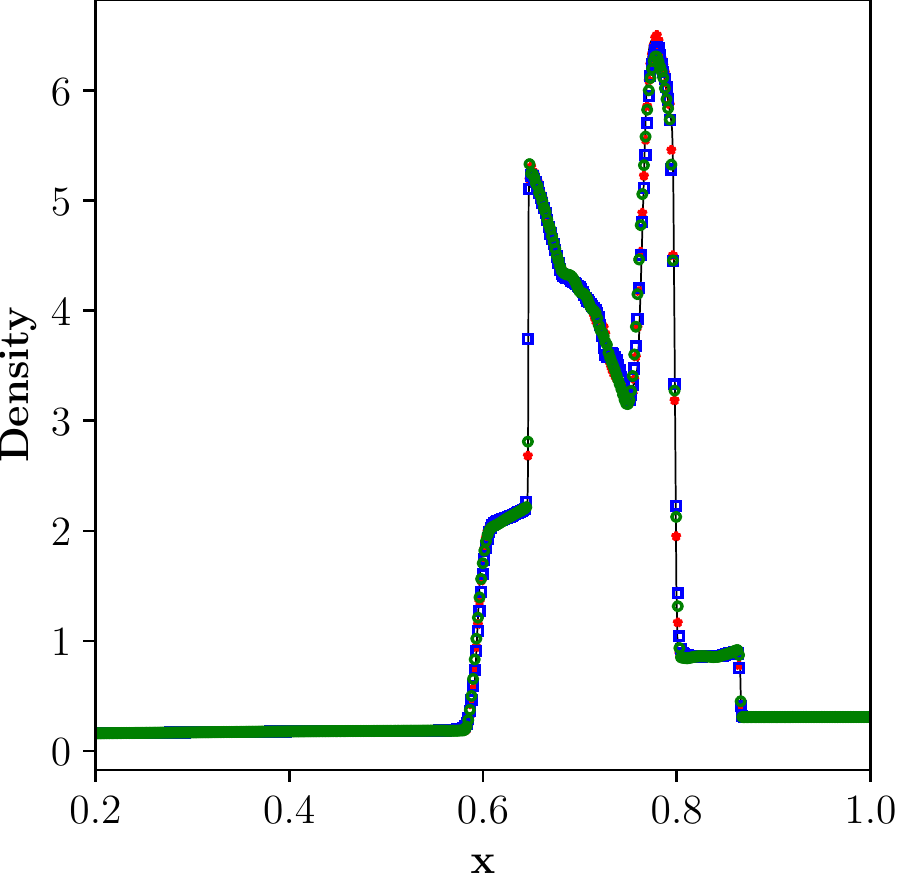}
\label{fig:blast-400}}
\subfigure[Global velocity profiles.]{\includegraphics[width=0.4\textwidth]{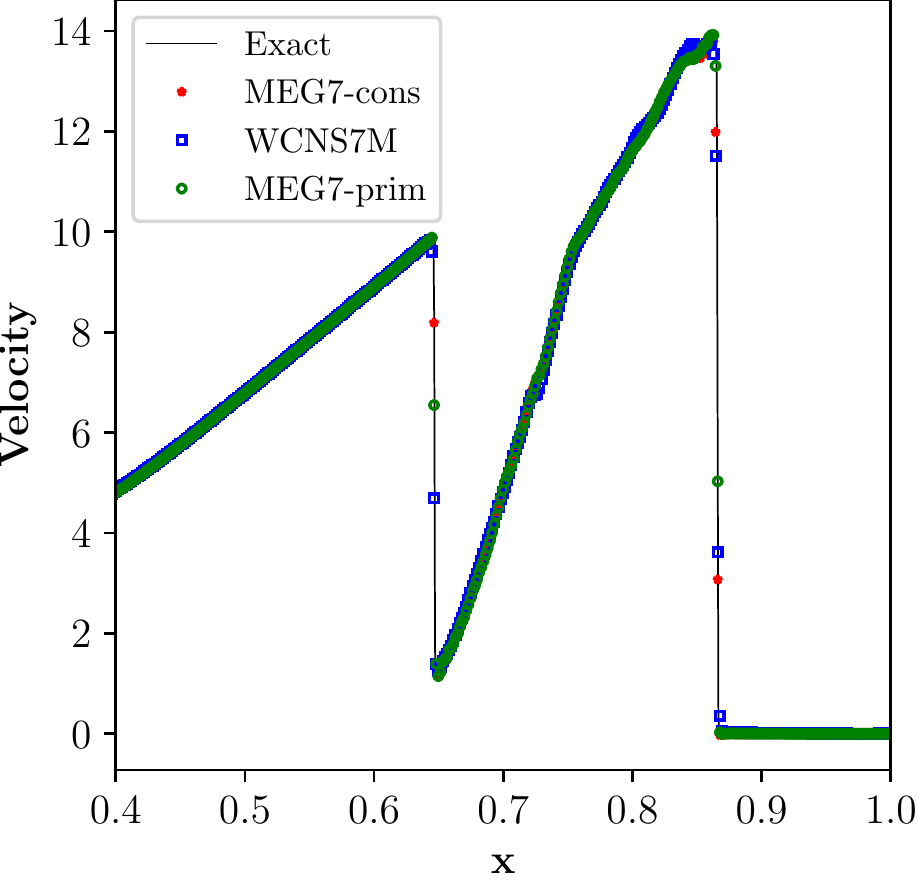}
\label{fig:blast-local}}
\caption{Numerical solution for blast wave problem using $N = 600$ grid points at $t = 0.038$, Example \ref{blast}. Solid line: exact solution; blue squares: WCNS7M; red stars: MEG7-cons and green circles: MEG7-prim.}
\label{fig_blast}
\end{figure}
 The discontinuity in velocity profile, Fig. \ref{fig:blast-local} at $x$$ \approx$ 0.87, is captured with fewer cells using the MEG7 schemes, whereas the WCNS7M scheme required more cells, indicating the low dissipation property of the current approach.

\begin{example}\label{Shu-Osher}{One-dimensional shock-entropy wave problem}
\end{example}
In this test, we consider the shock-entropy wave problem of Shu and Osher \cite{Shu1988}, and its modified version of Titarev and Toro \cite{titarev2004finite}. This test case is about shockwave interaction with high-frequency oscillating sinusoidal waves. These cases evaluate a scheme's capability to capture both shock and high-frequency oscillations at the same time. The initial conditions of Shu and Osher are as follows:
\begin{equation}
        \left( \rho,u,p \right) =
        \begin{cases}
            (3.857,2.629,10.333), & \text{if } -5 \leq x < -4, \\
            (1+0.2\sin(5(x-5)),0,1), & \text{if } -4 \leq x \leq 5.
        \end{cases}
\end{equation}
The computational domain of this test case is $x = [-5,5]$ and the final time is $t = 1.8$. Simulations are carried out on a grid size of $N = 200$. The \textit{exact solution} is obtained on a fine grid resolution of 2000 points using the MP5 scheme. As shown in Fig. \ref{fig:shu-global}, the proposed schemes match the reference solution well. Observing Fig. \ref{fig:shu-local}, MEG7-cons and MEG7-prim perform better than the WCNS7M scheme in capturing the density extrema.

\begin{figure}[H]
\centering 
\subfigure[Global density profiles.]{\includegraphics[width=0.41\textwidth]{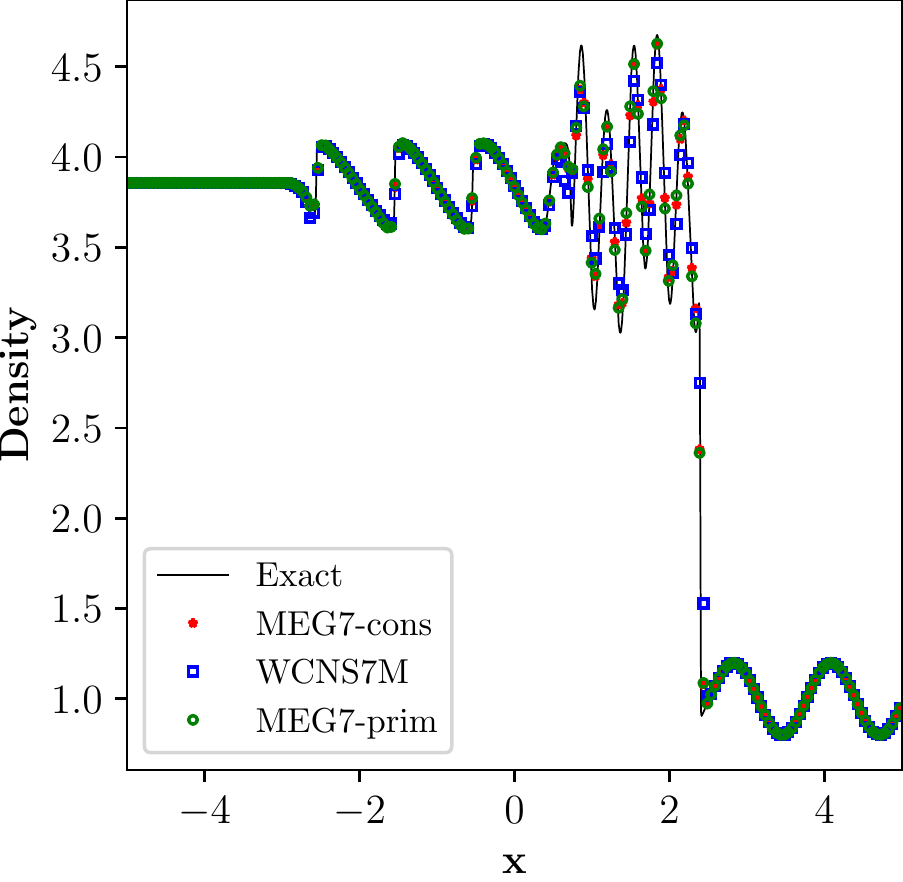}
\label{fig:shu-global}}
\subfigure[Local density profiles.]{\includegraphics[width=0.42\textwidth]{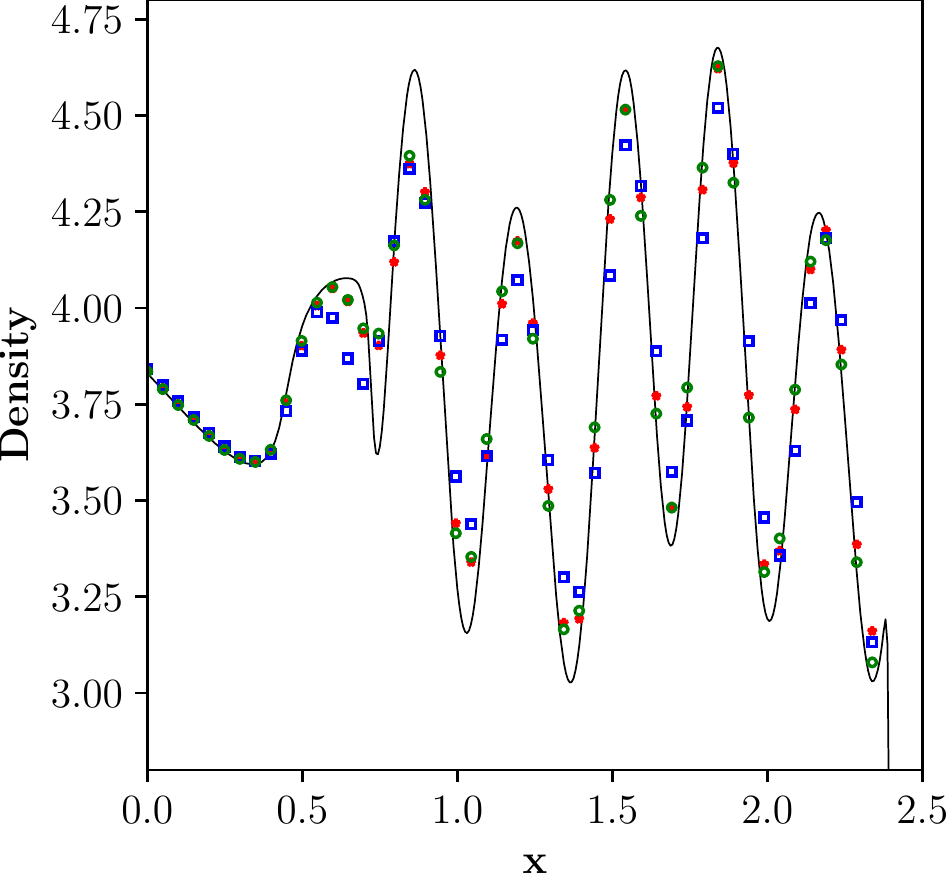}
\label{fig:shu-local}}
\caption{Numerical solution for Shu-Osher problem using $N = 200$ grid points at $t = 1.8$, Example \ref{Shu-Osher}. Solid line: \textit{exact solution}; blue squares: WCNS7M; red stars: MEG7-cons and green circles: MEG7-prim.}
\label{fig:1d-SO}
\end{figure}

The initial conditions of the Titarev and Toro's modified version of the above test case are follows:

\begin{equation}
        \left( \rho,u,p \right) =
        \begin{cases}
            (1.515695,0.523346,1.805), & \text{if } 0 \leq x < 0.5, \\
            (1+0.1\sin(20\pi(x-5)),0,1), & \text{if } 0.5 \leq x \leq 10.
        \end{cases}
\end{equation}

The computational domain of this test case is $x = [0,10]$ and the final time is $t = 5$. For this test case, Simulations are carried out with a uniform mesh of $N = 1000$. The numerical results obtained with MEG7 schemes and WCNS7M are shown in Fig. \ref{fig_tita}. We can observe that MEG7-cons and MEG7-prim schemes give better resolution by capturing the solution's fine-scale structures at the high-frequency waves than the WCNS7M scheme, Fig. \ref{fig:tita2}. In the following examples, the two-dimensional test cases will be considered.

\begin{figure}[H]
\centering
\subfigure[Global density profiles.]{\includegraphics[width=0.48\textwidth]{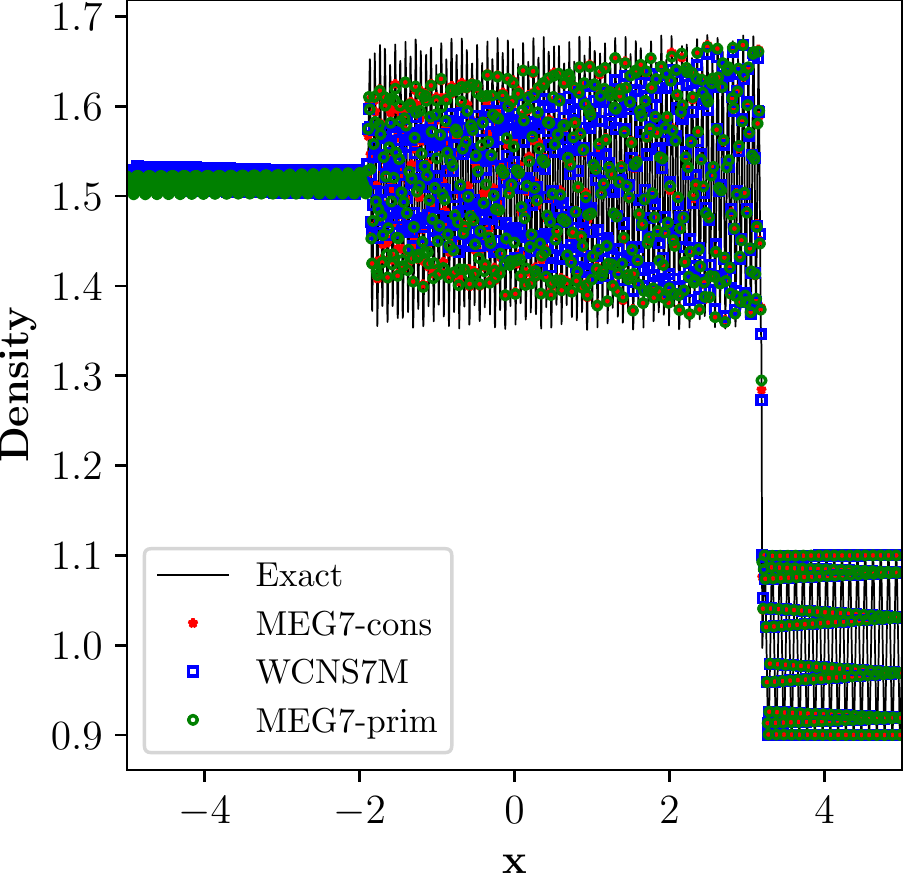}
\label{fig:tita1}}
\subfigure[Local density profiles.]{\includegraphics[width=0.48\textwidth]{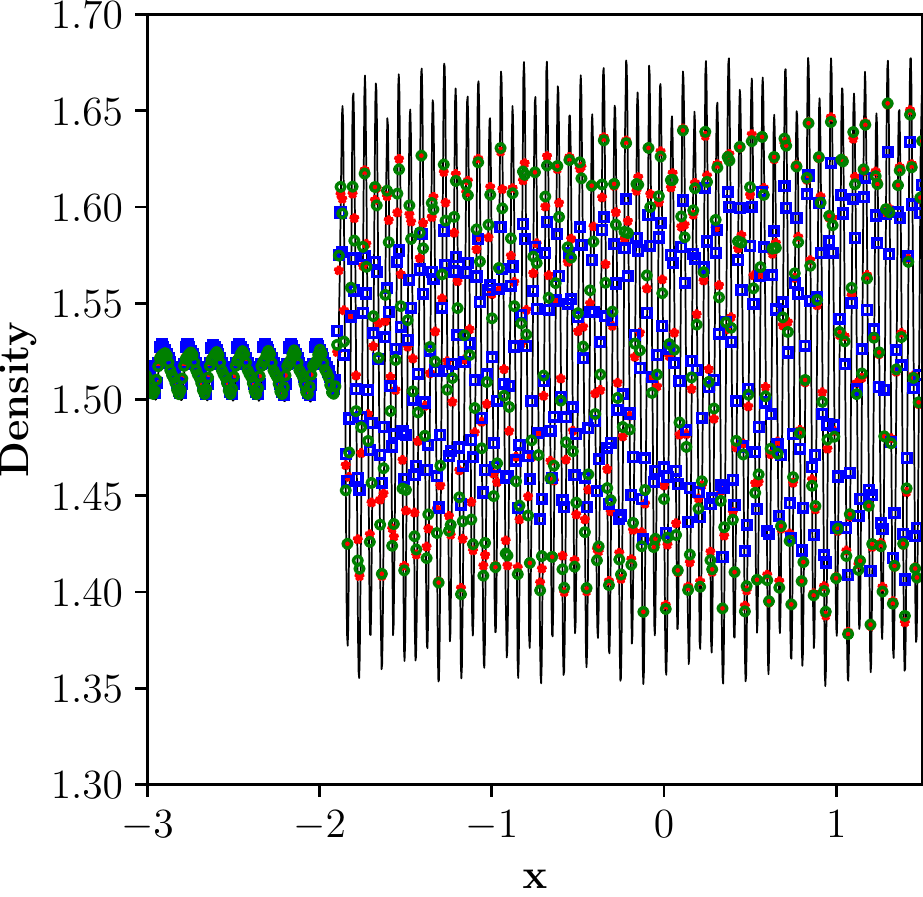}
\label{fig:tita2}}
\caption{Numerical solution for Titarev-Toro problem using $N = 1000$ grid points at $t = 5$. Solid line: \textit{exact solution}; blue squares: WCNS7M; red stars: MEG7-cons and green circles: MEG7-prim.}
\label{fig_tita}
\end{figure}

\begin{example}\label{ex:rm} {Richtmeyer - Meshkov instability}
\end{example}
In this first two-dimensional test case we consider the single-mode Richtmeyer--Meshkov instability problem \textcolor{black}{\cite{deng2019fifth,terashima2009front}}. This instability occurs when an incident shock accelerates a perturbed interface between two fluids of different densities. The computational domain of this test case is $\left[0, 4 \right] \times \left[ 0, 1 \right]$ and simulation is conducted until $t=9$. The test case is initialized with the following initial conditions:
\begin{equation*}
	\left( \rho, u, v, p \right)
 =
 \begin{cases}
 \left(5.04, 0, 0, 1 \right) , &\mbox{if x $<$ 2.9 - 0.1 sin(2$\pi$(y+0.25), perturbed interface}, \\
 	\left(1, 0, 0, 1 \right), &\mbox{if x $<$ 3.2}, \\
    \left(1.4112, -665/1556, 0, 1.628 \right), &\mbox{for $\mathrm{otherwise}$}. \\
 \end{cases}
\end{equation*}
Simulation is carried out on a uniform mesh size of $320 \times 80$, and periodic boundary conditions are used for upper and lower boundaries. The left and right boundary values are fixed to initial conditions. The density distribution contours computed using different schemes are shown in Fig. \ref{fig_RM}. Between WCNS7M and MEG7-cons schemes, the MEG7-cons have a thinner material interface indicating lower numerical dissipation. On the other hand, the MEG7-prim scheme has lower numerical dissipation and also produces small-scale roll-up vortices compared to the other schemes.  
\begin{figure}[H]
\centering\offinterlineskip
\subfigure[WCNS7M]{\includegraphics[width=0.23\textheight]{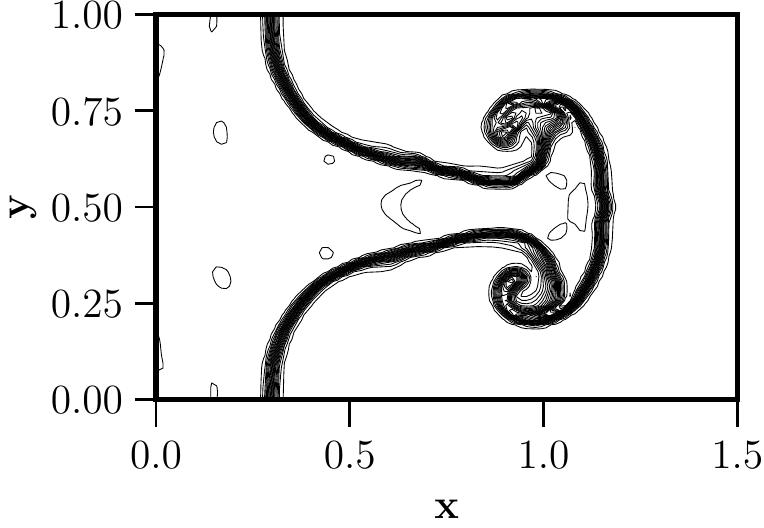}
\label{fig:TENO_RMp}}
\subfigure[MEG7-cons]{\includegraphics[width=0.23\textheight]{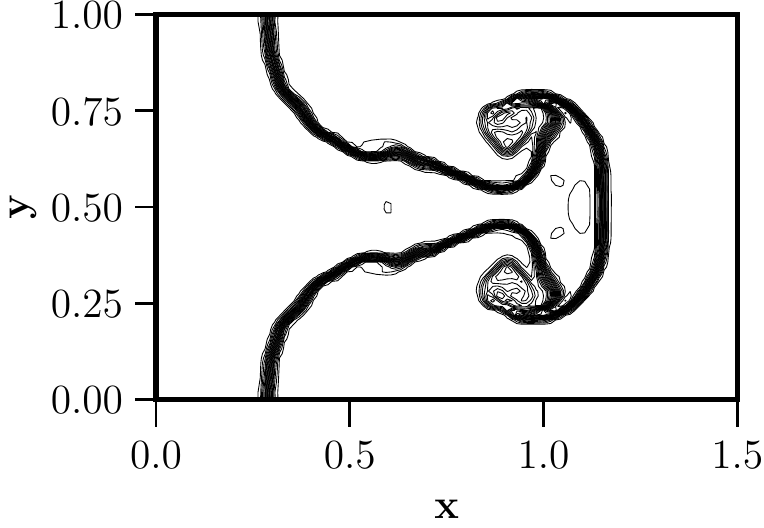}
\label{fig:MIGE_RMp-MP}}
\subfigure[MEG7-prim]{\includegraphics[width=0.23\textheight]{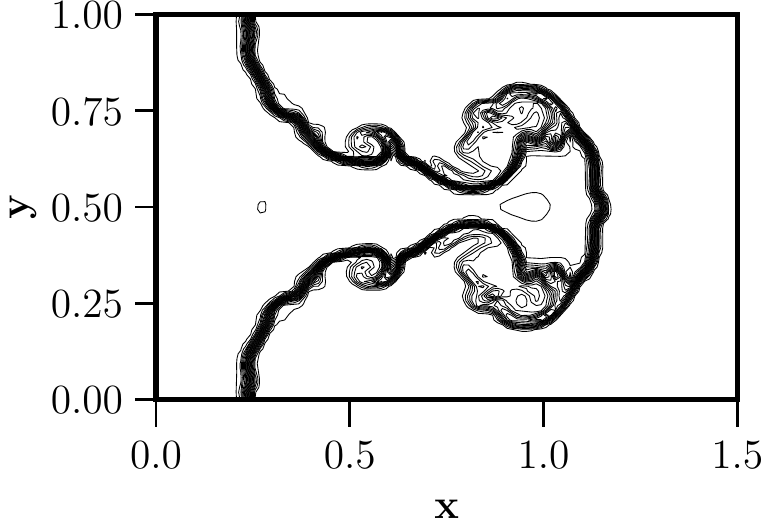}
\label{fig:MIGE_RMp}}
\caption{Numerical results of Richtmeyer - Meshkov instability described in Example \ref{ex:rm} for the considered schemes. The figures are drawn with 24 contours.}
\label{fig_RM}
\end{figure} 

\begin{example}\label{shock-entropy}{2D shock-entropy wave test} 
\end{example}

In this two-dimensional test case, the shock-entropy wave interaction problem \cite{chamarthi2021high,deng2019fifth,acker2016improved} is simulated using the proposed schemes. This test case can be considered an extension of the one-dimensional test case of Example \ref{Shu-Osher}. The computational domain of this test case is $[-5,5]\times [-1,1]$ and the final simulation time is $t$ = 1.8. The initial conditions for this test case are as follows: 
\begin{align}\label{shock_entropy}
(\rho,u,p)=
\begin{cases}
&(3.857143, \ \ 2.629369,\ \ 10.3333),\quad x<-4,\\
&(1+0.2\sin(10x \cos\theta+10y\sin\theta),\ \ 0,\ \ 1),\quad otherwise.
\end{cases}
\end{align}
The value of $\theta$ is taken as $\pi/6$. Simulations are carried out on a mesh size of $400\times 80$, which corresponds to $\Delta x= \Delta y =1/40$. For this test case, the shock-stable HLLC Riemann solver of Fleischmann et al. is used \cite{fleischmann2020shock} to prevent the carbuncle effect observed in \cite{chamarthi2023gradient}. The reference solution for this test case is computed using the MIG4 scheme \cite{chamarthi2018high} on a fine mesh of 1600 $\times$ 320. Figs. \ref{fig:SE-WCNS7M}, \ref{fig:SE-MEG7-cons} and \ref{fig:SE-MEG7-prim} shows the density contours of the simulations by using WCNS7M, MEG7-cons and MEG7-prim, respectively. These figures indicate the flow structures are better captured by the MEG7 schemes compared to the WCNS7 scheme even though all of them are seventh-order accurate.

The local density profile in the high-frequency region along $y$ = 0 is shown in Fig.\ref{fig:SSE-Compare}. It can be seen from the figure that the MEG7 approach predicts the density amplitudes with lower numerical dissipation than the WCNS7M scheme. In addition, MEG7 has less dispersion error around the region where the weak shock and entropy wave interacts, indicating superior resolution characteristics.
\begin{figure}[H]
\centering\offinterlineskip
\subfigure[WCNS7M]{\includegraphics[width=0.48\textwidth]{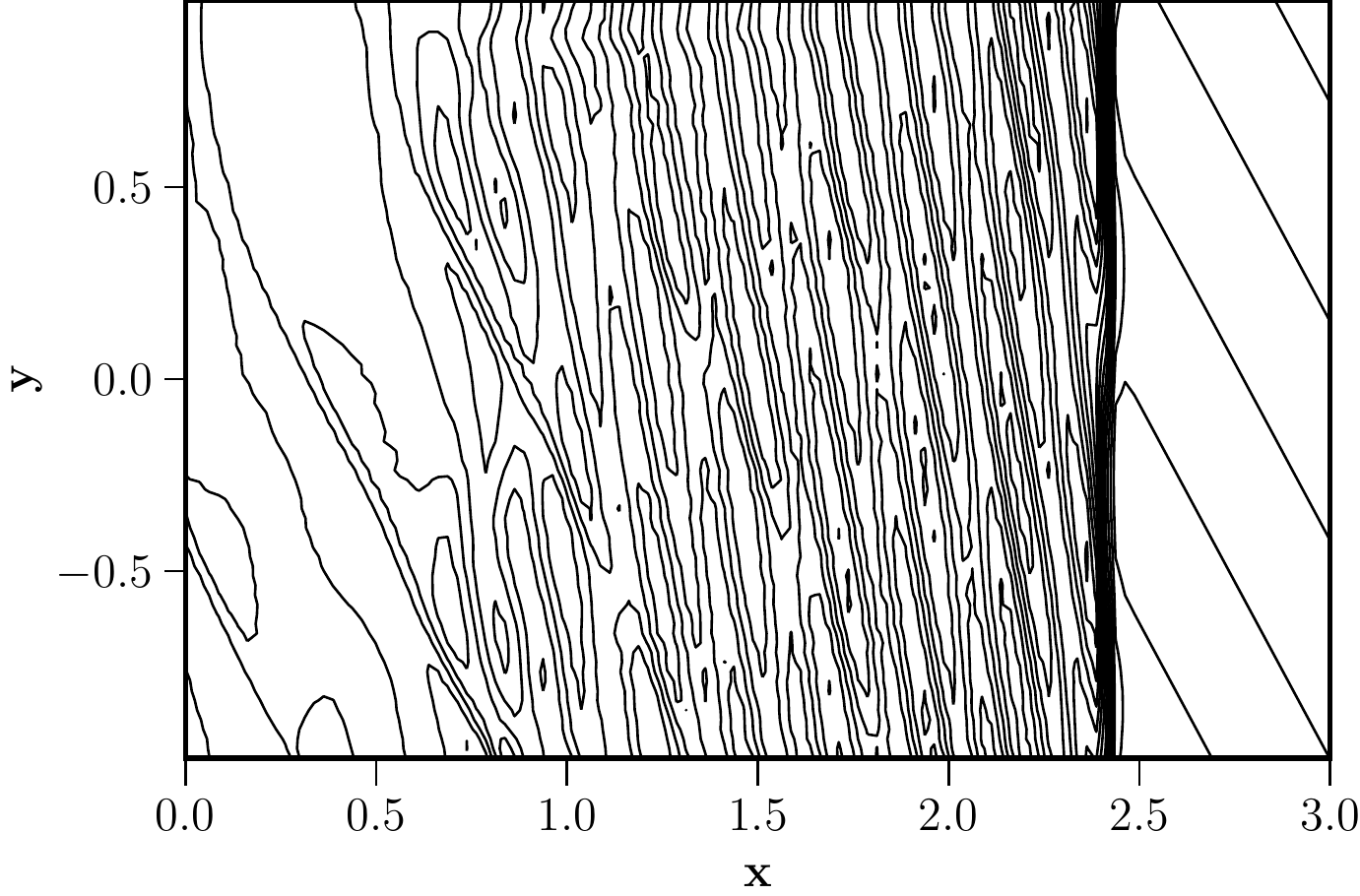}
\label{fig:SE-WCNS7M}}
\subfigure[\textcolor{black}{MEG7-cons}]{\includegraphics[width=0.48\textwidth]{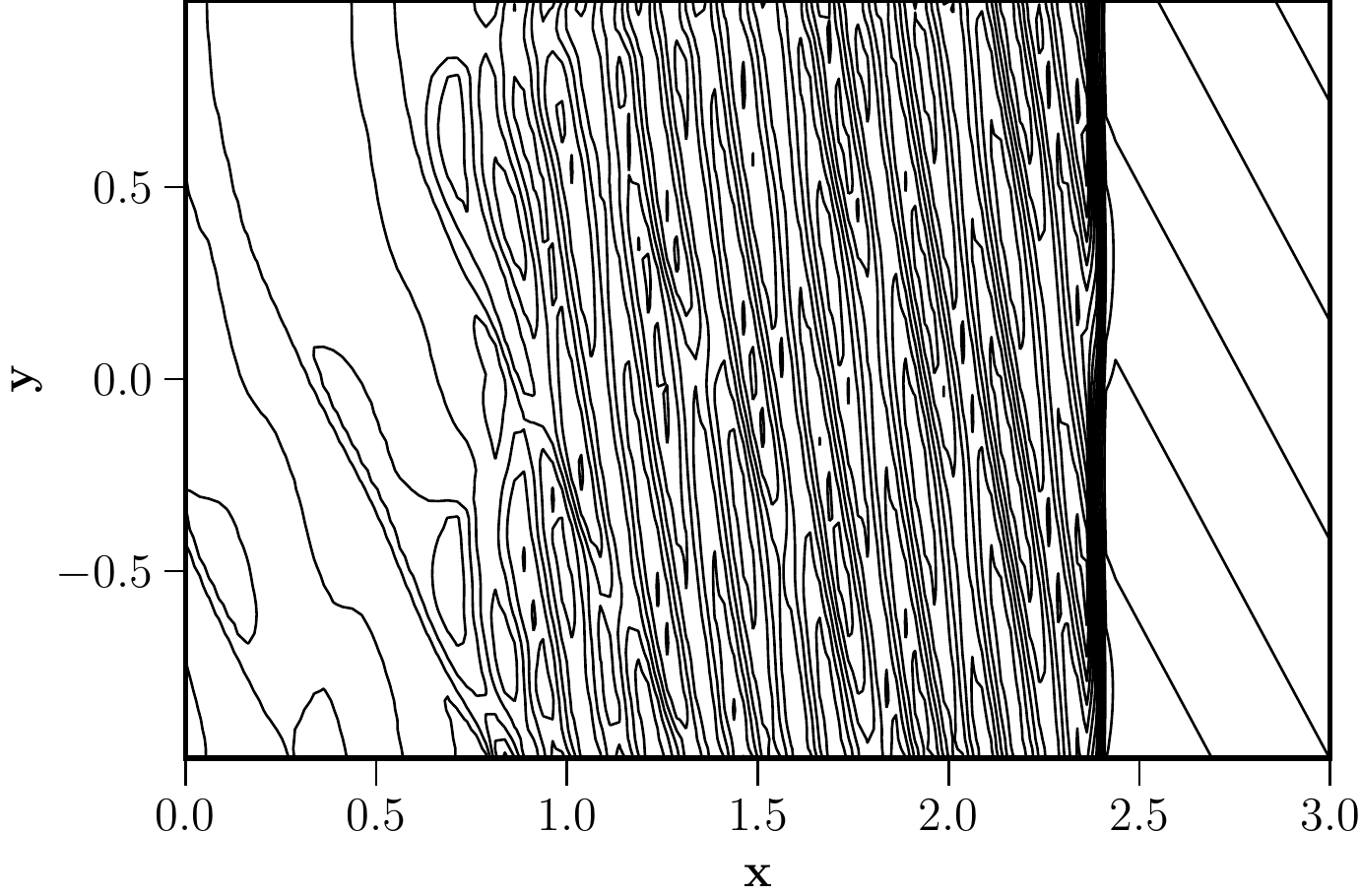}
\label{fig:SE-MEG7-cons}}
\subfigure[MEG7-prim]{\includegraphics[width=0.48\textwidth]{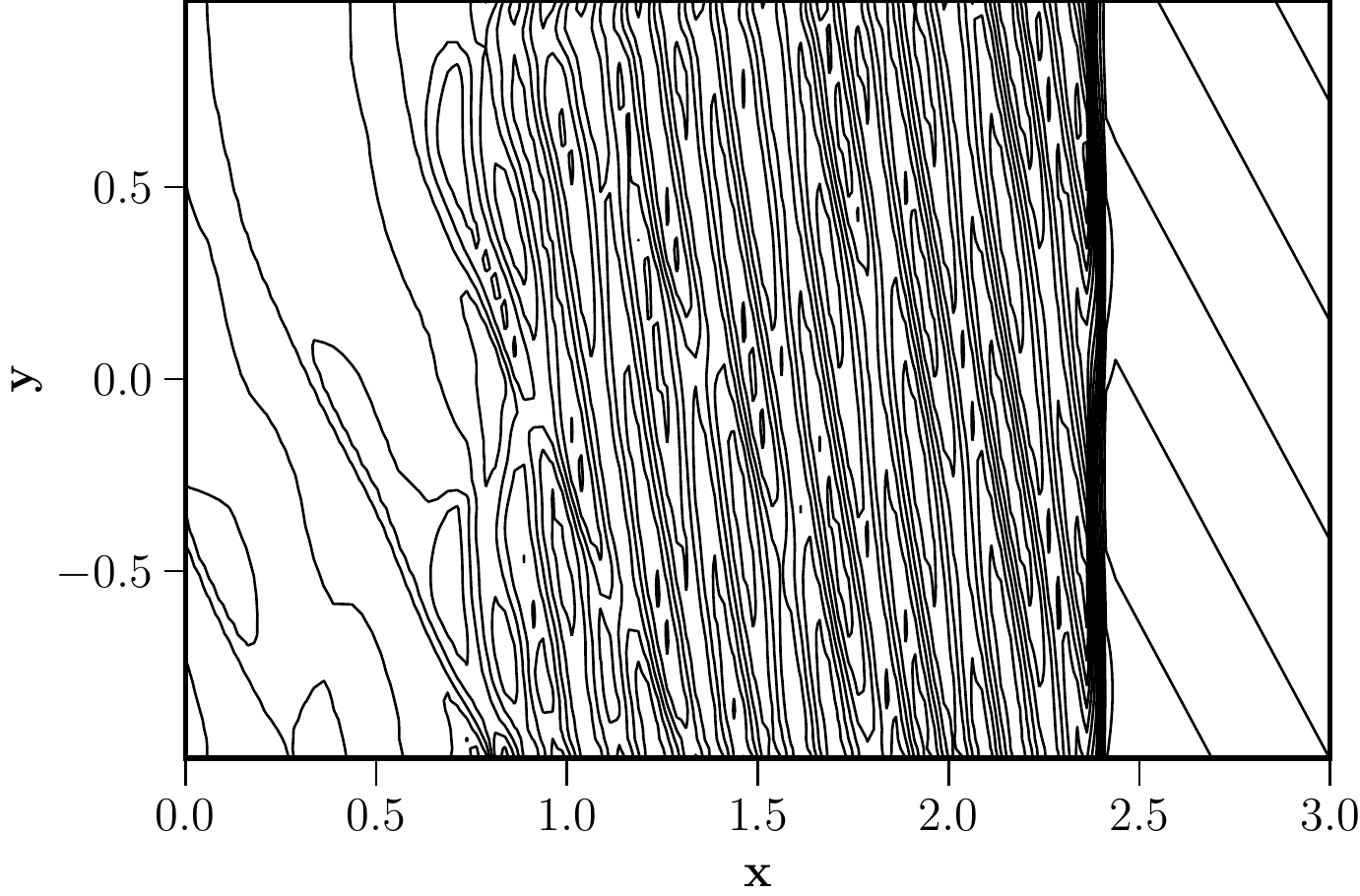}
\label{fig:SE-MEG7-prim}}
\subfigure[Local profile]{\includegraphics[width=0.495\textwidth]{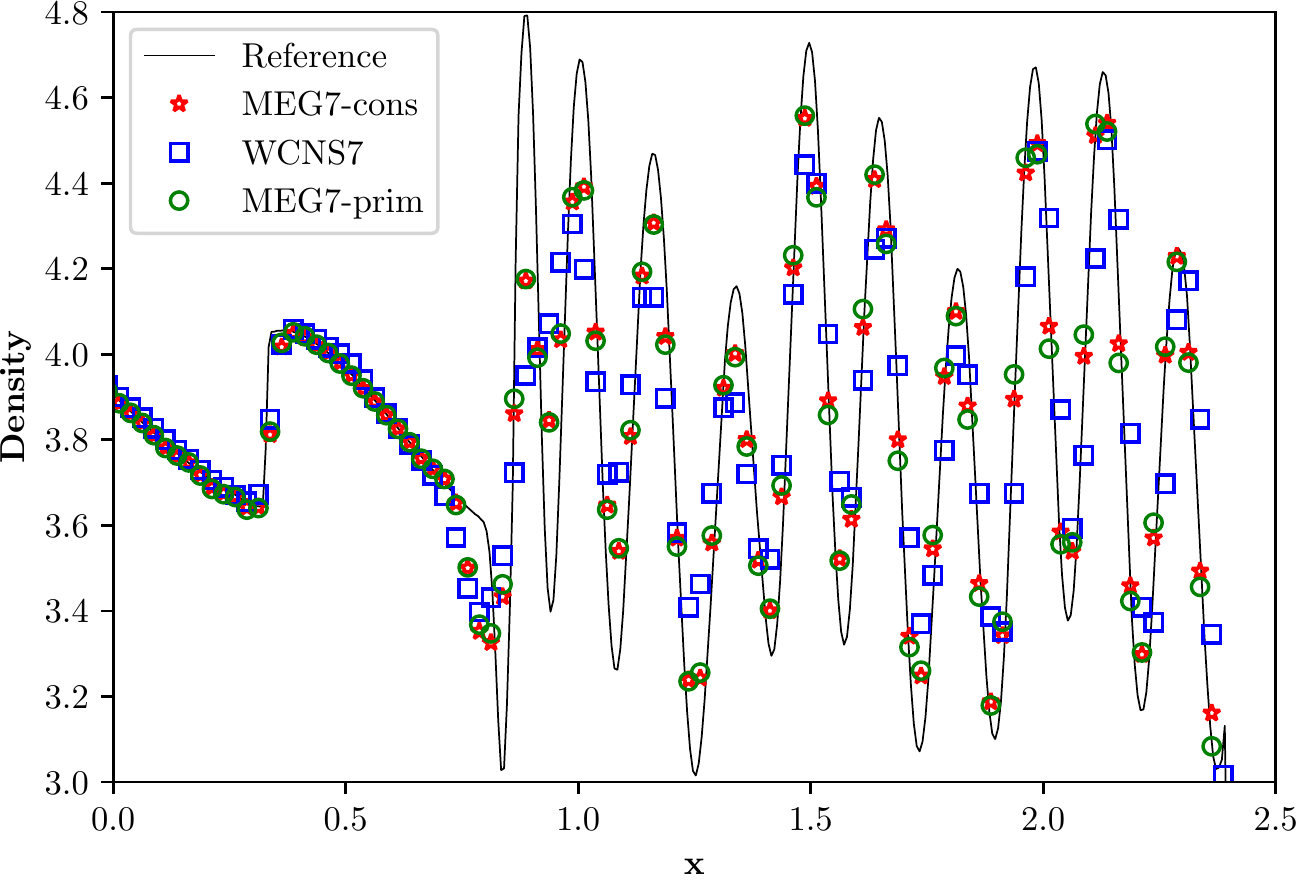}
\label{fig:SSE-Compare}}
\caption{15 density contours for the 2D shock-entropy wave test at $t=1.8$, Example \ref{shock-entropy}, for various schemes are shown in Figs. (a), (b) and (c), Solid line: \textit{Reference solution}; blue squares: WCNS7M; red stars: MEG7-cons and green circles: MEG7-prim.}
\label{fig_shock_entropy}
\end{figure}

\begin{example}\label{ex:dmr}{Double Mach reflection}
\end{example}

The proposed schemes are further assessed for the Double Mach Reflection case of Woodward and Collela \cite{woodward1984numerical}. In this test case, a Mach $10$ unsteady planar shock wave impinges on a 30-degree inclined surface. The computational domain for this test case is $[x,y] = [0, 4] \times [0, 1]$, and the case was run until a final time, $t = 0.2$. The initial conditions for this test case are as follows:

\begin{equation}
\begin{aligned}
(\rho,u,v,p)=
\begin{cases}
&(8,\ 8.25 \cos 30^\circ,\ -8.25 \sin 30^\circ,\
116.5),\quad x<1/6+\frac{y}{\tan 60^\circ},\\
&(1.4,\ 0,\ 0,\
1),\quad\quad\quad\quad\quad\quad\quad\quad\quad\quad\quad\ x>
1/6+\frac{y}{\tan 60^\circ}.
\end{cases}
\end{aligned}
\label{eu2D_mach}
\end{equation}
The boundary conditions are the same as that of Woodward and Collela \cite{woodward1984numerical}.  Simulations are carried out on a uniform grid of $1024 \times 256$ with all the considered schemes. For comparison, The reference result for this test case is computed on a grid size of about $4096 \times 1024$ using the HOUCS6 scheme \cite{chamarthi2021high} and is shown in Fig. \ref{fig:dmr-ref}. Observing Fig. \ref{fig_doublemach}, the vortices generated along the slip line by the Kelvin Helmholtz instability and near-wall jet can be seen for the proposed schemes. The MEG7-cons better resolve the vortices along the slip line than the WCNS7 approach. The results obtained by the MEG7-prim do not show vortices along the slip line, but the near wall jet is better resolved. It indicates that the primitive variable reconstruction is slightly more dissipative for this test case. Shock stable HLLC Riemann solver of Ref. \cite{fleischmann2020shock} is also used for this test case.

\begin{figure}[H]
\centering\offinterlineskip
\subfigure[WCNS7M]{\includegraphics[width=0.42\textwidth]{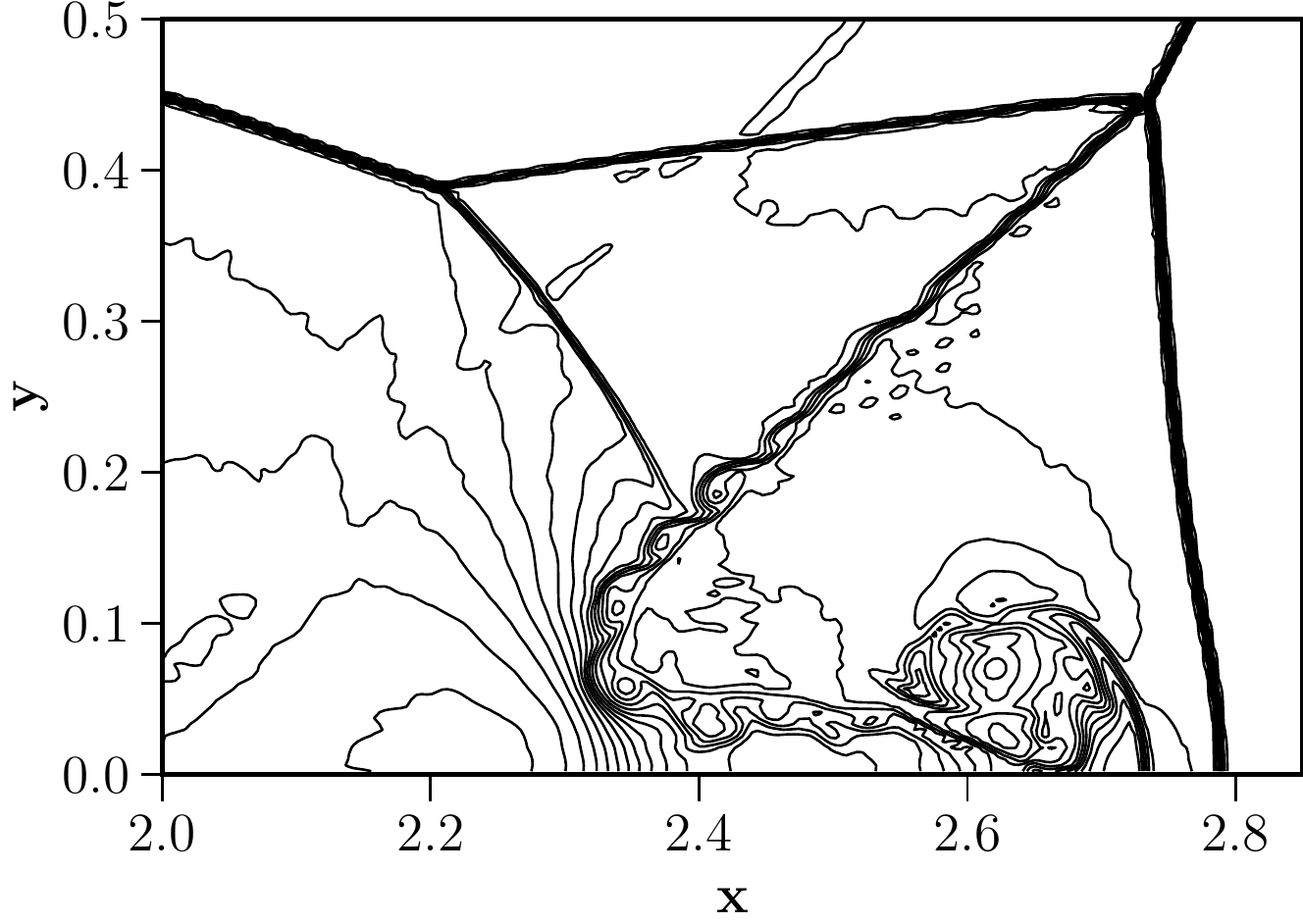}
\label{fig:wcns_DMR_prim}}
\subfigure[MEG7-cons]{\includegraphics[width=0.42\textwidth]{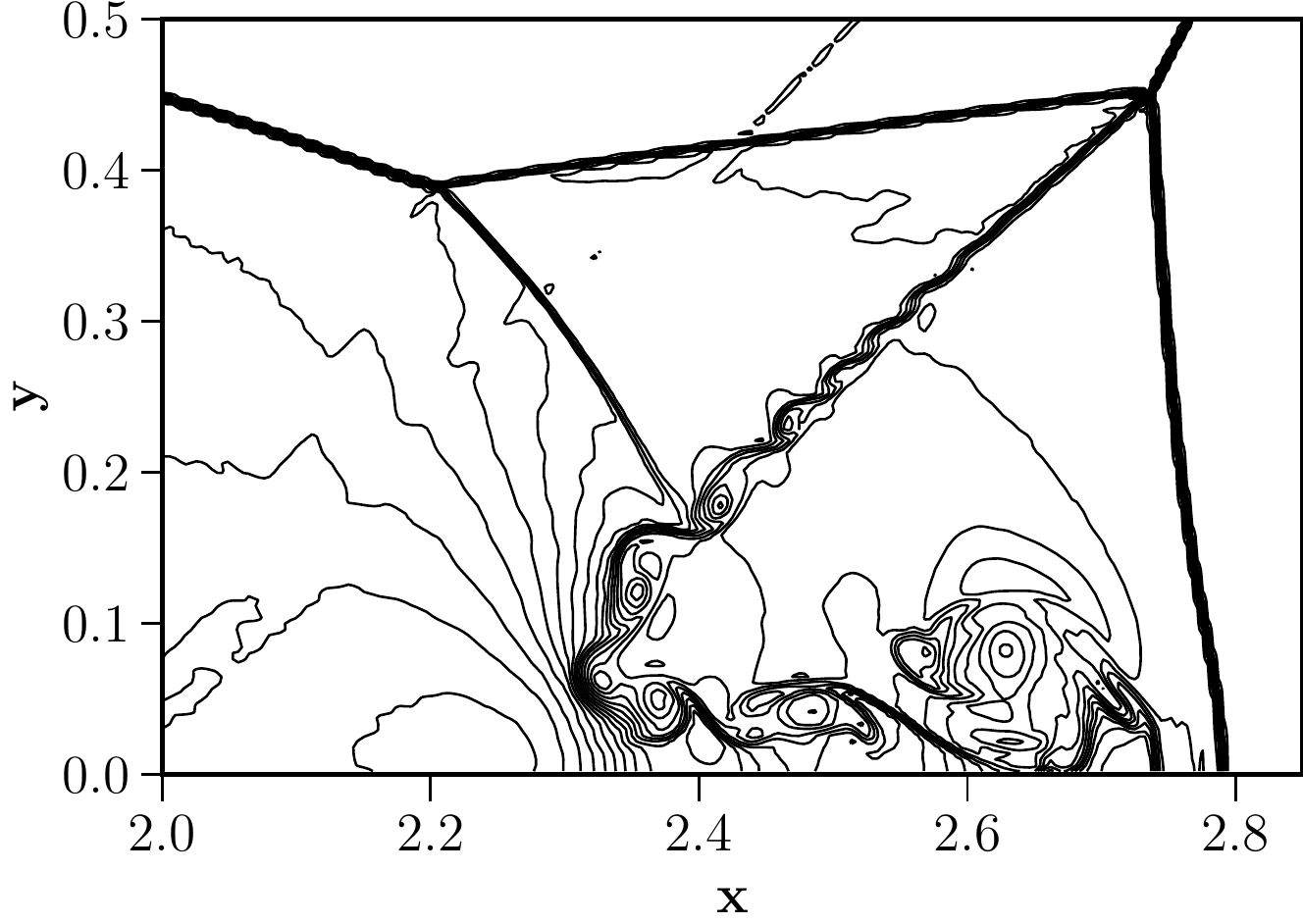}
\label{fig:Meg7_DBc}}
\subfigure[MEG7-prim]{\includegraphics[width=0.42\textwidth]{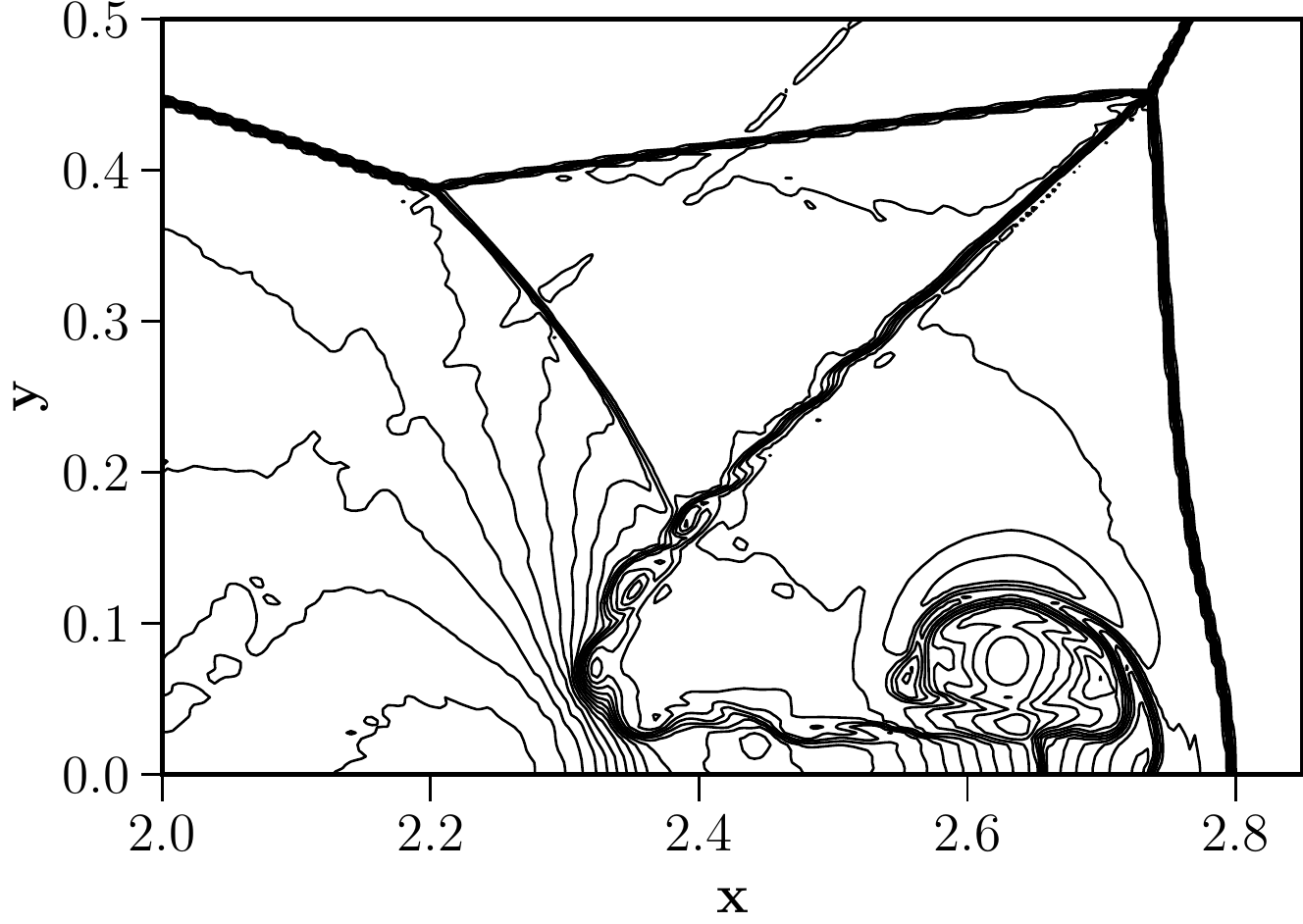}
\label{fig:Meg7_DBp}}
\subfigure[Reference]{\includegraphics[width=0.42\textwidth]{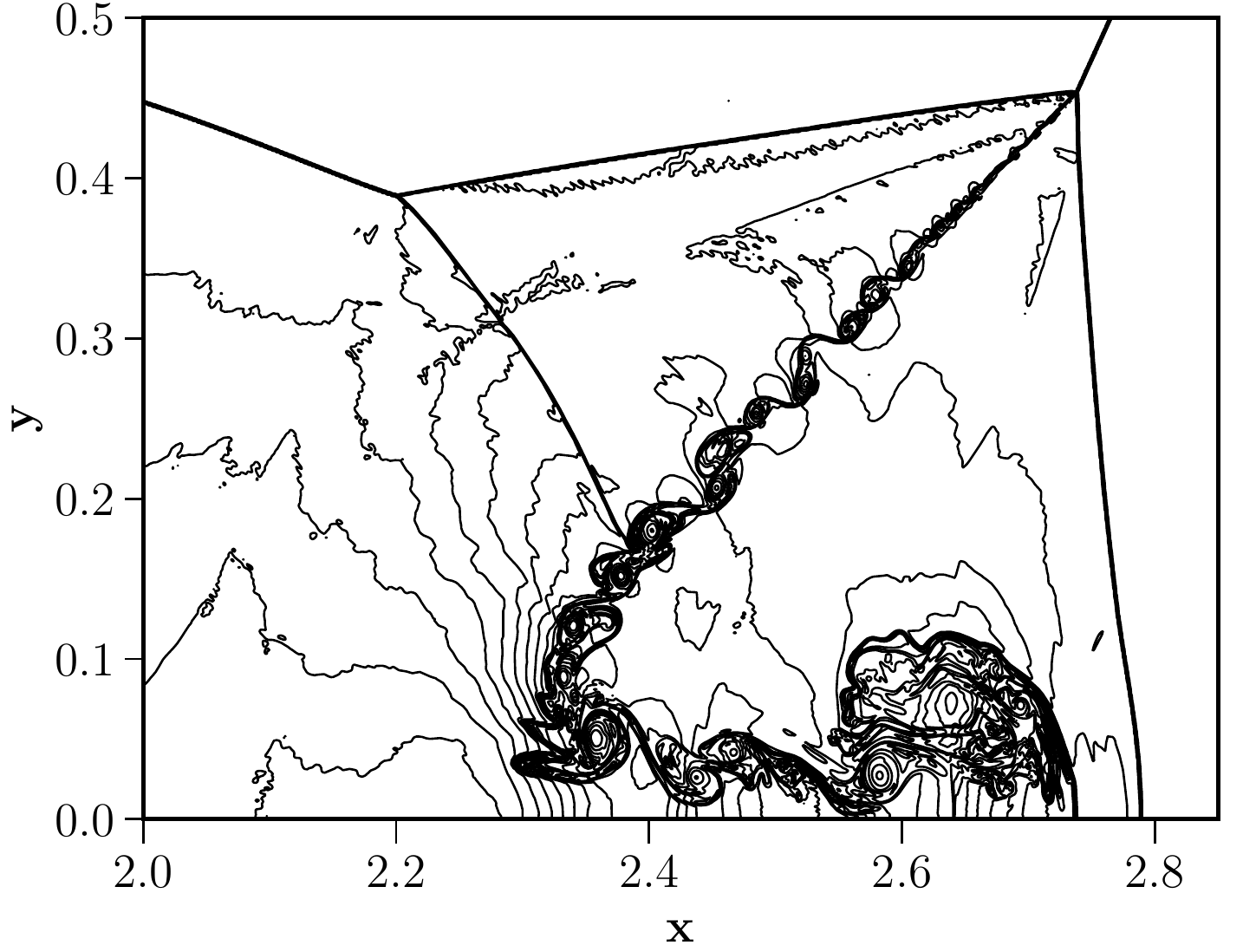}
\label{fig:dmr-ref}}
\caption{Computed density contours of the zoomed in Mach stem region of the Double Mach Reflection for the considered schemes. The figures are drawn with 38 contours.}
\label{fig_doublemach}
\end{figure}

\begin{example}\label{ex:rp}{Riemann Problem}
\end{example}

In this final inviscid test case, the two-dimensional Riemann problem of Schulz-Rinne et al. \cite{schulz1993numerical} is considered. Similar to the earlier test case, the small-scale structures formed along the slip lines due to Kelvin-Helmholtz instability indicate the numerical dissipation of the numerical scheme. The computational domain for this test case is $[x,y] = [0,1]\times [0,1]$ and the case is run until a final time, $t=0.8$. The initial conditions for the configuration considered are as follows:

\begin{equation}\label{ex:rp1}
        \left( \rho,u,v,p \right) =
        \begin{cases}
            (1.5,0,0,1.5), & \text{if } x > 0.8, y > 0.8, \\
            (33/62,4/\sqrt{11},0,0.3), & \text{if } x \leq 0.8, y > 0.8, \\
            (77/558,4/\sqrt{11},4/\sqrt{11},9/310), & \text{if } x \leq 0.8, y \leq 0.8, \\
            (33/62,0,4/\sqrt{11},0.3), & \text{if } x > 0.8, y \leq 0.8.
        \end{cases}
\end{equation}
Simulations are carried out on a uniform grid size of $400 \times 400$. Figs. \ref{fig_riemann} shows the density contours obtained by the considered schemes.
\begin{figure}[H]
\begin{onehalfspacing}
\centering\offinterlineskip
\subfigure[WCNS7M]{\includegraphics[width=0.45\textwidth]{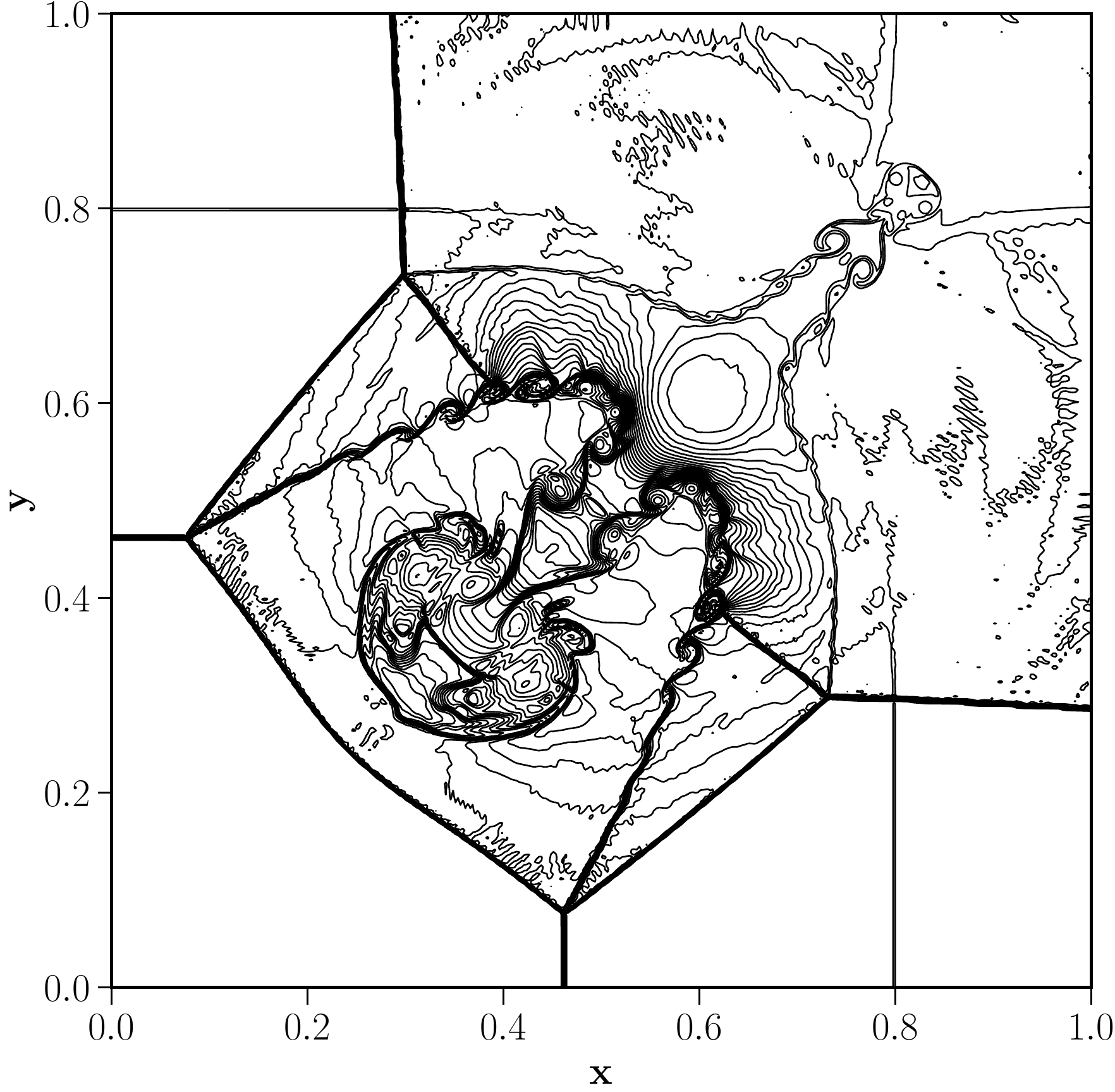}
\label{fig:wcns7_prim12}}
\subfigure[MEG7-cons]{\includegraphics[width=0.45\textwidth]{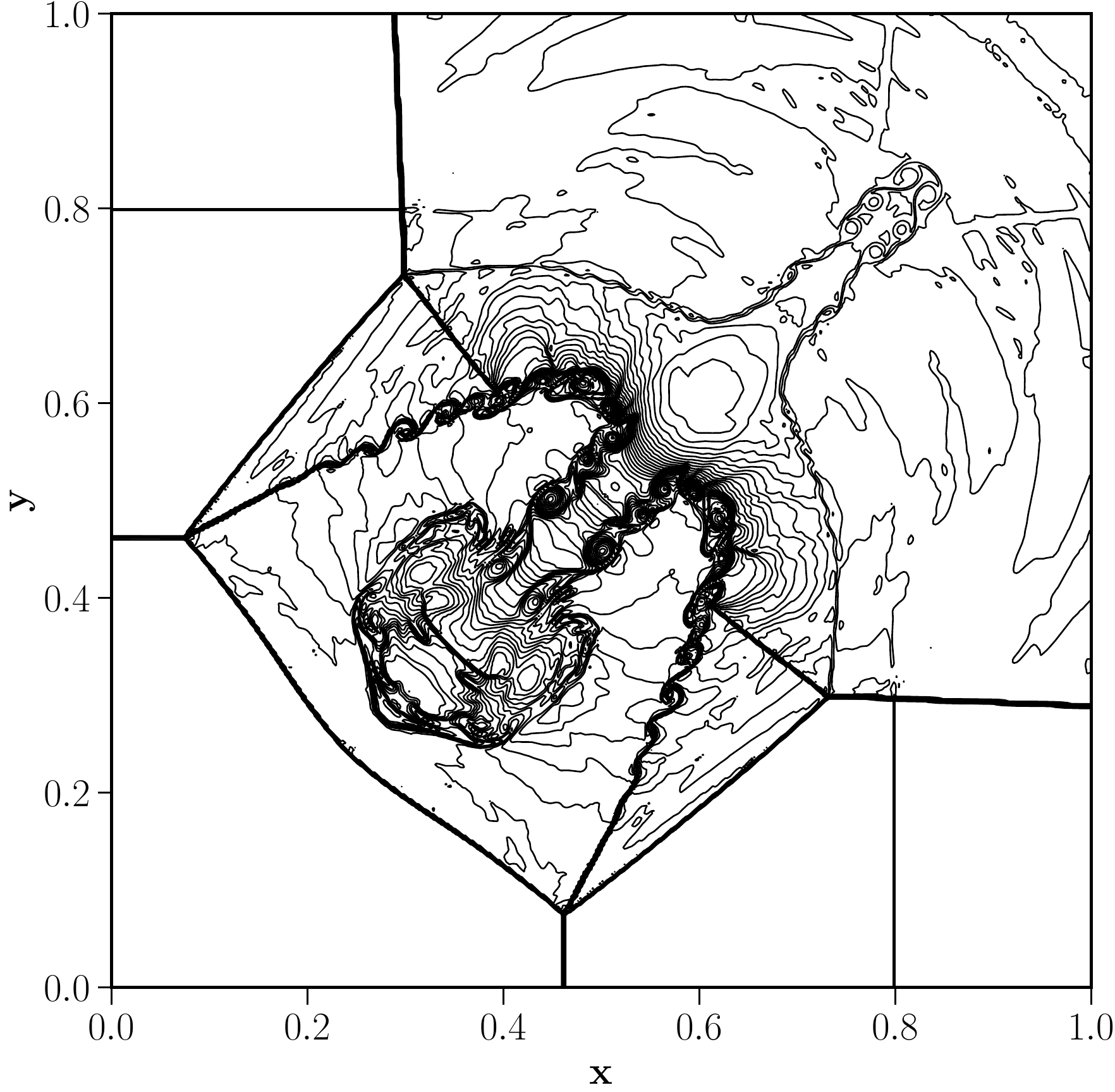}
\label{fig:Meg7_cons}}
\subfigure[MEG7-prim]{\includegraphics[width=0.45\textwidth]{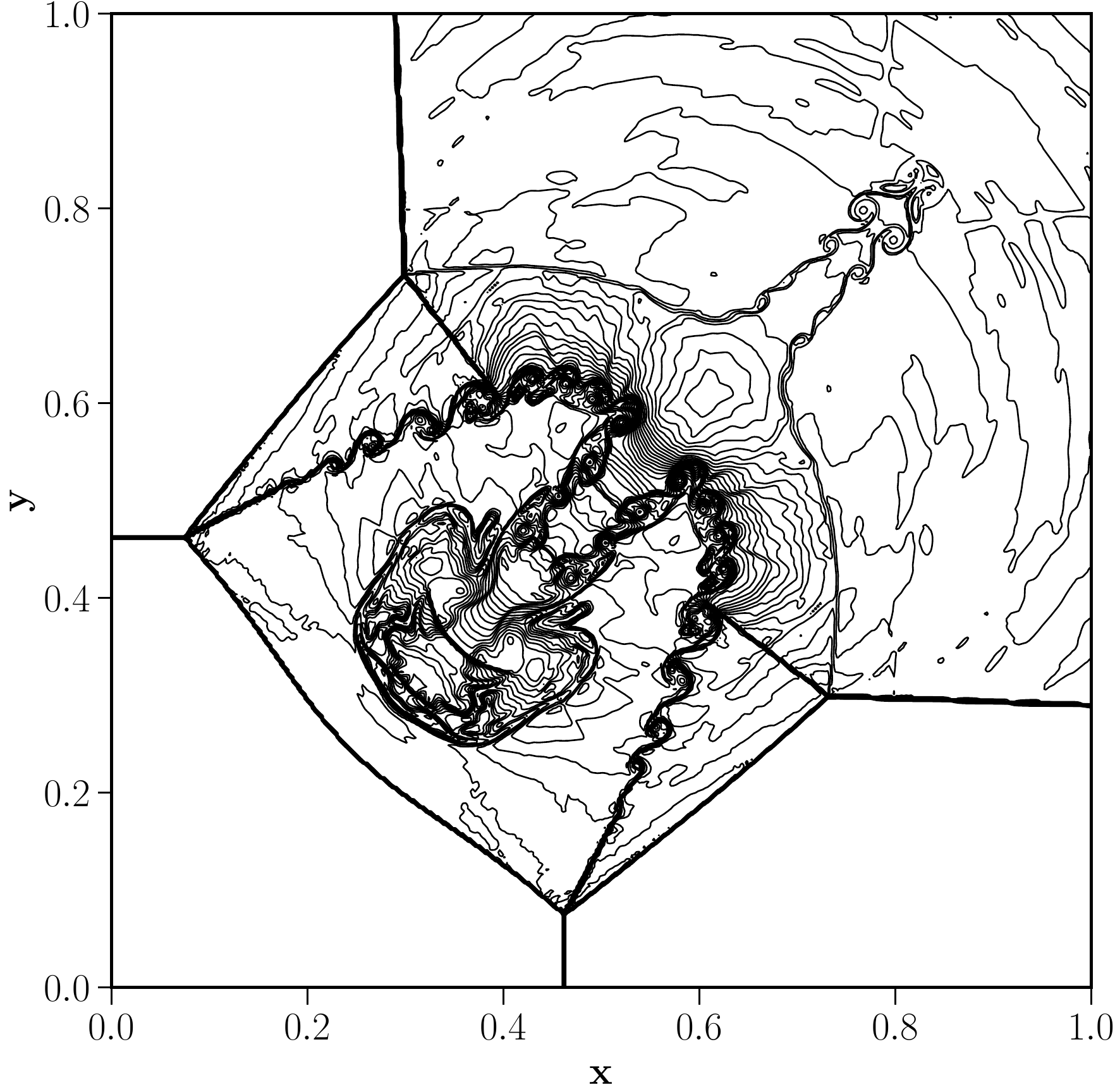}
\label{fig:Meg7_prim}}
\caption{Computed density contours of the Riemann problem described in Example \ref{ex:rp} for the considered schemes using WCNS7M, MEG7-cons, and MEG7-prim schemes, respectively. The figures are drawn with 40 contours.}
\label{fig_riemann}
\end{onehalfspacing}
\end{figure}

 Observing the figures \ref{fig:wcns7_prim12} and \ref{fig:Meg7_cons}, the MEG7-cons scheme better resolved the small-scale structures along the slip lines compared to the WCNS7M scheme. Density contours obtained using the MEG7-prim are shown in \ref{fig:Meg7_prim}. It also resolved the small-scale structures, and the difference between the conservative and primitive variable reconstruction is minor. These results indicate that the proposed approach has lower dissipation than the WCNS7M.
 
 \begin{example}\label{ex:vs}{Viscous Shock tube}
\end{example}

In this test case, we consider the viscous shock tube problem \cite{daru2009numerical} in order to evaluate the proposed scheme for viscous flow simulations. The computational domain for this test case is $[x,y] = [0,1]\times [0,0.5]$ and the case is run until a final time, $t=1.0$. The initial conditions for this test case are as follows:

\begin{equation}\label{vst}
\begin{aligned}
\left( {\rho , u,v, p} \right) = \left\{ \begin{array}{l}
\left( {120, 0 ,0,120/\gamma } \right),  \quad 0 < x < 0.5,\\
\left( {1.2 , 0 ,0, 1.2/\gamma } \right),  \quad 0.5 \le x < 1.
\end{array} \right.
\end{aligned}
\end{equation}
The propagation of the incident shock wave and contact discontinuity leads to a thin boundary layer at the bottom wall, and its reflection at the right wall interacts with this boundary layer. These interactions lead to a lambda-shaped shock pattern,  complex vortex system, and separation region, making it an ideal test case for evaluating high-resolution schemes. 
\begin{figure}[H]
\begin{onehalfspacing}
\centering\offinterlineskip
\subfigure[WCNS7M]{\includegraphics[width=0.45\textwidth]{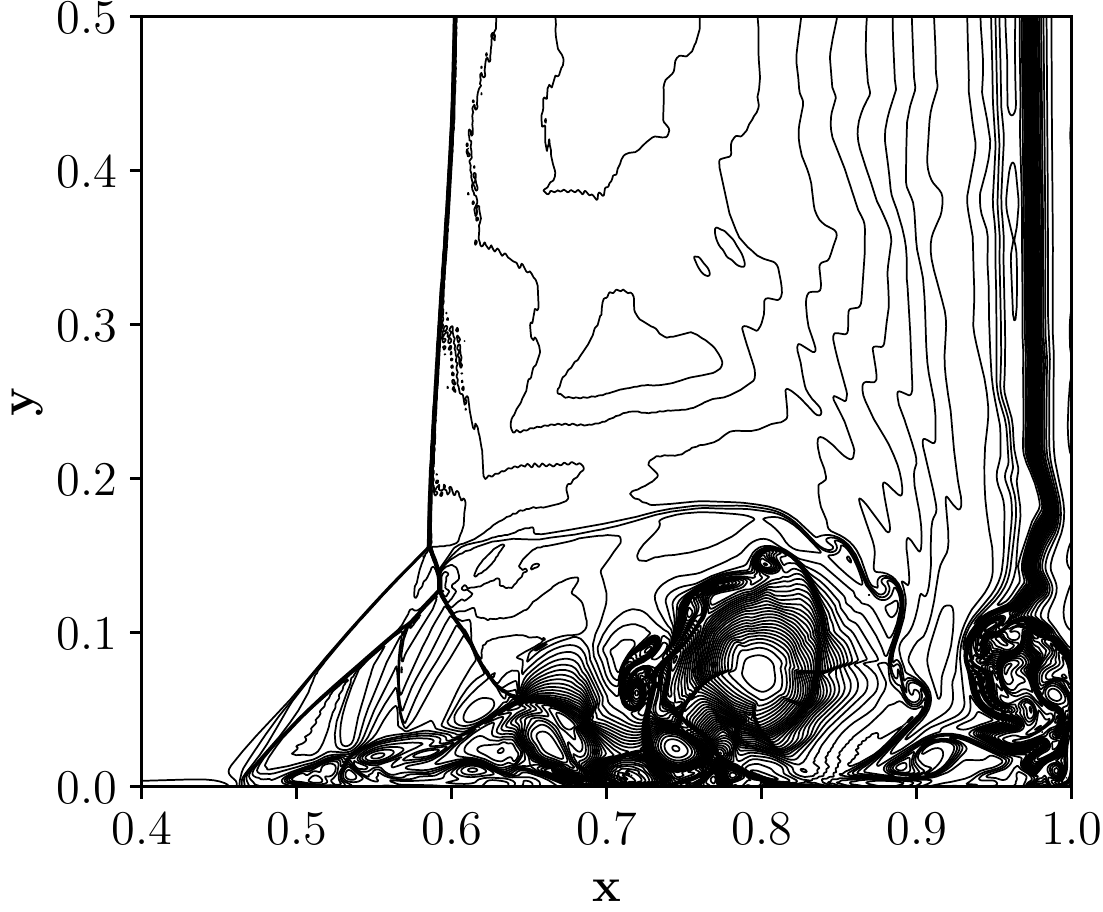}
\label{fig:VST_WCNS7M}}
\subfigure[MEG7-cons]{\includegraphics[width=0.45\textwidth]{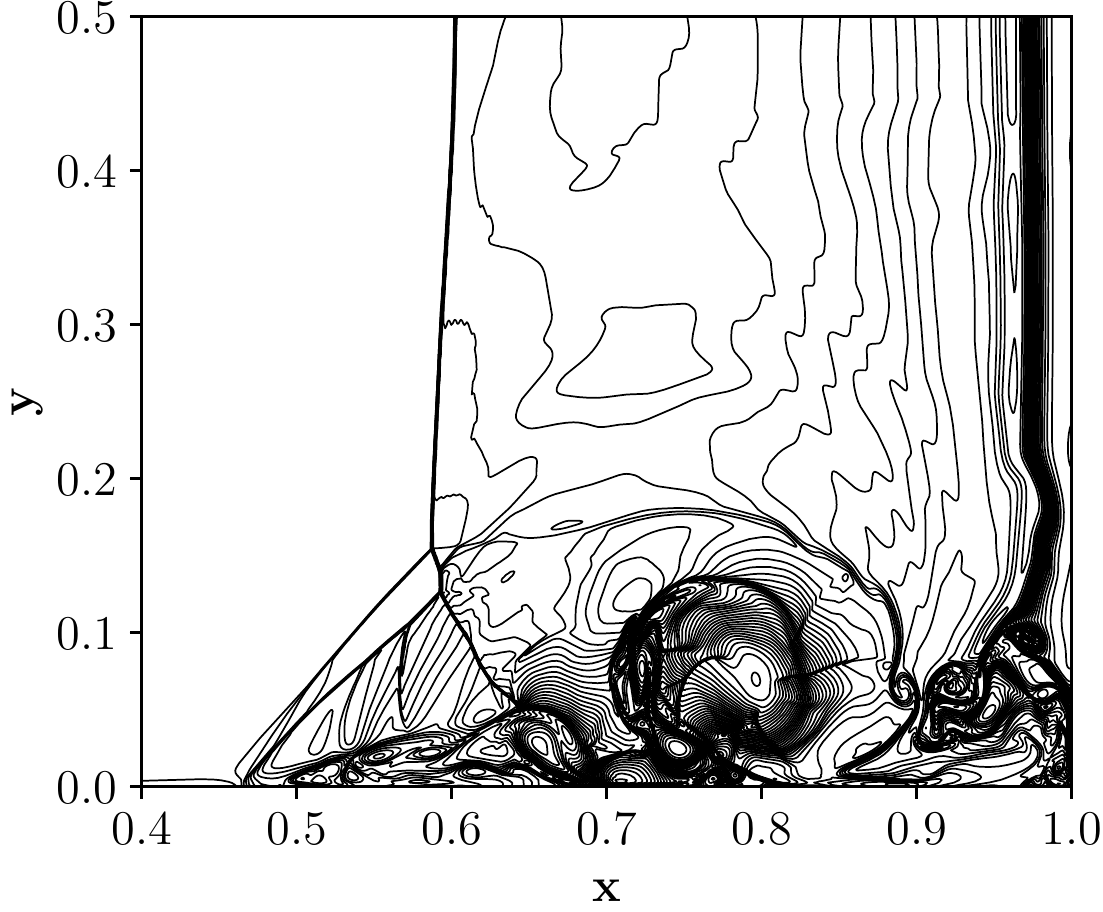}
\label{fig:VST_MEG7-cons}}
\subfigure[MEG7-prim]{\includegraphics[width=0.45\textwidth]{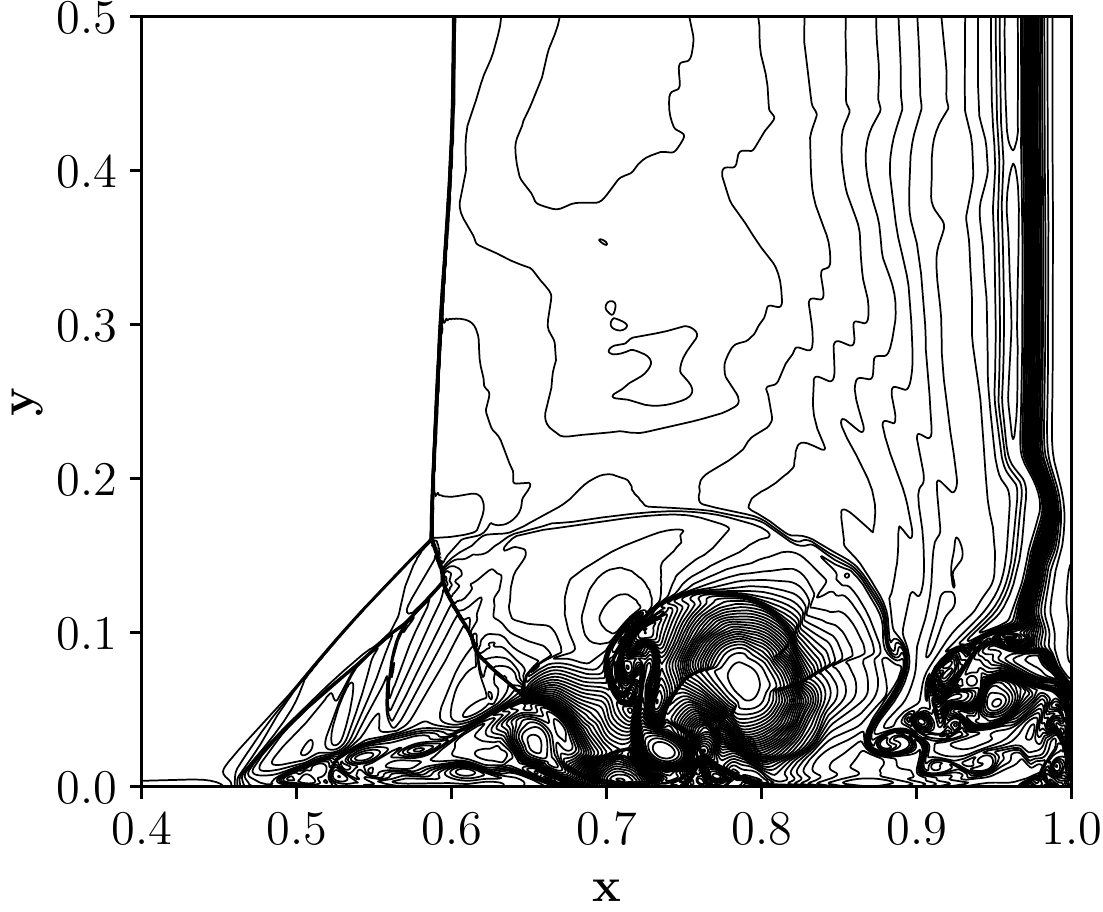}
\label{fig:VST_MEG7-prim}}
\subfigure[Reference solution]{\includegraphics[width=0.45\textwidth]{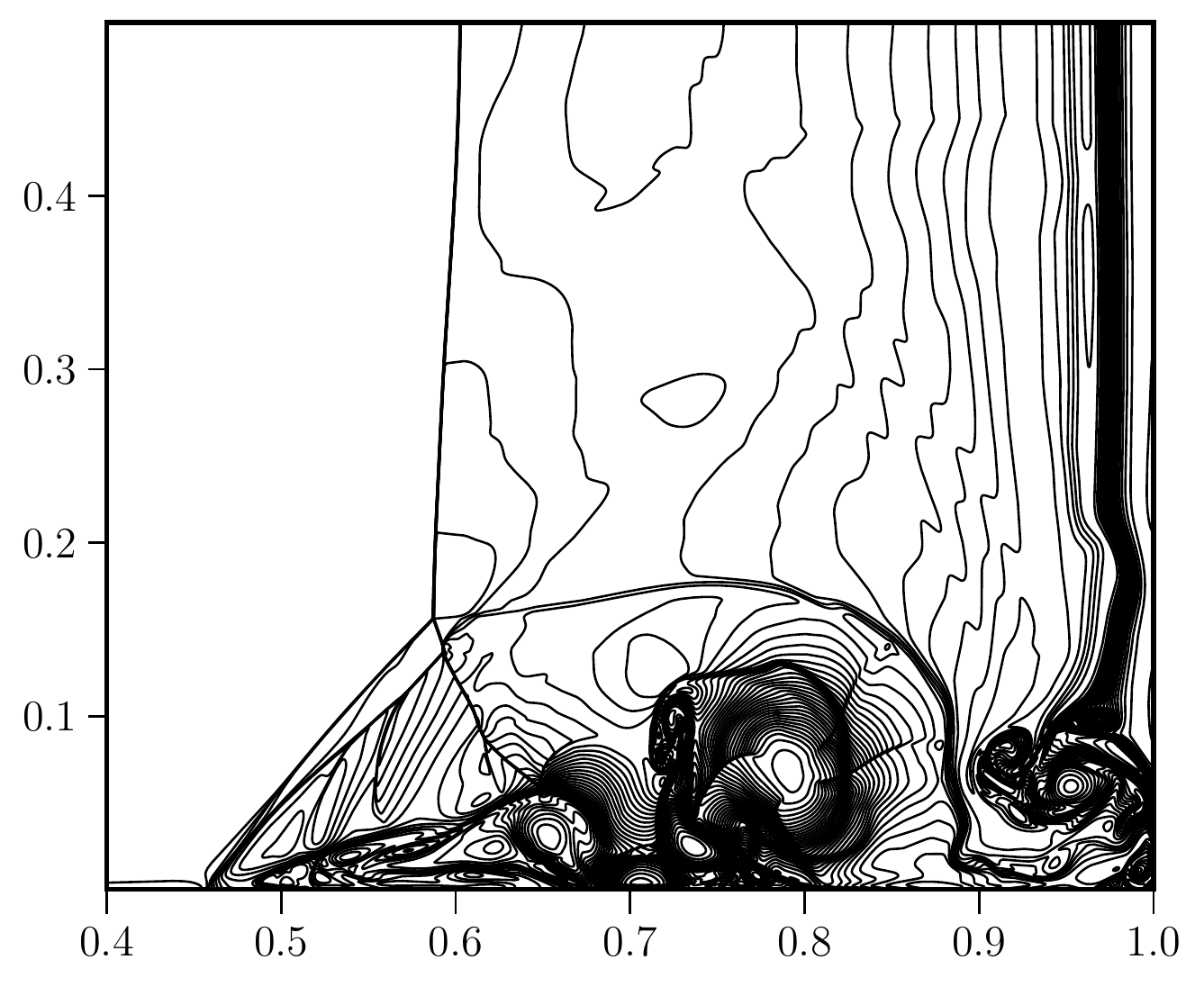}
\label{fig:VST_ref}}
\caption{Density contours obtained by using the different schemes for the Example \ref{ex:vs} for $Re=1000$ on a grid size of 1280 $\times$ 640 at $t$=1.0 are shown in Figs. \ref{fig:VST_WCNS7M}, \ref{fig:VST_MEG7-cons} and \ref{fig:VST_MEG7-prim}. Fig. \ref{fig:VST_ref} shows the reference solution.}
\label{fig_damp-vst}
\end{onehalfspacing}
\end{figure}

The simulations are carried out for a Reynolds number of 1000 as in Ref. \textcolor{black}{\cite{chamarthi2022,daru2009numerical}}. It has been shown that the $\alpha$-damping approach for viscous fluxes will prevent odd-even decoupling \cite{chamarthi2022}. Fig. \ref{fig_damp-vst} shows the density contours obtained with the proposed schemes and the WCNS7M approach for the inviscid fluxes on a grid size of 1280 $\times$ 640. The reference solution for this test case is carried out on a grid size of 4000 $\times$ 2000 and is shown in Fig. \ref{fig:VST_ref}. Compared with the reference solution, the WCNS7M's results show significant deviation in the primary vortex at location $x$=0.75 and the vortex system at the right wall. Figs. \ref{fig:VST_MEG7-cons}  and \ref{fig:VST_MEG7-prim} show the density contours with MEG7-cons and MEG7-prim schemes, respectively, and these results are much closer to the reference solution, which indicates that the current scheme can be used for complex multi-scale viscous simulations on a coarse grid. 

\begin{example}\label{ex:ssl}{Shock-shear layer}
\end{example}

In this example, the two-dimensional viscous shock-shear layer test case of Yee et al. \cite{yee1999low} is considered. The computational domain is $[x,y] = [0,200] \times [-20,20]$, the final time of simulation is $t$ =120. A 12-degree oblique shock propagates from the top boundary, impinges on a spatially evolving mixing layer, and later reflects from the bottom boundary. The Reynolds number for this test case is $Re = 500$, the initial convective Mach number of the mixing layer is $Ma = 0.6$, and the Prandtl number is set as $Pr = 0.72$. 
\begin{figure}[H]
\centering\offinterlineskip
\subfigure[WCNS7M, $320 \times 80$]{\includegraphics[width=0.8\textwidth]{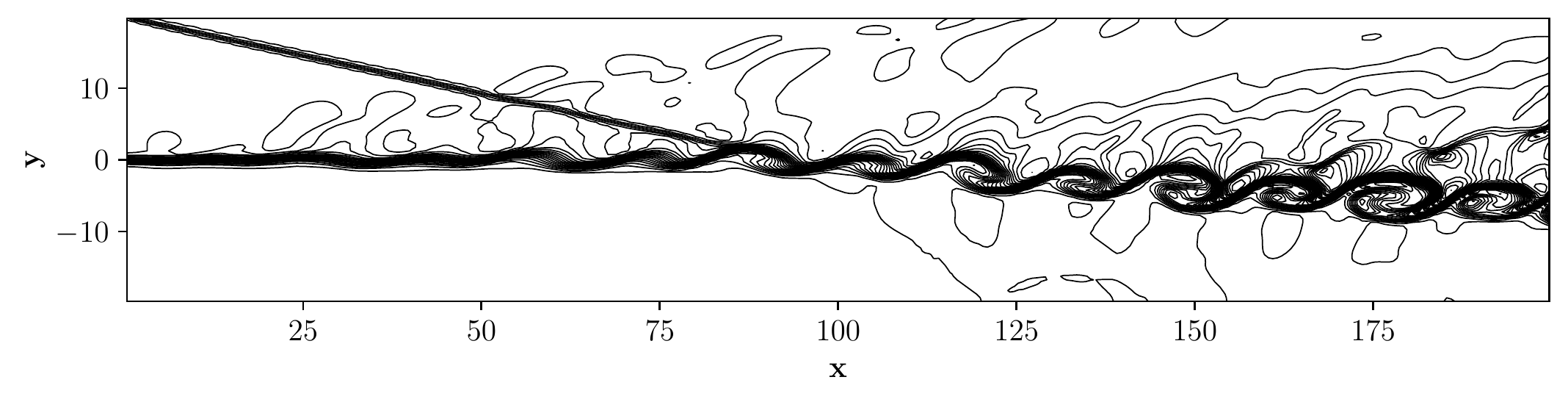}
\label{fig:TENO_SSL_fine}}
\subfigure[MEG7-cons, $320 \times 80$]{\includegraphics[width=0.8\textwidth]{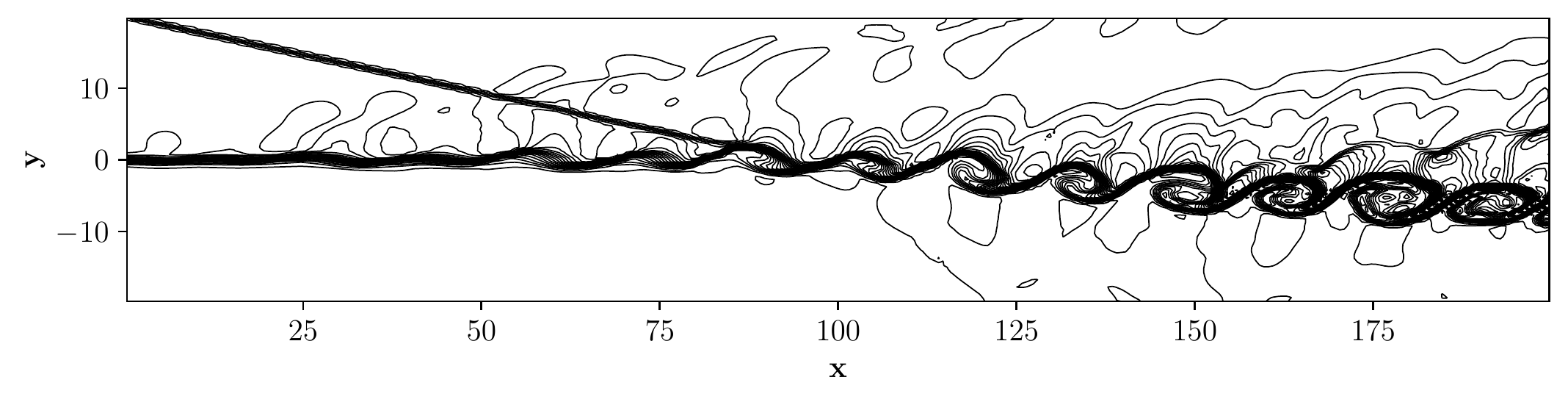}
\label{fig:MEG_SSL_fine}}
\subfigure[MEG7-prim, $320 \times 80$]{\includegraphics[width=0.8\textwidth]{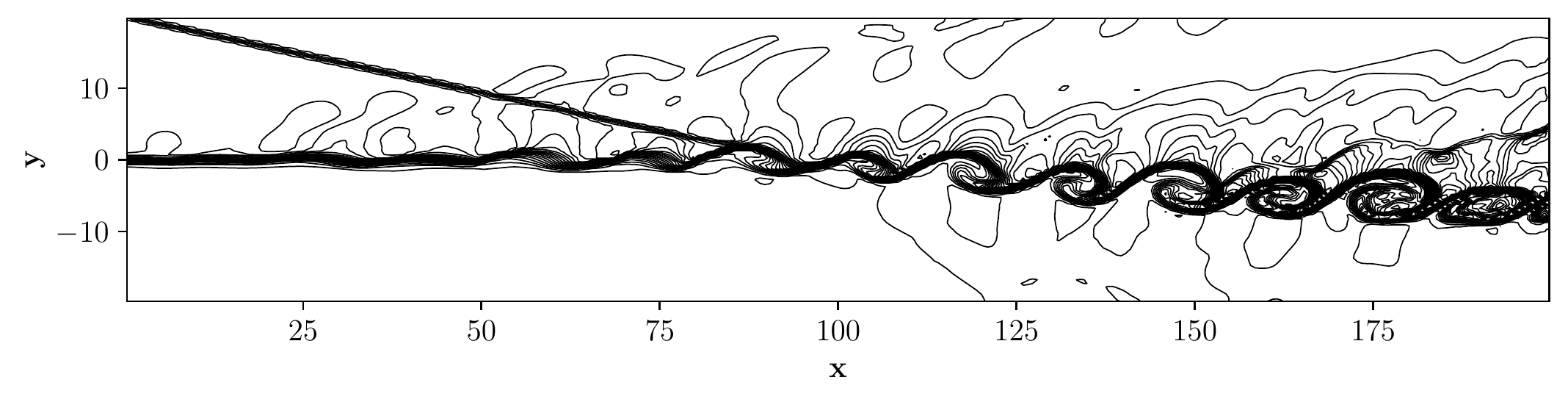}
\label{fig:MIG_SSLc}}
\caption{Computed density contours of the shock-shear layer interaction described in Example \ref{ex:ssl} on grid size of $320 \times 80$ for the proposed schemes and WCNS7M.}
\label{fig_SSL}
\end{figure} 
For the initial conditions, The initial stream-wise velocity is set as a hyperbolic tangent profile:

\begin{equation}
    u = 2.5 + 0.5 \tanh(2y)
\end{equation}
\noindent Fluctuations are added to the wall-normal velocity at the inlet:

\begin{equation}
    v' = \sum^{2}_{k=1} \alpha_k \cos(2 \pi k t/T + \phi_k) \exp(-y^2/b)
\end{equation}

\noindent where the period, $T = \lambda/u_c$, the wavelength, $\lambda = 30$, the convective velocity $u_c = 2.68$, and $b = 10$. When $k = 1$, $\alpha_1 = 0.05$ and $\phi = 0$. When $k = 2$, $\alpha_2 = 0.05$ and $\phi = \pi/2$. The boundary conditions for this test case are the same as that of Yee et al. \cite{yee1999low} and the simulations are carried out on a grid size of 320 $\times$ 80. Fig. \ref{fig_SSL} displays the density contours of the shock-shear layer case, and all schemes resolve the oblique shock without any oscillations. However, looking closely, the vortices in the shear layer are better resolved using the MEG7-cons and MEG7-prim schemes compared to the WCNS7M. It shows that the current approach and the hybrid algorithm for high-order accuracy can be used in various complex test cases.

\subsection{Computational efficiency}

Table \ref{tab:speed} shows the proposed scheme's computational efficiency compared with the WCNS7M scheme. Normalized values with respect to MEG7-prim are shown in the parenthesis. All the simulations are conducted on the same computer with the same compiler flags. The results indicate that the proposed scheme is at least 40$\%$ cheaper than the WCNS7M scheme, on average, and produces similar or improved results for all the test cases. Furthermore, for the viscous shock-tube test case, the proposed scheme is twice as fast where the sharing of gradients between the convective and viscous fluxes is possible. Given this superior performance and flexibility of the scheme regarding sharing of gradients and genuinely high-order accuracy, it is believed that the proposed method is a competitive scheme and can be used for many applications.

\begin{table}[H]
  \centering
  \caption{Computational efficiency of the proposed scheme for the test cases considered earlier.}
    \begin{tabular}{|c|c|c|c|}
    \hline
    Test case & MEG7-prim & MEG7-cons & WCNS7M \\
    \hline
    \hline
    Richtmeyer - Meshkov instability   & 224 s (1) & 205 s (0.915) & 345 s (1.54) \\
    \hline
    Riemann Problem & 1025 s (1) & 1031 \textcolor{black}{s} (1.005) & 1502 s (1.46) \\
    \hline
    Double Mach reflection   & 1806 s (1) & 1946 s (1.07) & 2493 s (1.38) \\
    \hline
    Viscous shock-tube   & 35455 s (1) & 38062 s (1.076) & 70082 s (1.97) \\
    \hline
    \end{tabular}%
  \label{tab:speed}%
\end{table}%

\textcolor{black}{\subsection{Comparison between fourth-order and seventh-order GRB schemes}\label{sec:comparison}}

\textcolor{black}{In Ref. \cite{chamarthi2023gradient}, Chamarthi has presented a fourth-order GRB scheme. Such a scheme is derived by assuming $\alpha$=4 and $\beta$=0 in Equations (\ref{good_6tth-order_scheme_flux}) and (\ref{alpha-beta}), respectively, and the gradients are still computed by sixth-order derivatives. Upon substituting the values and simplifications, one would obtain the following formula}

\textcolor{black}{
 \begin{equation}\label{fourth}
 \mathbf{F(f_L)} = \frac{1}{1440} (1050 \mathbf{f}_j-309 \mathbf{f}_{j-1}+64 \mathbf{f}_{j-2}-3 \mathbf{f}_{j-3}-\mathbf{f}_{j-4}+771 \mathbf{f}_{j+1}-152 \mathbf{f}_{j+2}+21 \mathbf{f}_{j+3}-\mathbf{f}_{j+4}),
 \end{equation}}

\textcolor{black}{It can be seen that the above formula has the same stencil as that of the seventh-order approach presented earlier in this paper, and the only difference is the values of the coefficients in the formula. It is possible that a higher-order scheme can have poor spectral properties and may lead to poor results. Fig. \ref{fig_disp_unlim} shows the dispersion and dissipation plots of the fourth- and seventh-order schemes. It can be seen that the seventh-order scheme has lower dissipation properties than the fourth-order scheme. Seventh-order scheme has two advantages over the fourth-order scheme, order of accuracy and low dissipation. The fourth-order scheme has a slightly better dispersion property.}

\begin{figure}[H]
\centering
\subfigure[]{\includegraphics[width=0.45\textwidth]{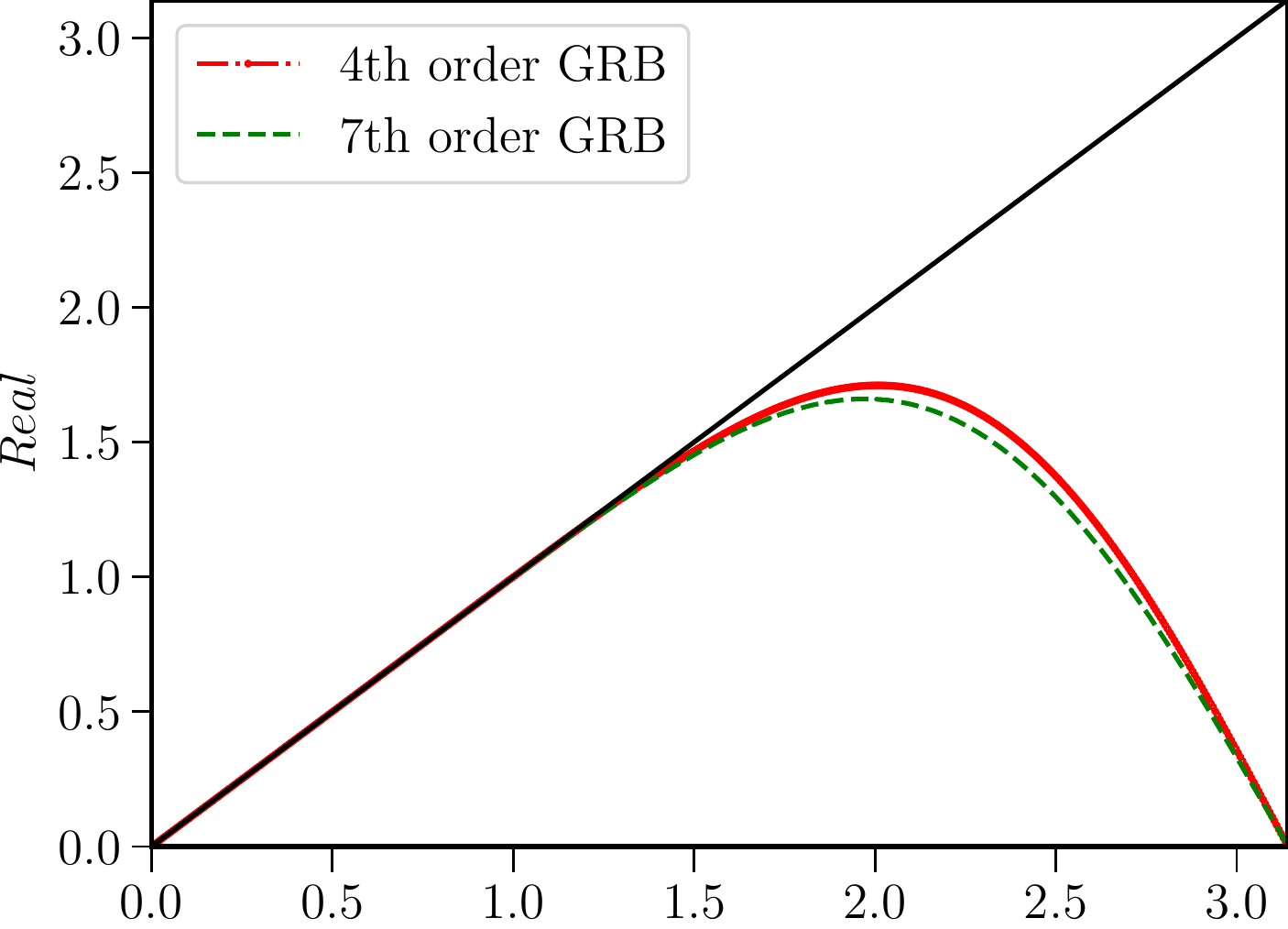}
\label{fig:dispersion_u}}
\subfigure[]{\includegraphics[width=0.47\textwidth]{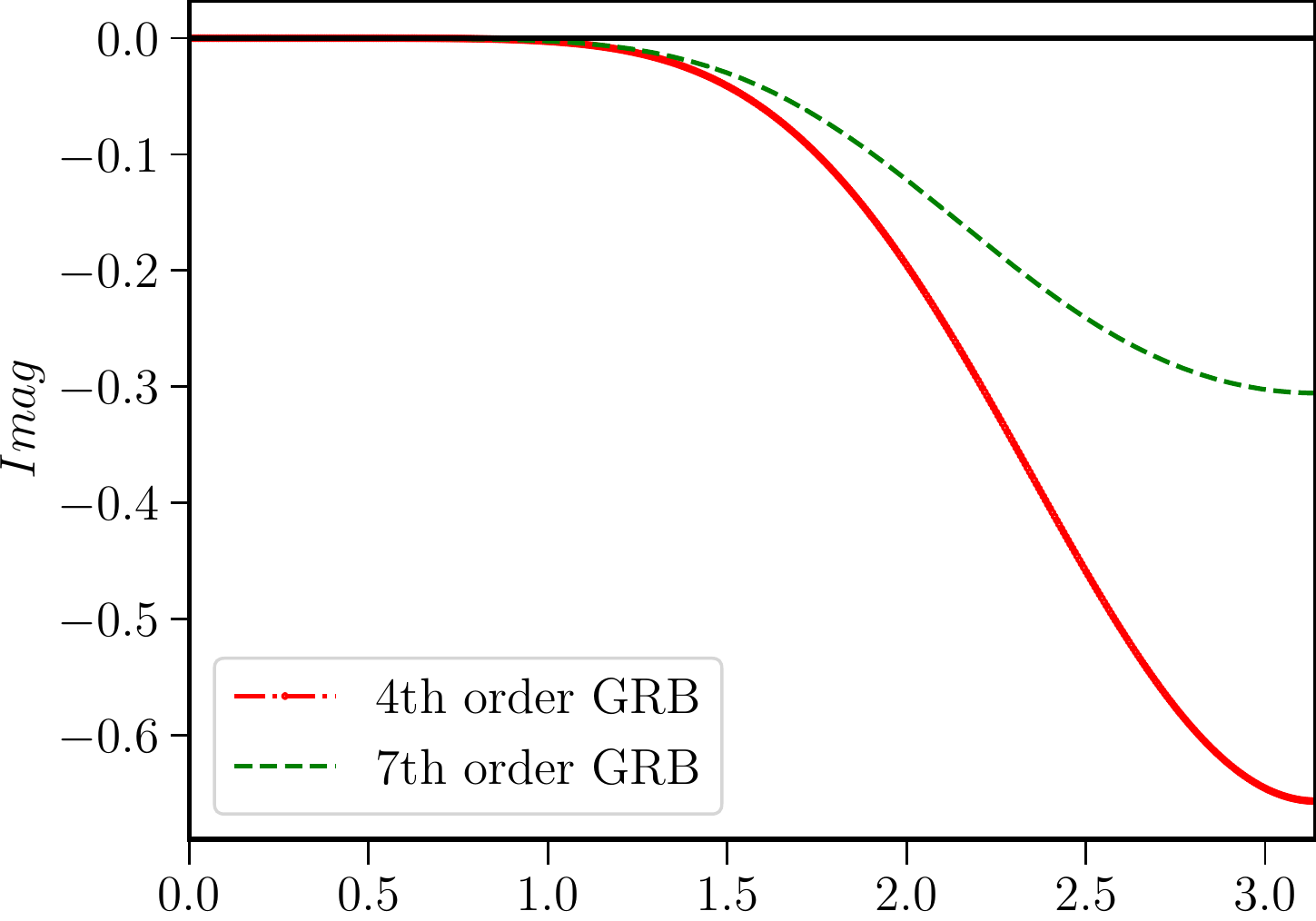}
\label{fig:dissipation_u}}
\caption{Dispersion and Dissipation properties of the unlimited schemes.}
\label{fig_disp_unlim}
\end{figure}

\textcolor{black}{To numerically analyze the dissipation properties of the fourth- and seventh-order order schemes, we consider the inviscid Taylor-Green vortex problem  with the following initial conditions:}

\textcolor{black}{\begin{equation}\label{itgv}
\begin{pmatrix}
\rho \\
u \\
v \\
w \\
p \\
\end{pmatrix}
=
\begin{pmatrix}
1 \\
\sin{x} \cos{y} \cos{z} \\
-\cos{x} \sin{y} \cos{z} \\
0 \\
100 + \frac{\left( \cos{(2z)} + 2 \right) \left( \cos{(2x)} + \cos{(2y)} \right) - 2}{16}
\end{pmatrix}.
\end{equation}}

\textcolor{black}{This test case is used to evaluate the different schemes' ability to preserve the kinetic energy and the growth of enstrophy in time.
Simulations are carried out on a domain of size $x,y,z \in [0,2\pi)$ with periodic boundary conditions applied for all boundaries until time $t=10$ The kinetic energy evolution and enstrophy evolution of both the numerical schemes are shown in Figs. \ref{fig:TGV_KE} and \ref{fig:TGV_ens}, respectively.}
\textcolor{black}{\begin{figure}[H]
\centering
\subfigure[Kinetic energy]{\includegraphics[width=0.45\textwidth]{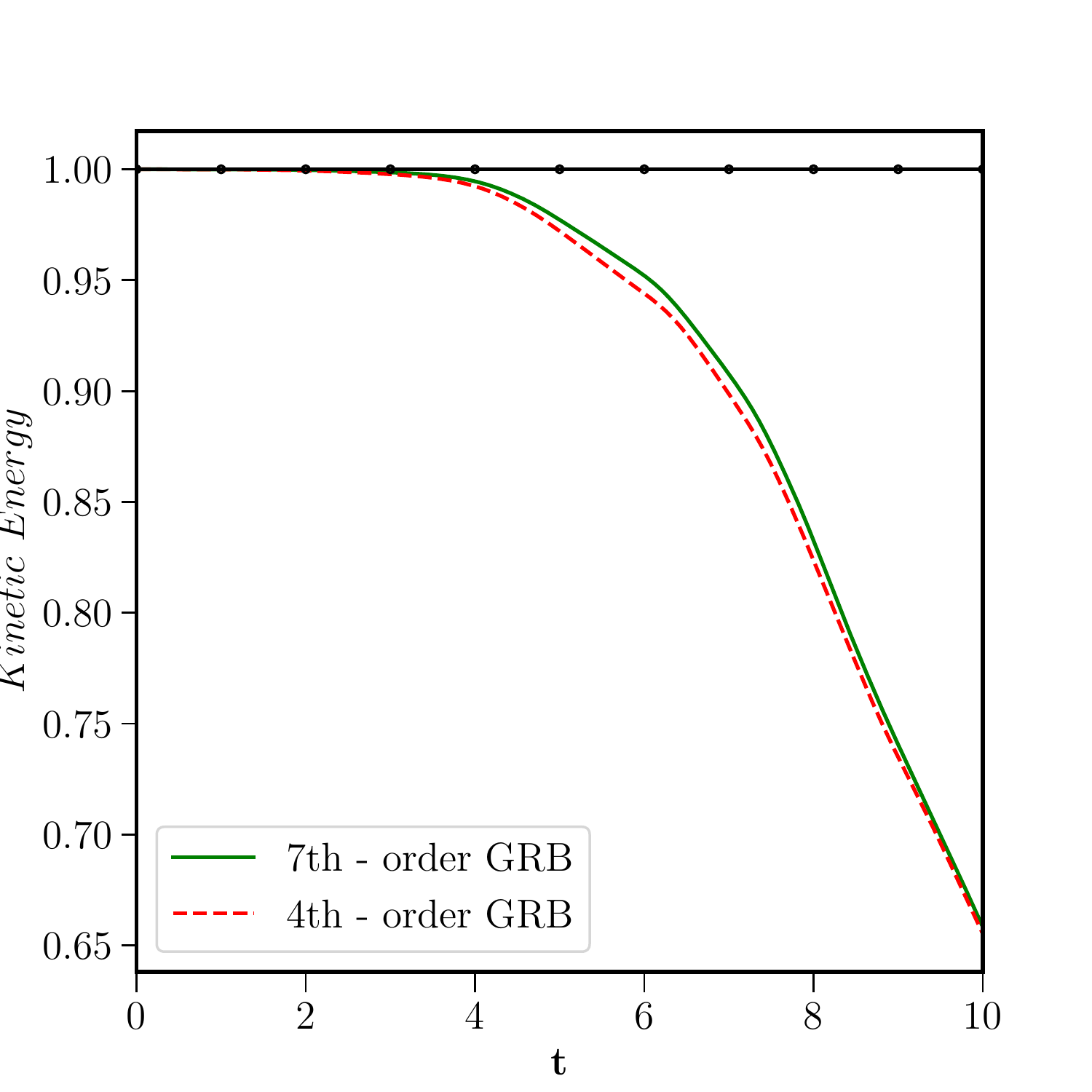}
\label{fig:TGV_KE}}
\subfigure[Enstrophy]{\includegraphics[width=0.45\textwidth]{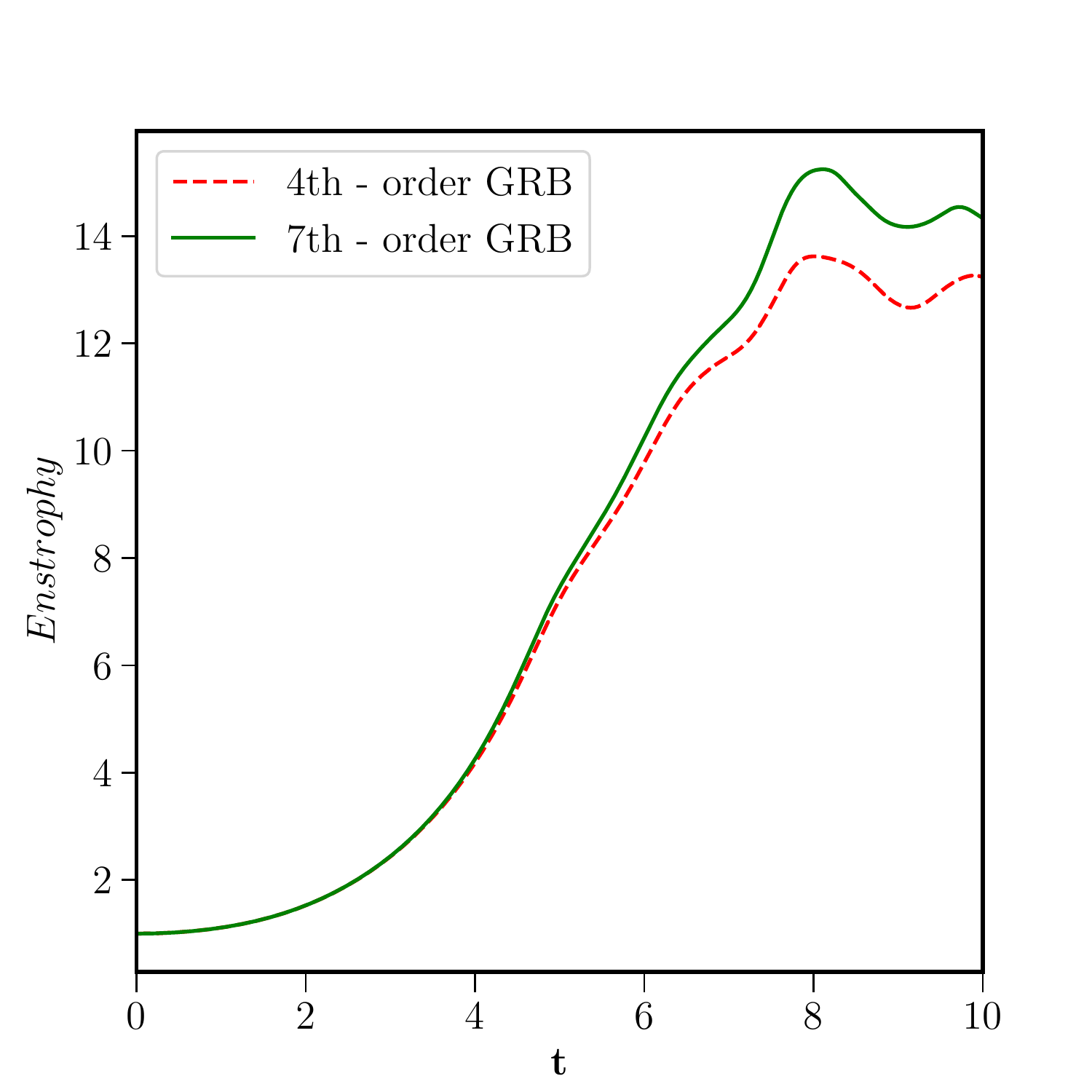}
\label{fig:TGV_ens}}
\caption{\textcolor{black}{Normalised kinetic energy and enstrophy for the fourth- and seventh-order GRB schemes on grid size of $64^3$. Solid line with circles: exact solution; solid green line: Seventh-order GRB; dashed red line: Fourth-order GRB.}}
\label{fig_TGV}
\end{figure}}

\textcolor{black}{It can be observed from these figures that the seventh-order scheme better preserves the kinetic energy and the values of the enstrophy are also better than the fourth-order scheme. Therefore, the proposed scheme is not inferior to the scheme presented in Ref. \cite{chamarthi2023gradient}.}\\

\section{Conclusions \textcolor{black} {and future work}}\label{sec:finish}

\textcolor{black}{The contributions in this paper are as follows:} 

\begin{itemize}
\item Under the assumption that all the cell-centre values are point values, It is shown that it is possible to obtain not only fourth-order accuracy can be obtained \textcolor{black} {by using the Equation (\ref{eqn:legendre})}, as in Ref. \cite{chamarthi2023gradient}, but also possible to obtain seventh-order accuracy. \textcolor{black}{The seventh-order scheme also has low dissipation properties compared to the fourth-order scheme in Ref. \cite{chamarthi2023gradient}.}
\item Unlike the earlier work \cite{chamarthi2023gradient} where the scheme is only second-order accurate for nonlinear problems, in the current paper, genuine high-order accuracy, i.e., seventh-order, is obtained. Such high-order accuracy is possible by flux reconstruction, which is carried out only in the smooth regions using a problem-independent discontinuity detector.
\item Sharing of gradients between the inviscid and viscous fluxes, which is the main objective of the gradient-based reconstruction, is still possible, which makes the proposed scheme very efficient.
\end{itemize}

\textcolor{black} {Viscous flux discretization considered in this paper is not truly high-order accurate for non-linear test cases, and such a scheme may be derived and its benefits need to be analyzed. The original objective of the GRB approach is to reuse the gradients in both viscous and inviscid fluxes. In the current approach, the viscous and inviscid fluxes are considered independently, but optimizing them together may be possible to improve the discretization further.} The extension of the present method to curvilinear coordinates is straightforward, facilitating simulations of complex geometries, and is presented elsewhere. \textcolor{black} {Furthermore,} the gradient-based reconstruction can be a promising approach for multi-component reacting flow simulations \cite{ziegler2011adaptive,cai2021mechanism}. With the sharing of the gradients between convective and viscous fluxes, the gradients of species mass fractions need to be computed only once, making it a very efficient approach to reacting flow simulations.

\section*{Appendix A}

Fig. \ref{fig:Mathe} shows the MATHEMATICA workbook detailing the derivation of the seventh-order gradient-based reconstruction. For $\alpha$ = 4 and $\beta$ = 0 one would obtain fourth-order gradient-based reconstruction presented in Ref. \cite{chamarthi2023gradient}. The author will make the original MATHEMATICA workbook available upon reasonable request.


\begin{figure}[H]
    \centering
    \includegraphics[width=1.0\textwidth]{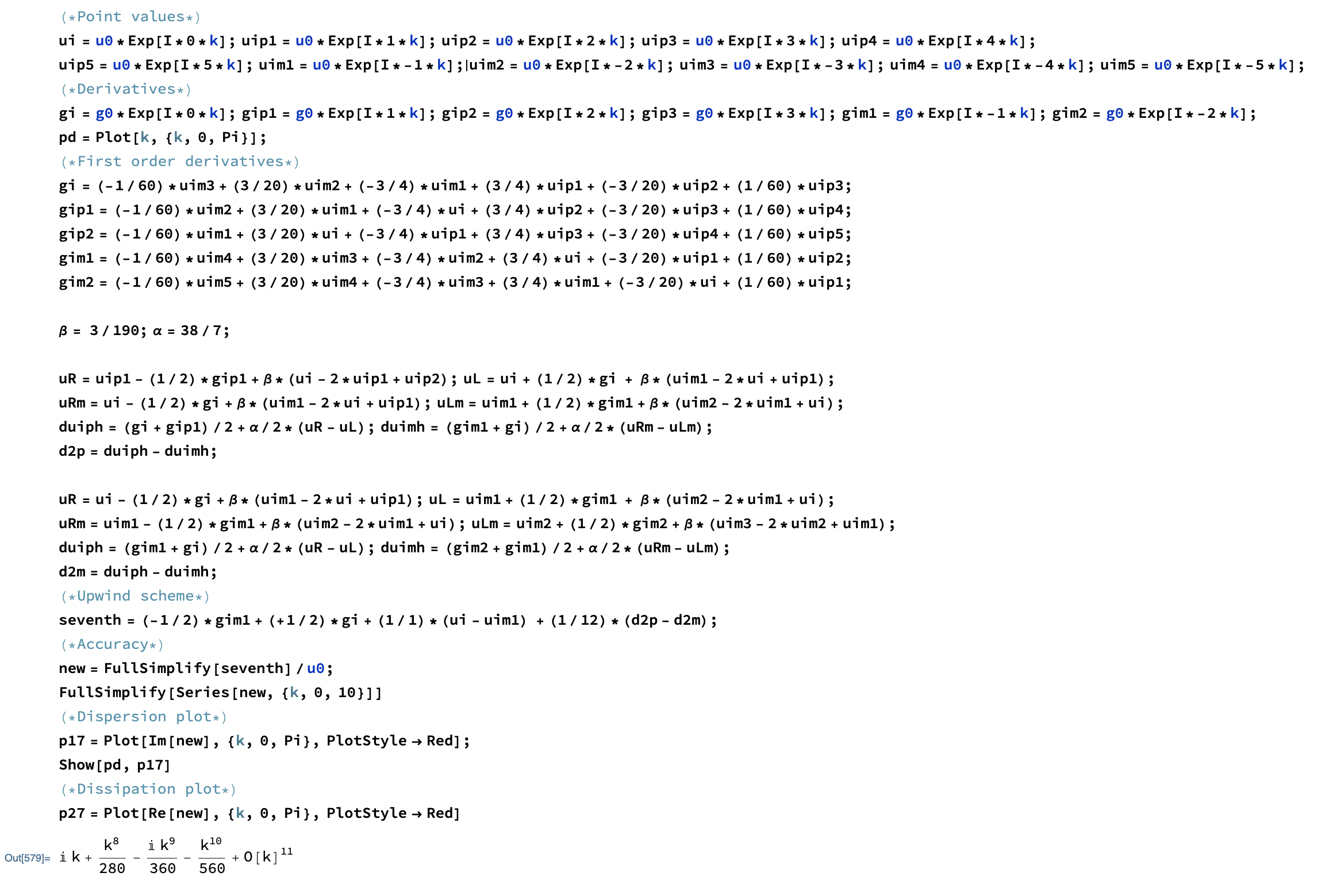}
    \caption{Mathematica Workbook showing order of accuracy of the seventh order scheme.}
    \label{fig:Mathe}
\end{figure}


\section*{Acknowledgements}
A.S. thanks his wife and kid for their support despite the difficult situations \textcolor{black}{(struggling with tuberculosis and subsequent lack of medical care, inability to go to school, near poverty, lack of food, death of family members, etc.)} they faced over the last eight years. \textcolor{black}{A.S. apologizes to anyone who finds any part of this paper inconvenient or distressing.}

\bibliographystyle{elsarticle-num}
\bibliography{bvd_ref}

\end{document}